\documentclass[11pt,a4paper]{article} 
\usepackage{jheppub} 
\usepackage{multirow} 

\newcommand\Ref[1] {Ref.\,\cite{#1}} 
\newcommand\Refs[1] {Refs.\,\cite{#1}}
 
\newcommand\eqn[1] {Eq.\,(\ref{#1})} 
\newcommand\eqns[2] {Eqs.\,(\ref{#1}) and~(\ref{#2})} 
\newcommand\eqnss[2] {Eqs.\,(\ref{#1})--(\ref{#2})} 

\newcommand\sect[1] {Section~\,{\ref{#1}}} 

\newcommand\App[1] {Appendix~\ref{#1}} 

\newcommand\tab[1] {Table~\ref{#1}} 
 
\newcommand\tabss[2] {Tables~\ref{#1}--\ref{#2}} 
 
\def\beq{\begin{equation}} 
\def\eeq{\end{equation}} 
\def\bsp#1\esp{\begin{split}#1\end{split}} 
\def\bal#1\eal{\begin{align}#1\end{align}} 
\newcommand\nt {\notag} 

\newcommand\tsig[1] {\sigma^{\mathrm{#1}}} 
\newcommand\dsig[1] {\rd\sigma^{{\rm #1}}} 
\newcommand\dsiga[2] {\rd\sigma^{{\rm #1,A}_{\scriptscriptstyle #2}}} 

\newcommand\la {\langle} 
\newcommand\ra {\rangle} 
\newcommand\bra[3] {\la {\cal M}_{#1}^{#2}#3|} 
\newcommand\ket[3] {|{\cal M}_{#1}^{#2}#3\ra} 
\newcommand{\cM} {{\cal M}} 
\newcommand\M[2] {\ensuremath{|{\cal{M}}_{#1}^{#2}|^2}} 
\newcommand\SME[3] {|{\cal M}_{#1}^{(#2)}{(#3)}|^2} 
 
\newcommand{\PS}[1] {\rd\phi_{#1}} 
\newcommand{\rd} {{\mathrm{d}}} 
\newcommand{\mom}[1] {\{p\}^{#1}} 
\newcommand{\momt}[1] {\{\ti{p}\}^{#1}} 
\newcommand{\momh}[1] {\{\ha{p}\}^{#1}} 
\newcommand{\cmap}[1] {\stackrel{{\cal C}_{#1}}{\longrightarrow}} 
\newcommand{\smap}[1] {\stackrel{{\cal S}_{#1}}{\longrightarrow}} 

\newcommand\tzz[2] {z_{#1,#2}}
\newcommand\tvv[2] {v_{#1,#2}}
\newcommand\kT[1] {k_{\perp,#1}} 
\newcommand\kTtt[1] {\tilde{k}_{\perp,#1}}

\newcommand{\calS} {{\cal S}} 
\newcommand{\cA}[1] {{\cal A}_{#1}}
\newcommand{\cC}[2] {{\cal C}_{#1}^{#2}} 
\newcommand{\cS}[2] {{\cal S}_{#1}^{#2}} 
\newcommand{\cSCS}[2] {{\cal C}\kern-2pt{\cal S}_{#1}^{#2}} 
 
\newcommand{\IcC}[2] {{\mathrm C}_{#1}^{#2}} 
\newcommand{\IcS}[2] {{\mathrm S}_{#1}^{#2}} 
\newcommand{\IcSCS}[2] {{\mathrm C}\!{\mathrm S}_{#1}^{#2}} 
\newcommand{\bI} {\bom{I}} 

\newcommand\as {\ensuremath{\alpha_{\mathrm{s}}}} 

\newcommand{\CF} {C_{\mathrm{F}}} 
\newcommand{\CA} {C_{\mathrm{A}}} 
\newcommand{\TR} {T_{\mathrm{R}}} 
\newcommand{\Nc} {N_{\mathrm{c}}} 
\newcommand{\Nf} {n_{\mathrm{f}}} 
 
\newcommand{\gam}[1] {\gamma_{#1}} 
\newcommand{\bT} {\bom{T}} 
\newcommand\qb {{\bar q}} 

\newcommand\Oe[1] {\ensuremath{\mathrm O(\ep^{#1})}} 
\newcommand{\cI} {{\cal I}} 
\def\hP{\hat{P}} 

\newcommand{\ep} {\ensuremath{\epsilon}} 
\newcommand{\eps} {\varepsilon} 
\newcommand\al {\ensuremath{\alpha}} 
\newcommand\be {\ensuremath{\beta}} 
\newcommand\vth {\ensuremath{\vartheta}} 
\newcommand\vph {\ensuremath{\varphi}} 

\newcommand\KKone  	{j} 
\newcommand\KKtwo  	{k} 
\newcommand\KKthree  {l} 
\newcommand\LL  {m} 
\newcommand\LLone  	{l} 
\newcommand\LLtwo  	{m} 

\newcommand\COEFFJ {} 
\newcommand\COEFFK {} 
\newcommand\COEFFIR {} 
 
\newcommand\bom[1] {{\mbox{\boldmath $#1$}}}  

\newcommand{\ti}[1] {\tilde{#1}} 
\newcommand{\wti}[1] {\widetilde{\,#1\,}} 
\newcommand{\ha}[1] {\hat{#1}} 
\newcommand{\wha}[1] {\widehat{\,#1\,}} 

\newcommand\msbar {\ensuremath{{\overline {\rm MS}}}} 
\def\ldot{\!\cdot\!} 
\def\AP{Altarelli--Parisi } 
\def\MB{Mellin--Barnes } 
\def\qqquad{\qquad\quad} 
\def\qqqquad{\qquad\qquad} 

\newcommand\ID {{\mathrm{id}}} 
\newcommand\AB {{\mathrm{ab}}} 
\newcommand\NAB {{\mathrm{nab}}} 
 
\newcommand\Yt[2]  {Y_{\wti{#1}\wti{#2},Q}}

\title{ 
Integration of collinear-type doubly unresolved \\
counterterms in NNLO jet cross sections 
} 
 
\author[a]{Vittorio Del Duca,} 
\author[b]{G\'abor Somogyi} 
\author[c]{and Zolt\'an Tr\'ocs\'anyi} 
 
\affiliation[a]{ 
Istituto Nazionale di Fisica Nucleare, 
Laboratori Nazionali di Frascati,\\ 
Via E. Fermi 40, I-00044 Frascati, Italy 
} 
\affiliation[b]{ 
PH Department, TH Unit, CERN, CH-1211 Geneva 23, Switzerland} 
\affiliation[c]{ 
University of Debrecen and MTA-DE Particle Physics Research Group 
\\ H-4010 Debrecen, PO Box 105, Hungary} 
 
\emailAdd{Vittorio.DelDuca@lnf.infn.it} 
\emailAdd{Gabor.Somogyi@cern.ch} 
\emailAdd{Zoltan.Trocsanyi@cern.ch} 
 
\abstract{In the context of a subtraction method for jet cross sections
at NNLO accuracy in the strong coupling, we perform the integration 
over the two-particle factorised phase space of 
the collinear-type contributions to the doubly unresolved counterterms. 
We present the final result as a convolution in colour space of the 
Born cross section and of an insertion operator, which is written in terms 
of master integrals that we expand in the dimensional regularisation
parameter.} 
 
 
\arxivnumber{arXiv:yymm.nnnn} 
 
\preprint{CERN-PH-TH/2013-005} 
 
\begin{document}

 
\maketitle 
\flushbottom 

 
\section{Introduction} 
\label{sec:introduction} 
 
The high precision of the experimental measurements at the LHC calls 
for an evaluation of the jet, heavy-quark, vector-boson and Higgs-boson 
production rates at hadron colliders which is at least as precise. In 
recent years, we have witnessed fast progress in the evaluation of the 
production rates mentioned above, in association with many jets, at 
next-to-leading order (NLO) accuracy in the strong coupling constant 
$\as$. That rested upon efficient methods to compute one-loop amplitudes 
with many legs, and upon subtraction algorithms to evaluate QCD cross 
sections at NLO at the partonic level~\cite{Frixione:1995ms,Catani:1996vz,
Nagy:1996bz,Somogyi:2006cz,Somogyi:2009ri}, 
or at the hadronic level, through interfaces, like MC@NLO~\cite{Frixione:2002ik} 
or POWHEG~\cite{Nason:2006hfa,Frixione:2007nw,Frixione:2007vw} with 
parton-shower event generators. Such algorithms are based on the 
universality --- i.e.,~on the independence from a specific scattering 
process --- of the infrared emissions.

Lately, the evaluation of production rates at next-to-next-to-leading 
order (NNLO) accuracy has received a lot of attention. Fully differential 
cross sections for vector-boson~\cite{Melnikov:2006kv,Catani:2009sm}, 
Higgs-boson~\cite{Anastasiou:2004xq,Catani:2007vq}, diphoton~\cite{Catani:2011qz} 
and Higgs--vector-boson~\cite{Ferrera:2011bk} production have been evaluated, 
and the computation of the NNLO corrections to top-pair production is 
well under way~\cite{Baernreuther:2012ws,Czakon:2012zr,Czakon:2012pz}.
Much work has gone into the lay-out of a general subtraction algorithm to 
compute cross sections at NNLO accuracy~\cite{Frixione:2004is,
Somogyi:2005xz,GehrmannDeRidder:2005cm,Somogyi:2006da,Somogyi:2006db,
Daleo:2006xa,Somogyi:2008fc,Aglietti:2008fe,Somogyi:2009ri,Bolzoni:2009ye,
Daleo:2009yj,Glover:2010im,Czakon:2010td,Bolzoni:2010bt,Abelof:2011jv,
Gehrmann:2011wi,GehrmannDeRidder:2011aa,GehrmannDeRidder:2012ja,Abelof:2012he,
Ridder:2012dg}.
The antenna scheme~\cite{GehrmannDeRidder:2005cm} has yielded the evaluation 
of total rates~\cite{GehrmannDeRidder:2007jk,GehrmannDeRidder:2008ug,
Weinzierl:2008iv,Weinzierl:2009nz} and event 
shapes~\cite{GehrmannDeRidder:2007bj,GehrmannDeRidder:2007hr,
GehrmannDeRidder:2009dp,Weinzierl:2009ms,Weinzierl:2009yz} in 
electron-positron annihilation, leading to precise determinations of
the strong coupling~\cite{Dissertori:2007xa,Dissertori:2009qa,
Dissertori:2009ik,Schieck:2012mp}.

No subtraction algorithm to compute cross sections at hadron colliders 
at NNLO accuracy, though, has been devised yet. In order to do that, one 
must define subtraction terms that properly regularise the real-emission 
phase-space integrals and then one must combine the integrated form of 
those counterterms with the virtual contributions, so as to cancel the 
infrared divergences of the loop amplitudes, in such a way that the 
cancellation of both the kinematic singularities in the real-emission 
pieces and the explicit $\ep$-poles in the virtual pieces be local. This 
implies that the subtraction terms and the real-emission contributions 
must tend to the same value in all kinematic limits where the latter diverge, 
and that the cancellation of explicit $\ep$-poles between the integrated 
subtraction terms and the virtual contributions must take place point-wise 
in phase space. In particular, that implies that it is possible to write 
the integrated counterterms in such a way that they can be explicitly combined 
with virtual contributions, before phase-space integration. Practically, 
the locality of the subtraction scheme is also important to ensure good 
numerical efficiency of the algorithm. In broad outline, we remind how this 
occurs at NLO. After fixing the leading-order $m$-jet cross section as the 
integral of the fully differential Born cross section $\dsig{B}_m$ of $m$ 
final-state patrons over the available $m$-parton phase space defined by 
the jet function $J_m$,
\beq 
\tsig{LO} = 
	\int_m\!\dsig{B}_m J_m\,,
\label{eq:sigmaLO} 
\eeq 
the NLO correction,
\beq 
\tsig{NLO} = 
	\int_{m+1}\!\dsig{R}_{m+1} J_{m+1} + \int_m\!\dsig{V}_m J_m\,,
\label{eq:sigmaNLO} 
\eeq 
can be rewritten as a sum of finite integrals,
\beq
\tsig{NLO} =  
	\int_{m+1}\dsig{NLO}_{m+1} + \int_m\dsig{NLO}_m\,,
\label{eq:sigmaNLOfin} 
\eeq
where each of the two terms on the right-hand side,
\bal
\dsig{NLO}_{m+1} &= 
	\left[ \dsig{R}_{m+1} J_{m+1} 
	- \dsiga{R}{1}_{m+1} J_{m}\right]_{\ep=0} \,,
\\
\dsig{NLO}_m &= 
	\left[ \left(\dsig{V}_m 
	+ \int_1\dsiga{R}{1}_{m+1}\right) J_m\right]_{\ep=0} \,,
\eal
is integrable in four dimensions by 
construction~\cite{Frixione:1995ms,Catani:1996vz,Nagy:1996bz,Somogyi:2006cz,
Somogyi:2009ri}.

Likewise, we can write the NNLO correction to the cross section as 
a sum of three contributions, the doubly unresolved, the one-loop
singly unresolved, and the two-loop doubly virtual terms,
\beq 
\tsig{NNLO} = 
	\int_{m+2}\dsig{RR}_{m+2} J_{m+2} 
	+ \int_{m+1}\dsig{RV}_{m+1} J_{m+1} 
	+ \int_m\dsig{VV}_m J_m\,,
\label{eq:sigmaNNLO} 
\eeq 
and rearrange it as follows,
\beq 
\tsig{NNLO} = 
	\int_{m+2}\dsig{NNLO}_{m+2} 
	+ \int_{m+1}\dsig{NNLO}_{m+1} 
	+ \int_m\dsig{NNLO}_m\,,
\label{eq:sigmaNNLOfin} 
\eeq 
where each of the three terms on the right-hand side,
\bal
\dsig{NNLO}_{m+2} &= 
	\Big\{\dsig{RR}_{m+2} J_{m+2} 
	- \dsiga{RR}{2}_{m+2} J_{m} 
	-\Big[\dsiga{RR}{1}_{m+2} J_{m+1} 
	- \dsiga{RR}{12}_{m+2} J_{m}\Big]\Big\}_{\ep=0}\,, 
\label{eq:sigmaNNLOm+2} 
\\ 
\dsig{NNLO}_{m+1} &= 
	\Big\{\Big[\dsig{RV}_{m+1} 
	+ \int_1\dsiga{RR}{1}_{m+2}\Big] J_{m+1}  
	-\Big[\dsiga{RV}{1}_{m+1} 
	+ \Big(\int_1\dsiga{RR}{1}_{m+2}\Big)\strut^{{\rm A}_{\scriptscriptstyle 1}} 
	\Big] J_{m} \Big\}_{\ep=0}\,,
\label{eq:sigmaNNLOm+1} 
\\
\dsig{NNLO}_{m} &= 
	\Big\{\dsig{VV}_m 
	+ \int_2\Big[\dsiga{RR}{2}_{m+2} 
	- \dsiga{RR}{12}_{m+2}\Big] 
	+\int_1\Big[\dsiga{RV}{1}_{m+1} 
	+ \Big(\int_1\dsiga{RR}{1}_{m+2}\Big) \strut^{{\rm A}_{\scriptscriptstyle 1}} 
	\Big]\Big\}_{\ep=0} J_{m}\,,
\label{eq:sigmaNNLOm}
\eal
is integrable in four dimensions by 
construction~\cite{Somogyi:2005xz,Somogyi:2006da,Somogyi:2006db}.

The counterterms which contribute to $\dsig{NNLO}_{m+2}$ and to 
$\dsig{NNLO}_{m+1}$ were introduced in \Refs{Somogyi:2006da} 
and \cite{Somogyi:2006db}, respectively. The integral of the real-virtual 
counterterms (the last two terms of \eqn{eq:sigmaNNLOm}) was performed in 
Refs.~\cite{Somogyi:2008fc,Aglietti:2008fe,Bolzoni:2009ye}. The integral 
of the iterated singly unresolved counterterm (the third term of 
\eqn{eq:sigmaNNLOm}) was performed in \Ref{Bolzoni:2010bt}. In this paper, 
we compute the integral of the collinear-type contributions to the doubly 
unresolved counterterm (the second term of \eqn{eq:sigmaNNLOm}). The 
soft-type contributions to the same integral are presented in a companion 
paper~\cite{Somogyi:inprep}.
 
 
\section{Notation} 
\label{sec:notation} 

In this paper, we use the notation introduced in Ref.~\cite{Bolzoni:2010bt}, 
which we recall in this Section.

%
%
 
\subsection{Matrix elements} 
\label{sec:ME} 
 
We consider processes with coloured particles (partons) in the final 
state, while the initial-state particles are colourless (typically 
electron-positron annihilation into hadrons).  Any number of additional 
non-coloured final-state particles are allowed, too, but they will be 
suppressed in the notation. Resolved partons in the final state are labeled  
with letters chosen from the middle of the alphabet, $i$, $j$, $k$, $l$, 
$\dots$, while letters  $r$, $s$ denote  unresolved final-state partons. 
 
We adopt the colour- and spin-state notation of \Ref{Catani:1996vz}.
In that notation the amplitude $\ket{}{}{}$ for a scattering
process is an abstract vector in colour and spin space, and its
normalisation is fixed such that the squared amplitude summed over
colours and spins is 
\beq 
|\cM|^2 = 
	\bra{}{}{}\ket{}{}{}\,. 
\label{eq:M2} 
\eeq 
In this paper we only need $\ket{}{}{}$ in the Born approximation,
denoted by $\ket{}{(0)}{}$.
 
Using the colour-state notation, we define the two-parton colour-correlated 
squared tree amplitudes as 
\beq 
|\cM^{(0)}_{(i,k)}|^2 \equiv 
	|\cM^{(0)}|^2 \otimes \bT_i \bT_k \equiv  
	\bra{}{(0)}{} \,\bT_i \ldot \bT_k \, \ket{}{(0)}{} 
\label{eq:colam2} 
\eeq 
and similarly the four-parton colour-correlated squared tree amplitudes,  
\beq 
|\cM^{(0)}_{(i,k),(j,l)}|^2 \equiv 
	|\cM^{(0)}|^2 \otimes \{\bT_i \ldot \bT_k, \bT_j \ldot \bT_l\} \equiv  
	\bra{}{(0)}{} \{\bT_i \ldot \bT_k, \bT_j \ldot \bT_l\} \ket{}{(0)}{}\,,
\label{eq:colam4} 
\eeq 
where we also introduced the $\otimes$ product notation to indicate  
the insertion of colour charge operators between $\bra{}{(0)}{}$ and
$\ket{}{(0)}{}$. The colour-charge algebra for the product  
$\sum_{n} (\bT_i)^n (\bT_k)^n \equiv \bT_i \ldot \bT_k$ is: 
\beq 
\bT_i \ldot \bT_k =
	\bT_k \ldot \bT_i \quad  {\rm if} \quad i \neq k; 
	\qquad \bT_i^2= C_{f_i}\,. 
\label{eq:colalg} 
\eeq 
Here $C_{f_i}$ is the quadratic Casimir operator in the representation 
of particle $i$ and we have 
$C_q \equiv \CF= \TR(\Nc^2-1)/\Nc= (\Nc^2-1)/(2\Nc)$ in the fundamental 
and $C_g \equiv \CA=2\,\TR \Nc=\Nc$ in the adjoint representation, 
with $\TR=1/2$.   
 
We also use squared colour charges with multiple indices, such as 
$\bT_{ir}^2\equiv C_{f_{ir}}$ and $\bT_{irs}^2\equiv C_{f_{irs}}$. 
In such cases the multiple index denotes a single parton with flavour 
obtained using the flavour summation rules: odd/even number of quarks 
plus any number of gluons gives a quark/gluon, or explicitly for the 
relevant cases at NNLO,
\bal
&q + g = q\,, &
&q + \qb = g\,, &
&g + g = g\,, &
\nt\\
&q + g + g = q\,, &
&q + q + \qb = q\,, &
&g + q + \qb = g\,, &
&g + g + g = g\,. 
\label{eq:flavoursummation} 
\eal 
%
 
%
%
 
\subsection{Cross sections} 
\label{sec:xsecs} 
 
In this paper we shall need to use only tree-level $n$-parton 
production cross sections, with $n=m$, the Born cross  
section, and $n=m+2$, the so-called doubly real correction.  
We have 
\beq 
\dsig{(0)}_{n}(\mom{}) = 
	{\cal N}\;\sum_{\{n\}}\PS{n}{(\mom{})}\frac{1}{S_{\{n\}}} 
	\SME{n}{0}{\mom{}}\,, 
\label{eq:dsig0n} 
\eeq 
where ${\cal N}$ includes all QCD-independent factors and 
$\PS{n}{(\mom{})}$ is the $d$-dimensional phase space for 
$n$ outgoing particles with momenta 
$\mom{} \equiv \{p_1,\dots,p_{n}\}$ and total incoming momentum  
$Q$, 
\beq 
\PS{n}(p_1,\ldots,p_n;Q) =  
	\prod_{i=1}^{n}\frac{\rd^d p_i}{(2\pi)^{d-1}}\,\delta_+(p_i^2)\, 
	(2\pi)^d \delta^{(d)}\bigg(Q-\sum_{i=1}^{n}p_i\bigg)\,. 
\label{eq:PSn} 
\eeq 
The symbol $\sum_{\{n\}}$ denotes summation over different  
subprocesses and $S_{\{n\}}$ is the Bose symmetry factor for identical 
particles in the final state. Then the Born cross section and the  
doubly real correction are simply 
\beq 
\dsig{B}_{m}(\mom{}) \equiv 
	\dsig{(0)}_{m}(\mom{})
\qquad\mbox{and}\qquad 
\dsig{RR}_{m+2}(\mom{}) \equiv 
	\dsig{(0)}_{m+2}(\mom{})\,. 
\eeq 

The final result will also contain the phase-space factor due to 
the integral over the $(d-3)$-dimensional solid angle, which is 
included in the definition of the running coupling in the \msbar\ 
renormalisation scheme\footnote{In the \msbar\ renormalisation scheme 
as often employed in the literature, the  definition of the running 
coupling includes the factor $S_\ep=(4\pi)^\ep \mathrm{e}^{-\ep\gam{E}}$. 
In a computation at NLO accuracy, the two definitions lead to the same 
expressions. At NNLO they  lead to slightly different bookkeeping of 
the IR and UV poles at intermediate steps of the computation, our definition 
leading to somewhat simpler expressions. Of course the physical cross section 
does not depend on these details.}, 
\beq 
S_\ep = 
	\int\frac{\rd^{(d-3)}\Omega}{(2\pi)^{d-3}} = 
	\frac{(4\pi)^\ep}{\Gamma(1-\ep)}\,. 
\label{eq:Seps} 
\eeq 
%
 
%
%
 
\subsection{Momentum mappings and phase-space factorisation} 
\label{sec:psfact} 
 
The subtraction terms are written in terms of the momenta obtained  
by the mappings of \Ref{Somogyi:2006da},
\beq 
\mom{}_{m+2} \stackrel{{\mathsf X}_{RR}}{\longrightarrow} 
	\momt{(RR)}_{m}\,, 
\eeq 
where $\stackrel{{\mathsf X}_{RR}}{\longrightarrow}$ may label 
a triple collinear, double collinear or soft collinear mapping%
\footnote{Integration of the subtraction terms obtained with the double 
soft mapping are presented in the companion paper~\cite{Somogyi:inprep}.}. 
In particular, the soft collinear mapping is obtained by the iterated 
application of a basic collinear mapping, followed by a basic soft mapping, 
\beq 
\mom{}_{m+2} \cmap{ir} \momh{(ir)}_{m+1} \smap{\ha{s}} \momt{(\ha{s},ir)}_{m}\,, 
\label{eq:CirSsmap} 
\eeq 
while the former two cases do not have such an iterated form. 
As the above notation suggests, the final set of $m$  
momenta are denoted by tildes, while hats indicate the intermediate  
set of $m+1$ momenta. In {\em kinematic} expressions where only  
the label of a momentum is displayed (we shall discuss several examples  
below), the tilde is inherited by the label, and we write for instance 
$\wti{i}$, $\wti{ir}$ and $\wti{irs}$, where the last two label 
{\em single momenta}. However, since these mappings affect only  
the momenta, but not the colour and flavour (apart from the flavour 
summation rules of \eqn{eq:flavoursummation}), we shall omit the tilde 
from flavour and colour indices. 
 
We also use labels such as $(ir)$ to denote a {\em single momentum} 
that is simply the sum of two momenta, $p_{(ir)}^\mu \equiv p_i^\mu + p_r^\mu$. 
 
The momentum mappings are chosen as to lead to an exact factorisation 
of the phase space. For the triple collinear phase-space mapping, we have,
\bal
&
\mom{}_{m+2} \cmap{irs} \momt{(irs)}_{m}\,: 
\nt\\ &\qquad\quad
	\PS{m+2}(\mom{}_{m+2};Q) = \PS{m}(\momt{(irs)}_{m},Q) 
	\,[\rd p_{2;m}^{(irs)}(p_r,p_s,\wti{p}_{irs};Q)]\,, 
\label{eq:PSfact_Cirs} 
\intertext{for the double collinear mapping,}
&
\mom{}_{m+2} \cmap{ir;js} \momt{(ir;js)}_{m}\,: 
\nt\\ &\qquad\quad
	\PS{m+2}(\mom{}_{m+2};Q) = \PS{m}(\momt{(ir;js)}_{m},Q) 
	\,[\rd p_{2;m}^{(ir;js)}(p_r,p_s,\wti{p}_{ir},\wti{p}_{js};Q)]\,, 
\label{eq:PSfact_Cirjs} 
\intertext{while for the iterated collinear and soft mapping appearing in
\eqn{eq:CirSsmap},}
&
\mom{}_{m+2} \cmap{ir} \momh{(ir)}_{m+1} \smap{\ha{s}} \momt{(\ha{s},ir)}_{m}\,: 
\nt\\ &\qquad\quad
	\PS{m+2}(\mom{}_{m+2};Q) = \PS{m}{}(\momt{(\ha{s},ir)}_{m}) 
	\,[\rd p_{1;m+1}^{(ir)}(p_{r},\ha{p}_{ir};Q)] 
	\,[\rd p_{1;m}^{(\ha{s})}(\ha{p}_s;Q)]\,.
\label{eq:PSfact_CSirs} 
\eal
%
 
%
%
 
\subsection{Kinematic variables} 
\label{sec:kinematic} 
 
The following types of kinematic variables are used to write the doubly 
unresolved subtraction  terms.
\begin{itemize} 
\item Two-particle invariants, such as  
\beq
s_{ir} = 2 p_i\ldot p_r
\qquad\mbox{or}\qquad
s_{iQ} = 2 p_i\ldot Q \,. 
\eeq
Two-particle invariants scaled with $Q^2$ are denoted by 
$y_{ij} = s_{ij}/Q^2$. 

\item Momentum fractions 
$z_{i,r}$  for the splittings $\ti{p}_{ir} \to p_i + p_r$ and 
$z_{i,rs}$  for the splittings $\ti{p}_{irs} \to p_i + p_r + p_s$, 
\beq
z_{i,r} = \frac{y_{iQ}}{y_{iQ}+y_{rQ}} 
\qquad\mbox{and}\qquad
z_{i,rs} = \frac{y_{iQ}}{y_{iQ}+y_{rQ}+y_{sQ}}\,,
\eeq
with $z_{i,r} + z_{r,i} = 1$ and $z_{i,rs} + z_{r,si} + z_{s,ir} = 1$ 
($z_{r,si}$ and $z_{s,ir}$ are computed by obvious permutations). 

\item We also use the eikonal factors,
\beq 
\calS_{jk}(s) = \frac{2s_{jk}}{s_{js} s_{ks}}\,. 
\label{eq:cSdef} 
\eeq 
\end{itemize}
 
As mentioned above, the sum of two momenta is often abbreviated with the  
two indices in parentheses e.g.,~$p_i^\mu + p_r^\mu = p_{(ir)}^\mu$, which 
is also used in other occurrences, such as 
\beq
s_{(ir)k} =  s_{ik}+s_{rk}\,,
\qquad
\calS_{(ir)k}(s) = \frac{2(s_{ik}+s_{rk})}{(s_{is}+s_{rs}) s_{ks}} \,.
\label{eq:softcolleikonal} 
\eeq

Finally, we express the integrated subtraction terms as 
functions of the following (combinations of) invariants:  
\beq
x_{\wti{i}} = y_{\wti{i}Q}
\qquad\mbox{and}\qquad
\Yt{i}{j} = \frac{y_{\wti{i}\wti{j}}}{y_{\wti{i}Q}\,y_{\wti{j}Q}}\,.
\label{eq:xi_YijQdef} 
\eeq
In the rest frame of $Q^\mu$, $y_{\wti{i}Q} = 2\ti{E}_i/Q$ is simply the scaled 
energy of parton $\ti{i}$, while  $\Yt{i}{j}$ is related to $\chi_{ij}$, the angle 
between momenta $\ti{p}_i^\mu$ and $\ti{p}_j^\mu$, by $\cos\chi_{ij} = 1-2\Yt{i}{j}$.
 
 
\section{Integrating the doubly unresolved approximate cross section} 
\label{sec:IntRRA2} 
 
%
%

\subsection{The integrated approximate cross section and insertion operator} 
 
The doubly unresolved approximate cross section times the jet function 
is defined as,
\beq 
\dsiga{RR}{2}_{m+2} \odot J_{m} =  
	{\cal N}\sum_{\{m+2\}} 
	\PS{m+2}(\mom{})\frac{1}{S_{\{m+2\}}} \cA{2}^{(0)} |{\cal 
	M}_{m+2}^{(0)}({p})|^2 \odot J_{m}({p}) \,,
\label{eq:dsigRRA2Jm} 
\eeq 
where the $\odot$ notation reminds us that the set of momenta entering  
$J_m$ is different from term to term,
\beq 
\bsp 
&\cA{2}^{(0)} |{\cal M}_{m+2}^{(0)}({p})|^2 \odot J_{m}({p}) = 
\\ &\quad 
	\sum_{r}\sum_{s}\Bigg\{ 
	\sum_{i\ne r,s}\Bigg[\frac16\, \cC{irs}{(0,0)}(\mom{}) J_{m}(\momt{(irs)}_m) 
	+ \sum_{j\ne i,r,s} \frac18\, \cC{ir;js}{(0,0)}(\mom{}) J_{m}(\momt{(ir;js)}_m) 
\\ &\qqquad 
	+ \frac12\,\Bigg( \cSCS{ir;s}{(0,0)}(\mom{}) 
	- \cC{irs}{}\cSCS{ir;s}{(0,0)}(\mom{}) 
	- \sum_{j\ne i,r,s} \cC{ir;js}{} \cSCS{ir;s}{(0,0)}(\mom{}) \Bigg) 
	J_{m}(\momt{(\ha{s},ir)}_m) 
\\ &\qqquad 
	+ \Bigg( - \cSCS{ir;s}{}\cS{rs}{(0,0)}(\mom{}) 
	- \frac12\, \cC{irs}{}\cS{rs}{(0,0)}(\mom{})
	+ \cC{irs}{}\cSCS{ir;s}{}\cS{rs}{(0,0)}(\mom{}) 
\\ &\qqquad 
	+ \sum_{j\ne i,r,s} \frac12\, \cC{ir;js}{}\cS{rs}{(0,0)}(\mom{}) 
	\Bigg) J_{m}(\momt{(rs)}_m) \Bigg] 
	+ \frac12\, \cS{rs}{(0,0)}(\mom{}) J_{m}(\momt{(rs)}_m) 
\Bigg\}\,.
\label{eq:dsigRRA2Jmfull} 
\esp 
\eeq 
All terms on the right-hand side of \eqn{eq:dsigRRA2Jmfull} are defined 
in \Ref{Somogyi:2006da}. 
 
Using that all momentum mappings lead to the exact factorisation  
of phase space, $\PS{m+2}(\mom{};Q)=\PS{m}(\momt{};Q) [\rd p_{2;m}]$, 
and that the jet function $J_m$ does not depend on the variables of the  
factorised two-parton measure, $[\rd p_{2;m}]$, we can compute the
integral of \eqn{eq:dsigRRA2Jmfull} over the two-parton factorised
phase space independently  of $J_m$,
\beq 
\bsp 
& \int_2 \dsiga{RR}{2}_{m+2} = 
	{\cal N}\sum_{\{m+2\}} \PS{m}(\momt{})\frac{1}{S_{\{m+2\}}} 
	\left[\frac{\as}{2\pi} S_\ep\left(\frac{\mu^2}{Q^2}\right)^{\ep}\right]^2 
\\ &\quad 
	\times \sum_{r}\sum_{s}\Bigg\{ 
	\sum_{i\ne r,s} \Bigg[ 
	\frac16\,[\IcC{irs}{(0)}]_{f_i f_r f_s} \,(\bT_{irs}^2)^2 
	+ \sum_{j\ne i,r,s} \frac18\,[\IcC{ir;js}{(0)}]_{f_i f_r;f_j f_s} 
	\,\bT_{ir}^2\,\bT_{js}^2  
\\ & \qqqquad\qqquad 
	+ \frac12\, \Bigg( \sum_{j\ne i,r,s}\sum_{k\ne j} 
	[\IcSCS{ir;s}{(0),(j,k)}]_{f_i f_r}\,\bT^2_{ir}\, \bT_{j} \bT_{k} 
	- [\IcC{irs}{}\!\IcSCS{ir;s}{(0)}]_{f_i f_r}\,(\bT^2_{ir})^2\, 
\\ & \qqqquad\qqqquad 
	- \sum_{j\ne i,r,s} 
	[\IcC{ir;js}{}\!\IcSCS{ir;s}{(0)}]_{f_i f_r}\,\bT^2_{ir}\, \bT^2_{j}\, \Bigg) 
	- \bT_{ir}^2\, 
	\sum_{j}\sum_{l\ne j} [\IcSCS{ir;s}{}\!\IcS{rs}{(0)}]^{(j,l)}\,\bT_{j} \bT_{l} 
\\ & \qqqquad\qqquad 
	+ \bigg([\IcC{irs}{}\!\IcSCS{ir;s}{}\!\IcS{rs}{(0)}]
	- \frac12 [\IcC{irs}{}\!\IcS{rs}{(0)}]_{f_r f_s}\bigg) \big(\bT_{irs}^2\big)^2
	+ \sum_{j\ne i,r,s} \frac12 
	[\IcC{ir;js}{}\!\IcS{rs}{(0)}]\, \bT_{ir}^2\, \bT_{js}^2 \Bigg] 
\\ & \qqquad 
	+ \frac12 \sum_{i,k} 
	\Bigg(\sum_{j,l}[\IcS{rs}{(0)}]^{(i,k)(j,l)}\{\bT_{i} \bT_{k},\bT_{j} \bT_{l}\}
	+ [\IcS{rs}{(0)}]_{f_r f_s}^{(i,k)}\,\CA \bT_{i} \bT_{k} \Bigg) \Bigg\} 
	\otimes  \SME{m}{0}{\momt{}} \,.
\label{eq:IntdsigRRA2} 
\esp 
\eeq 
In order to make the integrated subtraction terms dimensionless in
color  space, we have factored out quadratic Casimir operators in
\eqn{eq:IntdsigRRA2}. 
 
\eqn{eq:IntdsigRRA2} is not yet in the form of an $m$-parton 
contribution times a factor. In order to obtain such a form, we must 
still perform summation over unresolved flavours, rewriting the symmetry 
factor of an $m+2$-parton configuration to the symmetry factor of an 
$m$-parton configuration. The counting is performed as in the Appendix 
of \Ref{Bolzoni:2010bt}. As a result of the summation over the unresolved 
flavours, we obtain functions --- the flavour-summed integrated counterterms --- denoted by $\big(X^{(0)}\big)_{f_i\dots} ^{(j,l)\dots}$, which are specific 
sums of the non flavour-summed integrated subtraction terms,
$[X^{(0)}]_{f_k\dots} ^{(j,l)\dots}$. Schematically, we may write,
\beq 
\Big(X^{(0)}\Big)_{f_i\dots} ^{(j,l)\dots} =
	\sum \,[X^{(0)}]_{f_k\dots} ^{(j,l)\dots}\,. 
\eeq 
The {\em flavour indices} on the left- and right-hand sides of this 
equation need not match:  the non flavour-summed functions on the 
right-hand side carry dependence on unresolved flavours, while 
the flavour-summed functions on the left do not, by definition. 
 
After summing over unobserved flavours, the integrated doubly unresolved 
approximate cross section can be written as,
\beq 
\int_{2}\dsiga{RR}{2}_{m+2} = 
	\dsig{B}_{m}\otimes \bI^{(0)}_2(\mom{}_{m};\ep)\,, 
\label{eq:I1dsigRRA1} 
\eeq 
where the insertion operator (in colour space) has three contributions 
according to the possible colour structures,
\beq 
\bsp 
\bI^{(0)}_2(\mom{}_{m};\ep) = 
	\left[\frac{\as}{2\pi} S_\ep\left(\frac{\mu^2}{Q^2}\right)^{\ep}\right]^2 
	&\Bigg\{ \sum_{i} \Bigg[ 
	\IcC{2,i}{(0)} \, \bT_i^2 
	+ \sum_{j\ne i} \IcC{2,i j}{(0)}\, \bT_j^2 \Bigg] \bT_i^2 
\\ & 
	+ \sum_{l} \sum_{j\ne l} \Bigg[ 
	\IcS{2}{(0),(j,l)}\, \CA\,  
	+ \sum_{i} \IcSCS{2,i}{(0),(j,l)}\, \bT_i^2\, \Bigg] \bT_{j}\bT_{l} 
\\ & 
	+ \sum_{i,j,k,l} \IcS{2}{(0),(i,k)(j,l)} \{ \bT_{i}\bT_{k},\bT_{j}\bT_{l} \} 
	\Bigg\} \,, 
\label{eq:I2} 
\esp 
\eeq 
with $\bT_i^2 = C_{f_i}$ ($C_q = \CF$, $C_g = \CA$) as in \eqn{eq:colalg}. 
 
In terms of flavour-summed integrated counterterms discussed above, the 
kinematic functions of the insertion operator can be written as,
\beq 
\bsp 
\IcC{2,i}{(0)} &= 
	\Big(\IcC{irs}{(0)}\Big)_{f_i} 
	- \Big( \IcC{irs}{} \IcSCS{ir;s}{(0)} \Big)_{f_i} 
	- \Big( \IcC{irs}{} \IcS{rs}{(0)} \Big)_{f_i} 
	+ \Big( \IcC{irs}{} \IcSCS{ir;s}{} \IcS{rs}{(0)} \Big)_{f_i} \,,
\\
\IcC{2,i j}{(0)} &= 
	\Big(\IcC{ir;js}{(0)}\Big)_{f_if_j} 
	- \Big(\IcC{ir;js}{}\!\IcSCS{ir;s}{(0)}\Big)_{f_if_j} 
	+ \Big(\IcC{ir;js}{}\IcS{rs}{(0)}\Big)_{f_if_j}  \,,
\\ 
\IcSCS{2,i}{(0),(j,l)} &= 
	\Big(\IcSCS{ir;s}{(0)}\Big)^{(j,l)}_{f_i} 
	- \Big(\IcSCS{ir;s}{} \IcS{rs}{(0)}\Big)^{(j,l)}_{f_i}  \,,
\\
\IcS{2}{(0),(j,l)} &= 
	\Big(\IcS{rs}{(0)}\Big)^{(j,l)}  \,,
\\
\IcS{2}{(0),(i,k)(j,l)} &= 
	\Big(\IcS{rs}{(0)}\Big)^{(i,k)(j,l)} \,. 
\label{eq:cali} 
\esp 
\eeq 
On the right-hand side of \eqn{eq:cali}, the flavour-summed
functions depend on the kinematics through variables of the type $x_i$
and $Y_{ij,Q}$. The latter dependence stems from integrating an
eikonal factor which is always multiplied by a colour-connected
squared matrix element. In order to make the results more transparent,
we hid the arguments of the functions, but kept the relation to the  
colour-connected matrix elements, shown as upper indices. 
 
%
%
 
\subsection{Flavour-summed integrated counterterms} 
  
On the right-hand side of \eqn{eq:cali}, the flavour-summed functions 
are the following combinations of the integrated subtraction terms,
\begin{enumerate} 
\item Triple collinear: 
\beq 
\bsp 
\Big(\IcC{irs}{(0)}\Big)_q &=  
	\frac{1}{2} [ \IcC{irs}{(0)} ]_{qgg} 
	+ (\Nf-1)\,  [ \IcC{irs}{(0)} ]_{qq'\qb'} 
	+ \frac{1}{2} [ \IcC{irs}{(0)} ]_{qq\qb}\,, 
	\qquad q'\ne q\,, 
\\
\Big(\IcC{irs}{(0)}\Big)_g &=  
	\frac{1}{6} [ \IcC{irs}{(0)} ]_{ggg} 
	+ \Nf\, [ \IcC{irs}{(0)} ]_{gq\qb}\,. 
\esp 
\eeq
The factor $(\Nf-1)$, which appears after summing over the unresolved 
flavours, can be traded for $\Nf$ by introducing the integrated subtraction 
term $[ \IcC{irs}{(0)} ]_{q\qb q}^{(\ID)}$, defined by 
\beq 
[ \IcC{irs}{(0)} ]_{q\qb q} = 
	2 [ \IcC{irs}{(0)} ]_{q\qb' q'}  
	+ [ \IcC{irs}{(0)} ]_{q\qb q}^{(\ID)}\,. 
\label{eq:decomp}
\eeq 
This definition matches the decomposition of the same-flavour triple 
splitting function computed in \Ref{Catani:1999ss}.  Thus, we obtain 
\beq 
\Big(\IcC{irs}{(0)}\Big)_q =  
	\frac{1}{2} [ \IcC{irs}{(0)} ]_{qgg} 
	+ \Nf\,  [ \IcC{irs}{(0)} ]_{q\qb' q'} 
	+ \frac{1}{2} [ \IcC{irs}{(0)} ]_{q\qb q}^{(\ID)}\,, 
	\qquad q'\ne q\,. 
\label{eq:TC} 
\eeq

\item Triple collinear -- soft  collinear: 
\beq 
\bsp 
\Big( \IcC{irs}{} \IcSCS{ir;s}{(0)} \Big)_q &= 
	[ \IcC{irs}{}\!\IcSCS{ir;s}{(0)} ]_{qg}\,,  
\\
\Big( \IcC{irs}{} \IcSCS{ir;s}{(0)} \Big)_g &= 
	\frac{1}{2} [ \IcC{irs}{}\!\IcSCS{ir;s}{(0)} ]_{gg} 
	+ \Nf\, [ \IcC{irs}{}\!\IcSCS{ir;s}{(0)} ]_{q\qb}\,. 
\esp 
\eeq 

\item Double collinear: 
\beq 
\bsp 
\Big(\IcC{ir;js}{(0)}\Big)_{qq} &= 
	\frac{1}{2} [ \IcC{ir;js}{(0)} ]_{qg;qg}\,, 
\\
\Big(\IcC{ir;js}{(0)}\Big)_{qg} &= 
	\frac{1}{4} [ \IcC{ir;js}{(0)} ]_{qg;gg} 
	+ \frac{\Nf}{2} [ \IcC{ir;js}{(0)} ]_{qg;q\qb}\,, 
\\
\Big(\IcC{ir;js}{(0)}\Big)_{gq} &= 
	\frac{1}{4} [ \IcC{ir;js}{(0)} ]_{gg;qg} 
	+ \frac{\Nf}{2} [ \IcC{ir;js}{(0)} ]_{q\qb ;qg}\,, 
\\
\Big(\IcC{ir;js}{(0)}\Big)_{gg} &= 
	\frac{1}{8} [ \IcC{ir;js}{(0)} ]_{gg;gg} 
	+ \frac{\Nf^2}{2} [ \IcC{ir;js}{(0)} ]_{q\qb ;q'\qb'} 
\\ & 
	+ \frac{\Nf}{4}\Big\{ [ \IcC{ir;js}{(0)} ]_{gg;q\qb} 
	+ [ \IcC{ir;js}{(0)} ]_{q\qb ;gg}\Big\}\,, 
	\qquad\forall\, q,q' \,. 
\esp 
\eeq 

\item Double collinear -- soft collinear: 
\beq 
\bsp 
\Big(\IcC{ir;js}{}\!\IcSCS{ir;s}{(0)}\Big)_{q f} &= 
	[ \IcC{ir;js}{}\!\IcSCS{ir;s}{(0)} ]_{qg}\,, 
\\
\Big(\IcC{ir;js}{}\!\IcSCS{ir;s}{(0)}\Big)_{g f} &= 
	\frac{1}{2} [ \IcC{ir;js}{}\!\IcSCS{ir;s}{(0)} ]_{gg}  
	+ \Nf\, [ \IcC{ir;js}{}\!\IcSCS{ir;s}{(0)} ]_{q\qb}\,, 
\esp 
\eeq 
i.e.,~it is independent of the flavour $f$. 

\item Soft collinear: 
\beq 
\bsp 
\Big(\IcSCS{ir;s}{(0)}\Big)^{(j,l)}_q &= 
	[ \IcSCS{ir;s}{(0)} ]_{qg}^{(j,l)}\,, 
\\
\Big(\IcSCS{ir;s}{(0)}\Big)^{(j,l)}_g &= 
	\frac{1}{2} [ \IcSCS{ir;s}{(0)} ]_{gg}^{(j,l)}  
	+ \Nf\, [ \IcSCS{ir;s}{(0)} ]_{q\qb}^{(j,l)}\,. 
\esp 
\label{eq:SCS} 
\eeq 

\item Triple collinear -- double soft: 
\beq 
\Big( \IcC{irs}{} \IcS{rs}{(0)} \Big)_f =
	\frac12 [\IcC{irs}{} \IcS{rs}{(0)}]_{fgg}
	+ \Nf\,[ \IcC{irs}{} \IcS{rs}{(0)} ]_{f\qb q}\,.
\eeq

\item Triple collinear -- soft collinear -- double soft: 
\beq 
\Big( \IcC{irs}{} \IcSCS{ir;s}{} \IcS{rs}{(0)} \Big)_f = 
	[ \IcC{irs}{} \IcSCS{ir;s}{} \IcS{rs}{(0)} ]\,, 
\eeq 
i.e.,~it is independent of the flavour $f$. 

\item Double collinear -- double soft: 
\beq 
\Big(\IcC{ir;js}{}\IcS{rs}{(0)}\Big)_{f_1f_2} = 
	\frac{1}{2} [ \IcC{ir;js}{}\IcS{rs}{(0)} ]\,, 
\eeq 
i.e.,~it is independent of the flavours $f_1$ and $f_2$. 

\item Double soft: 
\beq 
\bsp 
\Big(\IcS{rs}{(0)}\Big)^{(i,k)(j,l)} &= 
	\frac{1}{2} [ \IcS{rs}{(0)} ]^{(i,k)(j,l)}\,, 
\\
\Big(\IcS{rs}{(0)}\Big)^{(j,l)} &= 
	\frac12 [ \IcS{rs}{(0)} ]^{(j,l)}_{gg} 
	+ \Nf\,[ \IcS{rs}{(0)} ]^{(j,l)}_{\qb q}\,. 
\esp 
\eeq 

\item Soft collinear -- double soft: 
\beq 
\Big(\IcSCS{ir;s}{} \IcS{rs}{(0)}\Big)^{(j,l)}_f = 
	[ \IcSCS{ir;s}{}\IcS{rs}{(0)} ]^{(j,l)}\,, 
\eeq 
i.e.,~it is independent of the flavour $f$. 
\end{enumerate}  
In this paper, we compute the collinear-type counterterms, which do not
involve $\IcS{rs}{(0)}$, i.e.,~those in \eqnss{eq:TC}{eq:SCS}. 
The integration of the subtraction terms which are obtained by the
double soft mapping is presented in a companion paper~\cite{Somogyi:inprep}.

 
\section{Integrated counterterms} 
\label{sec:ICTs} 
 
In this section we define the integrated counterterms and 
compute them in terms of master integrals. 

%
%

\subsection{Integrated triple collinear counterterm} 
\label{sec:Cirs} 
 
The integral of the triple collinear counterterm is 
\beq 
\bsp 
[ \IcC{irs}{(0)} ]_{f_i f_r f_s} &= 
	\left( \frac{(4\pi)^2}{S_\ep}Q^{2\ep}\right)^2  
	\int_2 [\rd p_{2;m}^{(irs)}(p_r,p_s,\ti{p}_{irs};Q)]\, 
	\frac{1}{s_{irs}^2}\, \frac{1}{(\bT_{irs}^2)^2}  
\\ &\times 
	P^{(0)}_{f_i f_r f_s}(\tzz{i}{rs}, \tzz{r}{is}, \tzz{s}{ir}, 
	s_{ir}, s_{is}, s_{rs};\ep)\, 
	f(\al_0, \al_{irs}, d(m,\ep))\,. 
\label{eq:ICirs0} 
\esp 
\eeq 
Here $P^{(0)}_{f_i f_r f_s}(\tzz{i}{rs}, \tzz{r}{is}, \tzz{s}{ir},
s_{ir}, s_{is}, s_{rs};\ep)$ are the complete\footnote{As opposed to 
abelian and non-abelian parts, see \App{app:P0_3}.} spin-averaged three-parton
splitting functions, recalled in \App{app:P0}, where we explain why
some of these are different from those presented in \Ref{Catani:1999ss}.
For consistency of the complete subtraction scheme one must use the forms
given in \App{app:P0}.  The function $f(\al_0,\al_{irs},d(m,\ep))$ is 
defined and its role is explained in \App{app:modsubterms}. The other 
subtraction terms discussed in this paper will also contain such harmless 
modifications as compared to the original ones in \Ref{Somogyi:2006da}.
 
Following the decomposition of the triple collinear splitting functions
into abelian and non-abelian pieces~\cite{Catani:1999ss}, we also
decompose the integrated functions $[ \IcC{irs}{(0)} ]_{qgg}$ and
$[ \IcC{irs}{(0)} ]_{gq\qb}$ likewise,
\beq 
[ \IcC{irs}{(0)} ]_{qgg} = 
	[ \IcC{irs}{(0)} ]_{qgg}^{(\AB)}
	+ [ \IcC{irs}{(0)} ]_{qgg}^{(\NAB)}
\qquad \mbox{and} \qquad 
[ \IcC{irs}{(0)} ]_{gq\qb} = 
	[ \IcC{irs}{(0)} ]_{gq\qb}^{(\AB)}
	+ [ \IcC{irs}{(0)} ]_{gq\qb}^{(\NAB)}\,.
\label{eq:ICirsAB-NAB}
\eeq 
They are computed in \App{app:Cirs}. The result can be written as
\beq 
[ \IcC{irs}{(0)} ]_{f_i f_r f_s} = 
	a_{f_i f_r f_s} \sum_{n=1}^5 \sum_{j,k,l,m} 
	c_{f_i f_r f_s;n}^{(0),j,k,l,m}\,  
	\cI_{2\cC{}{},n}^{(j,k,l,m)}(x_{\wti{irs}},\ep;\al_0,d_0)\,, 
\eeq 
where the constants $a_{f_i f_r f_s}$ denote colour-factor ratios as follows: 
\bal
a_{q\qb' q'} &= \frac{\TR}{\CF} \,,
&a_{q\qb q}^{(\ID)} &= 1-\frac{\CA}{2\CF} \,,
&a_{q g g}^{(\AB)} &= 1 \,,
&a_{q g g}^{(\NAB)} &= \frac{\CA}{\CF} \,,
\nt\\
a_{g q \qb}^{(\AB)} &= \frac{\CF\TR}{\CA^2} \,,
&a_{g q \qb}^{(\NAB)} &= \frac{\TR}{\CA} \,,
&a_{g g g} &= 1 \,. 
\label{eq:acoeffs}
\eal
The non-zero coefficients are listed in 
\tabss{tab:I2C1coeffs}{tab:I2C5coeffs}, while the integrals 
$\cI_{2\cC{}{},n}^{(j,k,l,m)}(x,\ep;\al_0,d_0)$ ($n = 1,\ldots, 5$) are 
defined and computed in \App{app:Cirs}. 

%
%

\begin{table}
\footnotesize
\renewcommand{\arraystretch}{2}
\begin{center}
\begin{tabular}{|rrrc|c|c|c|c|c|c|c|}
\hline 
\hline 
$j$ & 
$k$ & 
$l$ & 
$m$ & 
$c_{q \qb' q'; 1 }^{(0)}$ & 
$c_{q \qb q; 1 }^{(0),(\ID)}$ & 
$c_{q g g; 1}^{(0),(\AB)}$ & 
$c_{q g g; 1}^{(0),(\NAB)}$ & 
$c_{g q \qb; 1}^{(0),(\AB)}$ & 
$c_{g q \qb; 1}^{(0),(\NAB)}$ & 
$c_{g g g; 1}^{(0)}$ \\ 
\hline 
$\!\!0\!\!$ & $\!\!0\!\!$ & $\!\!0\!\!$ & $\!\!0\!$
 & ${\displaystyle -1+\ep}$
 & ${\displaystyle -4+2 \ep+2 \ep^2}$
 & ${\displaystyle 2 (1-\ep) \ep}$
 & ${\displaystyle 1-2 \ep+\ep^2}$
 & ${\displaystyle -2 \ep}$
 & ${\displaystyle -1+\ep}$
 & ${\displaystyle 6-6 \ep}$
\\
$\!\!0\!\!$ & $\!\!-1\!\!$ & $\!\!0\!\!$ & $\!\!0\!$
 & ${\displaystyle 0}$
 & ${\displaystyle 6-10 \ep-2 \ep^2}$
 & ${\displaystyle -2 (1-2 \ep) \ep}$
 & ${\displaystyle \frac{3}{2}+\frac{15 \ep}{2}-\ep^2}$
 & ${\displaystyle -4}$
 & ${\displaystyle \frac{4}{1-\ep}}$
 & ${\displaystyle 18+6 \ep}$
\\
$\!\!0\!\!$ & $\!\!-1\!\!$ & $\!\!0\!\!$ & $\!\!1\!$
 & ${\displaystyle 0}$
 & ${\displaystyle -4+4 \ep+2 \ep^2}$
 & ${\displaystyle 2 (1-\ep) \ep}$
 & ${\displaystyle \frac{5}{2}-\frac{9 \ep}{2}+\ep^2}$
 & ${\displaystyle 0}$
 & ${\displaystyle -\frac{4}{1-\ep}}$
 & ${\displaystyle 0}$
\\
$\!\!1\!\!$ & $\!\!-1\!\!$ & $\!\!0\!\!$ & $\!\!0\!$
 & ${\displaystyle -2}$
 & ${\displaystyle -4+4 \ep}$
 & ${\displaystyle -2+4 \ep-2 \ep^2}$
 & ${\displaystyle 2-2 \ep}$
 & ${\displaystyle 2-2 \ep}$
 & ${\displaystyle -2}$
 & ${\displaystyle 6-6 \ep}$
\\
$\!\!-1\!\!$ & $\!\!0\!\!$ & $\!\!0\!\!$ & $\!\!1\!$
 & ${\displaystyle -2 \ep}$
 & ${\displaystyle -2+6 \ep}$
 & ${\displaystyle 2 (1-\ep) \ep}$
 & ${\displaystyle 2-4 \ep}$
 & ${\displaystyle 4\frac{1+\ep}{1-\ep}}$
 & ${\displaystyle -4\frac{2+\ep}{1-\ep}}$
 & ${\displaystyle -12}$
\\
$\!\!0\!\!$ & $\!\!-2\!\!$ & $\!\!0\!\!$ & $\!\!0\!$
 & ${\displaystyle -2}$
 & ${\displaystyle 0}$
 & ${\displaystyle 0}$
 & ${\displaystyle 2-2 \ep}$
 & ${\displaystyle 0}$
 & ${\displaystyle -2}$
 & ${\displaystyle 6-6 \ep}$
\\
$\!\!1\!\!$ & $\!\!-2\!\!$ & $\!\!0\!\!$ & $\!\!0\!$
 & ${\displaystyle 4}$
 & ${\displaystyle 0}$
 & ${\displaystyle 0}$
 & ${\displaystyle -4+4 \ep}$
 & ${\displaystyle 0}$
 & ${\displaystyle 4}$
 & ${\displaystyle -12+12 \ep}$
\\
$\!\!2\!\!$ & $\!\!-2\!\!$ & $\!\!0\!\!$ & $\!\!0\!$
 & ${\displaystyle -2}$
 & ${\displaystyle 0}$
 & ${\displaystyle 0}$
 & ${\displaystyle 2-2 \ep}$
 & ${\displaystyle 0}$
 & ${\displaystyle -2}$
 & ${\displaystyle 6-6 \ep}$
\\
$\!\!-1\!\!$ & $\!\!-1\!\!$ & $\!\!0\!\!$ & $\!\!0\!$
 & ${\displaystyle 0}$
 & ${\displaystyle 2 \ep}$
 & ${\displaystyle 6}$
 & ${\displaystyle -5+2 \ep}$
 & ${\displaystyle -\frac{2 \ep}{1-\ep}}$
 & ${\displaystyle -\frac{7-4 \ep}{1-\ep}}$
 & ${\displaystyle 51}$
\\
$\!\!-1\!\!$ & $\!\!-1\!\!$ & $\!\!0\!\!$ & $\!\!1\!$
 & ${\displaystyle 0}$
 & ${\displaystyle 6-2 \ep}$
 & ${\displaystyle -6+2 \ep}$
 & ${\displaystyle 7-5 \ep}$
 & ${\displaystyle 0}$
 & ${\displaystyle 2\frac{7-\ep}{1-\ep}}$
 & ${\displaystyle -84}$
\\
$\!\!-1\!\!$ & $\!\!-1\!\!$ & $\!\!0\!\!$ & $\!\!2\!$
 & ${\displaystyle 0}$
 & ${\displaystyle 0}$
 & ${\displaystyle 0}$
 & ${\displaystyle 0}$
 & ${\displaystyle \frac{4}{1-\ep}}$
 & ${\displaystyle -\frac{6}{1-\ep}}$
 & ${\displaystyle 6}$
\\
$\!\!-1\!\!$ & $\!\!0\!\!$ & $\!\!-1\!\!$ & $\!\!1\!$
 & ${\displaystyle 0}$
 & ${\displaystyle \ep (1+\ep)}$
 & ${\displaystyle -\ep (1+\ep)}$
 & ${\displaystyle 2-\frac{3 \ep}{2}+\frac{\ep^2}{2}}$
 & ${\displaystyle \frac{2}{1-\ep}}$
 & ${\displaystyle \frac{4-\ep}{1-\ep}}$
 & ${\displaystyle -39}$
\\
$\!\!-1\!\!$ & $\!\!0\!\!$ & $\!\!-1\!\!$ & $\!\!2\!$
 & ${\displaystyle 0}$
 & ${\displaystyle 0}$
 & ${\displaystyle 0}$
 & ${\displaystyle 0}$
 & ${\displaystyle -\frac{2 \ep}{1-\ep}}$
 & ${\displaystyle -\frac{3-2 \ep}{1-\ep}}$
 & ${\displaystyle 9}$
\\[0.5em]
\hline
\hline
\end{tabular}
\end{center}
\normalsize 
\caption{\label{tab:I2C1coeffs} Non-zero coefficients of the  
$\cI_{2\cC{}{},1}^{(j,k,l,m)}(x,\ep;\al_0,d_0)$ integrals.  
} 
\end{table}

%
%

\begin{table}[ht]
\footnotesize
\renewcommand{\arraystretch}{2} 
\begin{center} 
\begin{tabular}{|rrrr|c|c|c|c|c|c|c|} 
\hline 
\hline 
{\normalsize $j$} & 
{\normalsize $k$} & 
{\normalsize $l$} & 
{\normalsize $m$} & 
{\normalsize $c_{q \qb' q'; 2 }^{(0)}$} & 
{\normalsize $c_{q \qb q; 2 }^{(0),(\ID)}$} & 
{\normalsize $c_{q g g; 2}^{(0),(\AB)}$} & 
{\normalsize $c_{q g g; 2}^{(0),(\NAB)}$} & 
{\normalsize $c_{g q \qb; 2}^{(0),(\AB)}$} & 
{\normalsize $c_{g q \qb; 2}^{(0),(\NAB)}$} & 
{\normalsize $c_{g g g; 2}^{(0)}$} \\ 
\hline 
$\!\!-1\!\!$ & $\!\!0\!\!$ & $\!\!-1\!\!$ & $\!\!0\!$
 & ${\displaystyle 0}$
 & ${\displaystyle 0}$
 & ${\displaystyle -8}$
 & ${\displaystyle 5-\ep}$
 & ${\displaystyle 0}$
 & ${\displaystyle 1}$
 & ${\displaystyle -12}$
\\
$\!\!-1\!\!$ & $\!\!0\!\!$ & $\!\!-1\!\!$ & $\!\!1\!$
 & ${\displaystyle 0}$
 & ${\displaystyle 0}$
 & ${\displaystyle 6-2 \ep}$
 & ${\displaystyle -3+3 \ep}$
 & ${\displaystyle 0}$
 & ${\displaystyle -\frac{3-\ep}{1-\ep}}$
 & ${\displaystyle 18}$
\\
$\!\!-1\!\!$ & $\!\!0\!\!$ & $\!\!-1\!\!$ & $\!\!2\!$
 & ${\displaystyle 0}$
 & ${\displaystyle 0}$
 & ${\displaystyle -2+2 \ep}$
 & ${\displaystyle 2-2 \ep}$
 & ${\displaystyle 0}$
 & ${\displaystyle \frac{4}{1-\ep}}$
 & ${\displaystyle -12}$
\\
$\!\!-1\!\!$ & $\!\!0\!\!$ & $\!\!-1\!\!$ & $\!\!3\!$
 & ${\displaystyle 0}$
 & ${\displaystyle 0}$
 & ${\displaystyle 0}$
 & ${\displaystyle 0}$
 & ${\displaystyle 0}$
 & ${\displaystyle -\frac{2}{1-\ep}}$
 & ${\displaystyle 6}$
\\
$\!\!0\!\!$ & $\!\!-1\!\!$ & $\!\!-1\!\!$ & $\!\!0\!$
 & ${\displaystyle 0}$
 & ${\displaystyle 0}$
 & ${\displaystyle -2}$
 & ${\displaystyle -\ep}$
 & ${\displaystyle 0}$
 & ${\displaystyle 1}$
 & ${\displaystyle -12}$
\\
$\!\!0\!\!$ & $\!\!-1\!\!$ & $\!\!-1\!\!$ & $\!\!1\!$
 & ${\displaystyle 0}$
 & ${\displaystyle 0}$
 & ${\displaystyle -2-2 \ep}$
 & ${\displaystyle 2 \ep}$
 & ${\displaystyle 0}$
 & ${\displaystyle -\frac{2}{1-\ep}}$
 & ${\displaystyle 6}$
\\
$\!\!0\!\!$ & $\!\!-1\!\!$ & $\!\!-1\!\!$ & $\!\!2\!$
 & ${\displaystyle 0}$
 & ${\displaystyle 0}$
 & ${\displaystyle 0}$
 & ${\displaystyle 0}$
 & ${\displaystyle 0}$
 & ${\displaystyle \frac{2}{1-\ep}}$
 & ${\displaystyle -6}$
\\
$\!\!-1\!\!$ & $\!\!-1\!\!$ & $\!\!-1\!\!$ & $\!\!0\!$
 & ${\displaystyle 0}$
 & ${\displaystyle 0}$
 & ${\displaystyle -8}$
 & ${\displaystyle 5-\ep}$
 & ${\displaystyle 0}$
 & ${\displaystyle 1}$
 & ${\displaystyle -12}$
\\
$\!\!-1\!\!$ & $\!\!-1\!\!$ & $\!\!-1\!\!$ & $\!\!1\!$
 & ${\displaystyle 0}$
 & ${\displaystyle 0}$
 & ${\displaystyle 6-2 \ep}$
 & ${\displaystyle -3+3 \ep}$
 & ${\displaystyle 0}$
 & ${\displaystyle -\frac{3-\ep}{1-\ep}}$
 & ${\displaystyle 18}$
\\
$\!\!-1\!\!$ & $\!\!-1\!\!$ & $\!\!-1\!\!$ & $\!\!2\!$
 & ${\displaystyle 0}$
 & ${\displaystyle 0}$
 & ${\displaystyle -2+2 \ep}$
 & ${\displaystyle 2-2 \ep}$
 & ${\displaystyle 0}$
 & ${\displaystyle \frac{4}{1-\ep}}$
 & ${\displaystyle -12}$
\\
$\!\!-1\!\!$ & $\!\!-1\!\!$ & $\!\!-1\!\!$ & $\!\!3\!$
 & ${\displaystyle 0}$
 & ${\displaystyle 0}$
 & ${\displaystyle 0}$
 & ${\displaystyle 0}$
 & ${\displaystyle 0}$
 & ${\displaystyle -\frac{2}{1-\ep}}$
 & ${\displaystyle 6}$
\\[0.5em]
\hline 
\hline 
\end{tabular} 
\end{center}
\caption{\label{tab:I2C2coeffs} Non-zero coefficients of the  
$\cI_{2\cC{}{},2}^{(j,k,l,m)}(x,\ep;\al_0,d_0)$ integrals.  
} 
\end{table} 

%
%

\begin{table}[ht] 
\footnotesize
\renewcommand{\arraystretch}{2} 
\begin{center}
\begin{tabular}{|rrrr|c|c|c|c|c|c|c|} 
\hline 
\hline 
{\normalsize $j$} & 
{\normalsize $k$} & 
{\normalsize $l$} & 
{\normalsize $m$} & 
{\normalsize $c_{q \qb' q'; 3 }^{(0)}$} & 
{\normalsize $c_{q \qb q; 3 }^{(0),(\ID)}$} & 
{\normalsize $c_{q g g; 3}^{(0),(\AB)}$} & 
{\normalsize $c_{q g g; 3}^{(0),(\NAB)}$} & 
{\normalsize $c_{g q \qb; 3}^{(0),(\AB)}$} & 
{\normalsize $c_{g q \qb; 3}^{(0),(\NAB)}$} & 
{\normalsize $c_{g g g; 3}^{(0)}$} \\ 
\hline 
$\!\!0\!\!$ & $\!\!-1\!\!$ & $\!\!0\!\!$ & $\!\!-1\!$
 & ${\displaystyle 2}$
 & ${\displaystyle 0}$
 & ${\displaystyle 0}$
 & ${\displaystyle -8}$
 & ${\displaystyle 0}$
 & ${\displaystyle 2}$
 & ${\displaystyle -24}$
\\
$\!\!0\!\!$ & $\!\!-1\!\!$ & $\!\!1\!\!$ & $\!\!-1\!$
 & ${\displaystyle -2}$
 & ${\displaystyle -4+4 \ep}$
 & ${\displaystyle 0}$
 & ${\displaystyle 2-2 \ep}$
 & ${\displaystyle 0}$
 & ${\displaystyle -2}$
 & ${\displaystyle 6-6 \ep}$
\\
$\!\!0\!\!$ & $\!\!-1\!\!$ & $\!\!2\!\!$ & $\!\!-1\!$
 & ${\displaystyle 2}$
 & ${\displaystyle 0}$
 & ${\displaystyle 0}$
 & ${\displaystyle -2+2 \ep}$
 & ${\displaystyle 0}$
 & ${\displaystyle \frac{8}{1-\ep}}$
 & ${\displaystyle -24}$
\\
$\!\!-1\!\!$ & $\!\!0\!\!$ & $\!\!0\!\!$ & $\!\!-1\!$
 & ${\displaystyle 0}$
 & ${\displaystyle 2-2 \ep}$
 & ${\displaystyle 0}$
 & ${\displaystyle -4}$
 & ${\displaystyle 0}$
 & ${\displaystyle 1}$
 & ${\displaystyle -12}$
\\
$\!\!-1\!\!$ & $\!\!0\!\!$ & $\!\!1\!\!$ & $\!\!-1\!$
 & ${\displaystyle 0}$
 & ${\displaystyle 4 \ep}$
 & ${\displaystyle 0}$
 & ${\displaystyle 3-\ep}$
 & ${\displaystyle 0}$
 & ${\displaystyle -\frac{3-\ep}{1-\ep}}$
 & ${\displaystyle 18}$
\\
$\!\!-1\!\!$ & $\!\!0\!\!$ & $\!\!2\!\!$ & $\!\!-1\!$
 & ${\displaystyle 0}$
 & ${\displaystyle 2-2 \ep}$
 & ${\displaystyle 0}$
 & ${\displaystyle -1+\ep}$
 & ${\displaystyle 0}$
 & ${\displaystyle \frac{4}{1-\ep}}$
 & ${\displaystyle -12}$
\\
$\!\!-1\!\!$ & $\!\!0\!\!$ & $\!\!3\!\!$ & $\!\!-1\!$
 & ${\displaystyle 0}$
 & ${\displaystyle 0}$
 & ${\displaystyle 0}$
 & ${\displaystyle 0}$
 & ${\displaystyle 0}$
 & ${\displaystyle -\frac{2}{1-\ep}}$
 & ${\displaystyle 6}$
\\
$\!\!0\!\!$ & $\!\!-2\!\!$ & $\!\!1\!\!$ & $\!\!-1\!$
 & ${\displaystyle 4}$
 & ${\displaystyle 0}$
 & ${\displaystyle 0}$
 & ${\displaystyle -4+4 \ep}$
 & ${\displaystyle 0}$
 & ${\displaystyle 4}$
 & ${\displaystyle -12+12 \ep}$
\\
$\!\!1\!\!$ & $\!\!-2\!\!$ & $\!\!1\!\!$ & $\!\!-1\!$
 & ${\displaystyle -4}$
 & ${\displaystyle 0}$
 & ${\displaystyle 0}$
 & ${\displaystyle 4-4 \ep}$
 & ${\displaystyle 0}$
 & ${\displaystyle -4}$
 & ${\displaystyle 12-12 \ep}$
\\
$\!\!-1\!\!$ & $\!\!-1\!\!$ & $\!\!0\!\!$ & $\!\!-1\!$
 & ${\displaystyle 0}$
 & ${\displaystyle -4}$
 & ${\displaystyle 0}$
 & ${\displaystyle 2}$
 & ${\displaystyle 0}$
 & ${\displaystyle 0}$
 & ${\displaystyle 0}$
\\
$\!\!-1\!\!$ & $\!\!-1\!\!$ & $\!\!1\!\!$ & $\!\!-1\!$
 & ${\displaystyle 0}$
 & ${\displaystyle -2-2 \ep}$
 & ${\displaystyle 0}$
 & ${\displaystyle -2}$
 & ${\displaystyle 0}$
 & ${\displaystyle 1}$
 & ${\displaystyle -12}$
\\
$\!\!-1\!\!$ & $\!\!-1\!\!$ & $\!\!2\!\!$ & $\!\!-1\!$
 & ${\displaystyle 0}$
 & ${\displaystyle -2+2 \ep}$
 & ${\displaystyle 0}$
 & ${\displaystyle 1-\ep}$
 & ${\displaystyle 0}$
 & ${\displaystyle -\frac{2}{1-\ep}}$
 & ${\displaystyle 6}$
\\
$\!\!-1\!\!$ & $\!\!-1\!\!$ & $\!\!3\!\!$ & $\!\!-1\!$
 & ${\displaystyle 0}$
 & ${\displaystyle 0}$
 & ${\displaystyle 0}$
 & ${\displaystyle 0}$
 & ${\displaystyle 0}$
 & ${\displaystyle \frac{2}{1-\ep}}$
 & ${\displaystyle -6}$
\\
$\!\!0\!\!$ & $\!\!-2\!\!$ & $\!\!2\!\!$ & $\!\!-2\!$
 & ${\displaystyle -2}$
 & ${\displaystyle 0}$
 & ${\displaystyle 0}$
 & ${\displaystyle 2-2 \ep}$
 & ${\displaystyle 0}$
 & ${\displaystyle -2}$
 & ${\displaystyle 6-6 \ep}$
\\
\hline 
\hline 
\end{tabular} 
\end{center}
\caption{\label{tab:I2C3coeffs} Non-zero coefficients of the  
$\cI_{2\cC{}{},3}^{(j,k,l,m)}(x,\ep;\al_0,d_0)$ integrals.  
} 
\end{table} 

%
%

\begin{table}[ht] 
\footnotesize
\renewcommand{\arraystretch}{2} 
\begin{center}
\begin{tabular}{|rrrr|c|c|c|c|c|c|c|} 
\hline 
\hline 
{\normalsize $j$} & 
{\normalsize $k$} & 
{\normalsize $l$} & 
{\normalsize $m$} & 
{\normalsize $c_{q \qb' q'; 4 }^{(0)}$} & 
{\normalsize $c_{q \qb q; 4 }^{(0),(\ID)}$} & 
{\normalsize $c_{q g g; 4}^{(0),(\AB)}$} & 
{\normalsize $c_{q g g; 4}^{(0),(\NAB)}$} & 
{\normalsize $c_{g q \qb; 4}^{(0),(\AB)}$} & 
{\normalsize $c_{g q \qb; 4}^{(0),(\NAB)}$} & 
{\normalsize $c_{g g g; 4}^{(0)}$} \\ 
\hline 
$\!\!-1\!\!$ & $\!\!0\!\!$ & $\!\!-1\!\!$ & $\!\!-1\!$
 & ${\displaystyle 0}$
 & ${\displaystyle 0}$
 & ${\displaystyle 4}$
 & ${\displaystyle 0}$
 & ${\displaystyle 0}$
 & ${\displaystyle 0}$
 & ${\displaystyle 12}$
\\
$\!\!0\!\!$ & $\!\!-1\!\!$ & $\!\!-1\!\!$ & $\!\!-1\!$
 & ${\displaystyle 0}$
 & ${\displaystyle 0}$
 & ${\displaystyle 4}$
 & ${\displaystyle -2}$
 & ${\displaystyle 0}$
 & ${\displaystyle 0}$
 & ${\displaystyle 6}$
\\
$\!\!-1\!\!$ & $\!\!-1\!\!$ & $\!\!-1\!\!$ & $\!\!-1\!$
 & ${\displaystyle 0}$
 & ${\displaystyle 0}$
 & ${\displaystyle 4}$
 & ${\displaystyle -2}$
 & ${\displaystyle 0}$
 & ${\displaystyle 0}$
 & ${\displaystyle 6}$
\\
\hline 
\hline 
\end{tabular} 
\end{center}
\caption{\label{tab:I2C4coeffs} Non-zero coefficients of the  
$\cI_{2\cC{}{},4}^{(j,k,l,m)}(x,\ep;\al_0,d_0)$ integrals.  
} 
\end{table} 

%
%

\begin{table}[ht] 
\footnotesize
\renewcommand{\arraystretch}{2} 
\begin{center}
\begin{tabular}{|rrrr|c|c|c|c|c|c|c|} 
\hline 
\hline 
{\normalsize $j$} & 
{\normalsize $k$} & 
{\normalsize $l$} & 
{\normalsize $m$} & 
{\normalsize $c_{q \qb' q'; 5 }^{(0)}$} & 
{\normalsize $c_{q \qb q; 5 }^{(0),(\ID)}$} & 
{\normalsize $c_{q g g; 5}^{(0),(\AB)}$} & 
{\normalsize $c_{q g g; 5}^{(0),(\NAB)}$} & 
{\normalsize $c_{g q \qb; 5}^{(0),(\AB)}$} & 
{\normalsize $c_{g q \qb; 5}^{(0),(\NAB)}$} & 
{\normalsize $c_{g g g; 5}^{(0)}$} \\ 
\hline 
$\!\!-1\!\!$ & $\!\!-1\!\!$ & $\!\!-1\!\!$ & $\!\!-1\!$
 & ${\displaystyle 0}$
 & ${\displaystyle 2}$
 & ${\displaystyle 0}$
 & ${\displaystyle 0}$
 & ${\displaystyle 0}$
 & ${\displaystyle 0}$
 & ${\displaystyle 3}$
\\
\hline 
\hline 
\end{tabular} 
\end{center}
\caption{\label{tab:I2C5coeffs} Non-zero coefficients of the  
$\cI_{2\cC{}{},5}^{(j,k,l,m)}(x,\ep;\al_0,d_0)$ integrals.  
} 
\end{table} 

%
%
 
\subsection{Integrated double collinear counterterm} 
\label{sec:Cirjs} 
 
The integral of the double collinear counterterm is 
\beq 
\bsp 
[ \IcC{ir;js}{(0)} ]_{f_if_r;f_jf_s} &= 
	\left( \frac{(4\pi)^2}{S_\ep}Q^{2\ep}\right)^2  
	\int [\rd p_{2;m}^{(ir;js)}(p_r,p_s,\ti{p}_{ir},\ti{p}_{js};Q)]  
	\frac{1}{s_{ir}s_{js}} \, \frac{1}{\bT_{ir}^2\bT_{js}^2}\,  
\\ &\times 
	P^{(0)}_{f_i f_r} (\tzz{r}{i};\ep) 
	P^{(0)}_{f_j f_s} (\tzz{s}{j};\ep)\, 
	f(\al_0,\al_{ir}+\al_{js},d(m,\ep)) \,.
\label{eq:ICirjs} 
\esp 
\eeq 
In \eqn{eq:ICirjs}, $P^{(0)}_{f_i f_r} (\tzz{r}{i};\ep)$ and
$P^{(0)}_{f_i f_r} (\tzz{r}{i};\ep)$  are the spin-averaged \AP
splitting functions in $d$ dimensions, as recalled in \App{app:P0}. 
 
The integrated counterterm is computed in \App{app:Cirjs}. The result is 
\beq 
\bsp 
[ \IcC{ir;js}{(0)} ]_{f_if_r;f_jf_s} &= 
	a_{f_i f_r} a_{f_j f_s} \sum _{k,l=-1}^2
	c^{(0),k}_{f_i f_r}\, c^{(0),l}_{f_j f_s}\, 
	{\cal I}_{2\cC{}{},6}^{(k,l)}(x_{\wti{ir}},x_{\wti{js}};\ep,\al_0,d_0)\,, 
\label{eq:ICirjs0result} 
\esp 
\eeq 
where the constants $a_{f_i f_r}$ denote the colour factor ratios
\beq
a_{q\qb} = \frac{\TR}{\CA} 
\,, \qquad
a_{q g} = a_{g g} = 1 
\,, 
\label{eq:acoeffs2}
\eeq
and the coefficients $c_{f_i f_r}$ are those already defined in 
\Ref{Bolzoni:2009ye} stripped of their colour factors, as presented 
in \tab{tab:I2C6coeffs}.

%
%

\begin{table}[ht] 
\footnotesize
\renewcommand{\arraystretch}{2} 
\begin{center}
\begin{tabular}{|r|c|c|c|} 
\hline 
\hline 
{\normalsize $k$} & 
{\normalsize $c_{q g}^{(0),k}$} & 
{\normalsize $c_{q \qb}^{(0),k}$} & 
{\normalsize $c_{g g}^{(0),k}$}\\ 
\hline 
$-1$ & $2$ & 0 & $4$ \\ 
$ 0$ & $-2$ & 1 & $-4$ \\ 
$ 1$ & $1-\ep$ & $\displaystyle -\frac{2}{1-\ep}$ & $2$ \\ 
$ 2$ & $0$ & $\displaystyle \frac{2}{1-\ep}$ & $-2$ \\[0.5em]
\hline 
\hline 
\end{tabular} 
\normalsize
\caption{\label{tab:I2C6coeffs} Values of the $c_{f_i f_r}^{(0),k}$ 
coefficients that appear in Eqs.~(\ref{eq:ICirjs0result}), 
(\ref{eq:ICSirsresult}), (\ref{eq:ICSirs-MI2}), (\ref{eq:ICCSirsresult}) 
and (\ref{eq:ICirjsCSirsresult}).
} 
\end{center}
\end{table} 

%
%
 
\subsection{Integrated soft collinear counterterms} 
\label{sec:CSirs} 

There are three types of soft collinear counterterms. 
\begin{enumerate} 
\item Soft collinear: 
\beq 
\bsp 
[ \IcSCS{ir;s}{(0)} ]_{f_if_r}^{(j,k)} &= 
	- \left( \frac{(4\pi)^2}{S_\ep}Q^{2\ep}\right)^2 
	\int_2 [\rd p_{1;m+1}^{(ir)}(p_{r},\ha{p}_{ir};Q)]\, 
	[\rd p_{1;m}^{(\ha{s})}(\ha{p}_s;Q)] 
\\ &\times 
	\frac1{2} \calS_{jk}(s)\, 
	\frac{1}{s_{ir}} \frac{1}{\bT_{ir}^2}\, P^{(0)}_{f_i f_r}(\tzz{r}{i};\ep)\, 
	f(\al_0,\al_{ir},d(m,\ep)) \,
	f(y_0,y_{\ha{s}Q},d'(m,\ep)) \,. 
\label{eq:ICSirs} 
\esp 
\eeq 
The factor $\calS_{jk}(s)$ is the eikonal factor, given by \eqn{eq:cSdef} 
if $j,k \ne (ir)$, and by \eqn{eq:softcolleikonal} if e.g.,~$j = (ir)$.  

The integrated counterterm is computed in \App{app:CSirs}. The result is 
\beq 
[\IcSCS{ir;s}{(0)}]_{f_i f_r}^{(j,k)} = 
	a_{f_i f_r} \sum_{l=-1}^2 c_{f_i f_r}^{(0),l}\, 
	\cI_{2\cSCS{}{},1}^{(l)}(x_{\wti{ir}},\Yt{ir}{j},
	\Yt{ir}{k},\Yt{j}{k}; \ep,\al_{0},y_0,d_0,d'_0)\,, 
\label{eq:ICSirsresult} 
\eeq 
if $j,k\ne (ir)$, and  
\beq 
\bsp
[\IcSCS{ir;s}{(0)}]_{f_i f_r}^{(ir,k)} =  
	a_{f_i f_r} \sum_{l=-1}^2 c_{f_i f_r}^{(0),l} \Big[ 
	&\cI_{2\cSCS{}{},2}^{(l)}(x_{\wti{ir}},\Yt{ir}{k};
	\ep,\al_{0},y_0,d_0,d'_0) 
\\ &
	+ \cI_{2\cSCS{}{},3}^{(l)}(x_{\wti{ir}};\ep,\al_{0},y_0,d_0,d'_0)\Big]\,,
\label{eq:ICSirs-MI2} 
\esp
\eeq 
if e.g.,~$j = (ir)$, where $a_{f_i f_r}$ and the coefficients are given 
in \eqn{eq:acoeffs2} and \tab{tab:I2C6coeffs} respectively. 
 
\item Triple collinear -- soft collinear: 
\beq 
\bsp 
[ \IcC{irs}{}\!\IcSCS{ir;s}{(0)} ]_{f_if_r} &= 
	\left( \frac{(4\pi)^2}{S_\ep}Q^{2\ep}\right)^2 
	\int_2 [\rd p_{1;m+1}^{(ir)}(p_{r},\ha{p}_{ir};Q)]\, 
	[\rd p_{1;m}^{(\ha{s})}(\ha{p}_s;Q)] 
\\ &\times 
	\frac{2}{s_{(ir)s}}\frac{1-\tzz{s}{ir}}{\tzz{s}{ir}}\, 
	\frac{1}{s_{ir}}\, \frac{1}{\bT^2_{ir}}\, P^{(0)}_{f_i f_r}(\tzz{r}{i};\ep)\, 
	f(\al_0,\al_{ir},d(m,\ep)) \,
	f(y_0,y_{\ha{s}Q},d'(m,\ep)) \,,
\label{eq:ICCSirs} 
\esp 
\eeq 
which is computed in \App{app:CSirs}. The result is 
\beq 
[\IcC{irs}{}\IcSCS{ir;s}{(0)}]_{f_i f_r} = 
	a_{f_i f_r} \sum_{l=-1}^2 c^{(0),l}_{f_i f_r}\,
	\cI_{2\cSCS{}{},4}^{(l)}(x_{\wti{ir}};\ep,\al_{0},y_0,d_0,d'_0)\,. 
\label{eq:ICCSirsresult}
\eeq 

\item Double collinear -- soft collinear: 
\beq 
\bsp 
[ \IcC{ir;js}{}\!\IcSCS{ir;s}{(0)} ]_{f_if_r} &= 
	\left( \frac{(4\pi)^2}{S_\ep}Q^{2\ep}\right)^2 
	\int_2 [\rd p_{1;m+1}^{(ir)}(p_{r},\ha{p}_{ir};Q)]\, 
	[\rd p_{1;m}^{(\ha{s})}(\ha{p}_s;Q)] 
\\ &\times 
	\frac{2}{s_{js}} \frac{\tzz{j}{s}}{\tzz{s}{j}}\, 
	\frac{1}{s_{ir}}\, \frac{1}{\bT^2_{ir}}\, P^{(0)}_{f_i f_r}(\tzz{r}{i};\ep)\, 
	f(\al_0,\al_{ir},d(m,\ep)) \,
	f(y_0,y_{\ha{s}Q},d'(m,\ep)) \,, 
\label{eq:ICirjsCSirs} 
\esp 
\eeq 
which is also computed in \App{app:CSirs}, as 
\beq 
[\IcC{ir;js}{}\IcSCS{ir;s}{(0)}]_{f_i f_r} = 
	a_{f_i f_r} \sum_{l=-1}^2 c^{(0),l}_{f_i f_r}\,
	\cI_{2\cSCS{}{},5}^{(l)}(x_{\wti{ir}},\Yt{ir}{j};\ep,\al_{0},y_0,d_0,d'_0)\,. 
\label{eq:ICirjsCSirsresult}
\eeq 
\end{enumerate} 
 

\section{Results} 
\label{sec:results} 

After evaluating the integrals as explained in the Appendices, we obtain
the kinematic functions defined in \eqn{eq:cali} as Laurent expansions 
in $\ep$. We compute those expansions analytically up to $\Oe{-2}$, while 
the remaining coefficients up to $\Oe{0}$ are calculated numerically via 
sector decomposition, see \cite{Heinrich:2008si} and references therein.  
In obtaining our results, we use $d_0=D_0+d_1 \ep$ and $d'_0=D'_0+d'_1 \ep$, 
see \eqn{eq:d0s}. The parameters $D_0$, $d_1$, $D'_0$ and $d'_1$, as well 
as $\al_0$ and $y_0$ are left symbolic throughout the analytic computation 
and sector decomposition. 
 
To present analytic results for the non-soft counterterms up to $\Oe{-2}$,
we use the following abbreviations,
\beq 
\gam{q}(\Nf) = \frac{3}{2}\,,\quad 
\gam{g}(\Nf) = \frac{11}{6} - \frac{2\TR}{3\CA}\Nf 
\quad\mbox{and}\quad 
\Sigma(z,N) = \ln z - \sum_{i=1}^{N} \frac{1-(1-z)^k}{k}\,,
\eeq 
where $\gam{q}(\Nf)$ is actually independent of the number of light 
flavours, but introducing the flavour dependence formally makes possible 
a flavour-independent notation. Furthermore, the kinematic functions
are dimensionless in colour space, hence these $\Nf$-dependent 
$\gam{f}(\Nf)$ constants are related to the constans $\gam{f}$ often used
in the literature \cite{Kunszt:1992tn} by $\gam{f} \to C_f \gam{f}(\Nf)$.
 
\begin{enumerate} 
\item Triple collinear: 
\bal 
\Big(\IcC{irs}{(0)}\Big)_{f_i}(x_i) &=  
	{1 \over 2}\bigg(1 + {\CA \over 2 C_{f_i}}\bigg)  
	\bigg[ 
	{1 \over \ep^4} - {1 \over \ep^3} \bigg(4 \ln x_i - \gam{f_i}(\Nf)
	- \gam{f_i}\bigg(\frac{\CF}{C_{f_i}}\Nf\bigg) \bigg) \bigg] 
\nt\\ &
	+ {1 \over \ep^3} {\CA \over 4 C_{f_i}} 
	\gam{g}\bigg(\frac{\CF}{C_{f_i}}\Nf\bigg) 
	+ \Oe{-2} \,. 
\eal 

\item Triple collinear -- soft collinear: 
\bal 
\Big(\IcC{irs}{}\IcSCS{ir;s}{(0)}\Big)_{f_i}(x_i) &= 
	{2 \over 3} \bigg[{1 \over \ep^4} - {2 \over \ep^3}  
	\bigg(\ln x_i + \Sigma(y_0, D'_0 - 1)\bigg)\bigg]  
	+ {1 \over 2\ep^3} \gam{f_i}(\Nf) 
\nt\\ &
	+ \Oe{-2} \,. 
\eal 

\item Double collinear: 
\bal 
\Big(\IcC{ir;js}{(0)}\Big)_{f_if_j}(x_i,x_j) &= 
	{1 \over 2 \ep^4}  
	- {1 \over 2 \ep^3} \bigg[ 2 ( \ln x_i + \ln x_j ) 
	- \gam{f_i}(\Nf) - \gam{f_j}(\Nf) \bigg] 
\nt\\ &
	+ \Oe{-2} \,. 
\eal 

\item Double collinear -- soft collinear: 
\bal 
\Big(\IcC{ir;js}{}\IcSCS{ir;s}{(0)}\Big)_{f_i f}(x_i,Y_{ij;Q}) &= 
	{1 \over \ep^4}  
	- {2 \over \ep^3}\bigg(\ln x_i + \Sigma(y_0,D'_0-1)\bigg) 
	+ {1 \over \ep^3} \gam{f_i}(\Nf) 
\nt\\ &
	+ \Oe{-2} \,. 
\eal 

\item Soft collinear: 
\bal 
\Big(\IcSCS{ir;s}{(0)}\Big)_{f_i}^{(j,l)}(x_i,Y_{jl;Q}) &= 
	- {1 \over \ep^4}  
	+ {2 \over \ep^3}\bigg(\ln x_i + \Sigma(y_0,D'_0-1)\bigg) 
	+ {1 \over \ep^3} \big(\ln Y_{jl;Q} - \gam{f_i}(\Nf)\big) 
\nt\\ &
	+ \Oe{-2}
\intertext{for $j,l \ne i$, and}
\Big(\IcSCS{ir;s}{(0)}\Big)_{f_i}^{(i,l)}(x_i,Y_{il;Q}) &= 
	{5 \over 6} \bigg[- {1 \over \ep^4}  
	+ {2 \over \ep^3} \bigg(\ln x_i  + \Sigma(y_0,D'_0-1)\bigg)\bigg] 
\nt\\ &
	+ {1 \over \ep^3} \bigg( \ln Y_{il;Q}- {3 \over 4} \gam{f_i}(\Nf)\bigg) 
	+ \Oe{-2}
\eal 
for e.g.,~$j=i$.
\end{enumerate} 
 
The remaining coefficients in the Laurent expansion are computed numerically. 
By way of illustration, we present results for the flavour-summed counterterms 
in \tabss{tab:Cirs0q}{tab:Cirjs0gg} for two kinematic points: one relevant for 
$2$-jet production and the other corresponding to the fully symmetric configuration 
of final state momenta in $3$-jet production. In terms of the invariants 
introduced in \eqn{eq:xi_YijQdef}, these two phase-space points correspond to
\bal
\mbox{$2$-jet} \;\;:\;\;
	x_1 &= x_2 = 1
	&\mbox{and}\qquad
	&Y_{12,Q} = 1\,,
\\
\mbox{$3$-jet symmetric} \;\;:\;\;
	x_1 &= x_2 = x_3 = \frac{2}{3}
	&\mbox{and}\qquad
	&Y_{12,Q} = Y_{13,Q} = Y_{23,Q} = \frac{3}{4}\,.
\eal
In the numerical computations, we chose the following values for the 
phase-space cut parameters: $\al_0 = 1$, $y_0 = 1$,  $D_0 = D'_0 = 3$ 
and $d_1 = d'_1 = -3$. 
In \tabss{tab:Cirs0q}{tab:Cirjs0gg}, a displayed error estimate of 
$\pm 0.000$ implies that the numerical uncertainty for that particular 
value is smaller than $10^{-3}$.

%
%

\begin{table}[htbp]
\renewcommand{\arraystretch}{1.2} 
\begin{center}
\begin{tabular}{|c|c|c|c|c|}
\hline\hline
colour & $x$ & $1/\ep^{2}$ &  $1/\ep^{1}$ &  finite
\\
\hline\hline
\multirow{2}{*}{$1$} 
 & 1.000
 & -1.870 $\pm$ 0.003
 & -22.187 $\pm$ 0.020
 & -76.907 $\pm$ 0.155
\\ 
 & 0.667
 & 2.371 $\pm$ 0.004
 & -15.435 $\pm$ 0.023
 & -93.639 $\pm$ 0.156
\\ 
\hline 
\multirow{2}{*}{$\frac{\CA}{\CF}$} 
 & 1.000
 & 2.632 $\pm$ 0.002
 & 4.595 $\pm$ 0.013
 & 13.916 $\pm$ 0.095
\\ 
 & 0.667
 & 6.080 $\pm$ 0.002
 & 21.854 $\pm$ 0.014
 & 84.489 $\pm$ 0.100
\\ 
\hline 
\multirow{2}{*}{$\frac{\Nf \TR}{\CF}$} 
 & 1.000
 & -0.611 $\pm$ 0.001
 & 1.159 $\pm$ 0.005
 & 16.629 $\pm$ 0.037
\\ 
 & 0.667
 & -0.881 $\pm$ 0.001
 & -1.016 $\pm$ 0.006
 & 7.092 $\pm$ 0.046
\\ 
\hline\hline
\end{tabular}
\caption{Coefficients of the Laurent expansion of 
$\Big(\IcC{irs}{(0)}\Big)_q(x)$ for $d_0=3-3\ep$ and $\al_0=1$.}
\label{tab:Cirs0q}
\end{center}
\end{table}

%
%

\begin{table}[htbp]
\renewcommand{\arraystretch}{1.2} 
\begin{center}
\begin{tabular}{|c|c|c|c|c|}
\hline\hline
colour & $x$ & $1/\ep^{2}$ &  $1/\ep^{1}$ &  finite
\\
\hline\hline
\multirow{2}{*}{$1$} 
 & 1.000
 & 3.664 $\pm$ 0.003
 & -8.987 $\pm$ 0.029
 & -42.269 $\pm$ 0.207
\\ 
 & 0.667
 & 11.993 $\pm$ 0.004
 & 19.817 $\pm$ 0.032
 & 30.542 $\pm$ 0.224
\\ 
\hline 
\multirow{2}{*}{$\frac{\Nf \TR}{\CA}$} 
 & 1.000
 & -2.614 $\pm$ 0.005
 & -5.230 $\pm$ 0.047
 & -3.361 $\pm$ 0.345
\\ 
 & 0.667
 & -3.424 $\pm$ 0.006
 & -11.873 $\pm$ 0.049
 & -32.821 $\pm$ 0.372
\\ 
\hline 
\multirow{2}{*}{$\frac{\CF \Nf \TR}{\CA^2}$} 
 & 1.000
 & -4.798 $\pm$ 0.000
 & -17.623 $\pm$ 0.003
 & -45.473 $\pm$ 0.015
\\ 
 & 0.667
 & -5.540 $\pm$ 0.000
 & -24.021 $\pm$ 0.004
 & -74.365 $\pm$ 0.020
\\ 
\hline\hline
\end{tabular}
\caption{Coefficients of the Laurent expansion of 
$\Big(\IcC{irs}{(0)}\Big)_g(x)$ for $d_0=3-3\ep$ and $\al_0=1$.}
\label{tab:Cirs0g}
\end{center}
\end{table}

%
%

\begin{table}[htbp]
\renewcommand{\arraystretch}{1.2} 
\begin{center}
\begin{tabular}{|c|c|c|c|c|c|}
\hline\hline
colour & $x$ & $Y$ & $1/\ep^{2}$ &  $1/\ep^{1}$ &  finite
\\
\hline\hline
\multirow{2}{*}{$1$} 
 & 1.000 & 1.000
 & -7.311 $\pm$ 0.003
 & -3.228 $\pm$ 0.009
 & -9.300 $\pm$ 0.027
\\ 
 & 0.667 & 0.750
 & -13.938 $\pm$ 0.004
 & -20.393 $\pm$ 0.019
 & -27.573 $\pm$ 0.066
\\  
\hline\hline
\end{tabular}
\caption{Coefficients of the Laurent expansion of 
$\Big(\IcSCS{ir;s}{(0)}\Big)^{(j,l)}_q(x,Y)$ for $d_0=d'_0=3-3\ep$ 
and $\al_0=y_0=1$.}
\end{center}
\end{table}

%
%

\begin{table}[htbp]
\renewcommand{\arraystretch}{1.2} 
\begin{center}
\begin{tabular}{|c|c|c|c|c|c|}
\hline\hline
colour & $x$ & $Y$ & $1/\ep^{2}$ &  $1/\ep^{1}$ &  finite
\\
\hline\hline
\multirow{2}{*}{$1$} 
 & 1.000 & 1.000
 & -9.205 $\pm$ 0.003
 & -8.103 $\pm$ 0.009
 & -16.717 $\pm$ 0.028
\\ 
 & 0.667 & 0.750
 & -16.113 $\pm$ 0.004
 & -27.124 $\pm$ 0.019
 & -40.577 $\pm$ 0.067
\\ 
\hline 
\multirow{2}{*}{$\frac{\Nf \TR}{\CA}$} 
 & 1.000 & 1.000
 & 4.454 $\pm$ 0.001
 & 14.204 $\pm$ 0.004
 & 29.039 $\pm$ 0.020
\\ 
 & 0.667 & 0.750
 & 5.017 $\pm$ 0.001
 & 18.479 $\pm$ 0.008
 & 44.490 $\pm$ 0.046
\\ 
\hline\hline
\end{tabular}
\caption{Coefficients of the Laurent expansion of 
$\Big(\IcSCS{ir;s}{(0)}\Big)^{(j,l)}_g(x,Y)$ for $d_0=d'_0=3-3\ep$ 
and $\al_0=y_0=1$.}
\end{center}
\end{table}

%
%

\begin{table}[htbp]
\renewcommand{\arraystretch}{1.2} 
\begin{center}
\begin{tabular}{|c|c|c|c|c|c|}
\hline\hline
colour & $x$ & $Y$ & $1/\ep^{2}$ &  $1/\ep^{1}$ &  finite
\\
\hline\hline
\multirow{2}{*}{$1$} 
 & 1.000 & 1.000
 & -4.874 $\pm$ 0.002
 & 2.910 $\pm$ 0.007
 & 5.559 $\pm$ 0.024
\\ 
 & 0.667 & 0.750
 & -10.330 $\pm$ 0.003
 & -8.326 $\pm$ 0.018
 & 9.292 $\pm$ 0.087
\\ 
\hline\hline
\end{tabular}
\caption{Coefficients of the Laurent expansion of 
$\Big(\IcSCS{ir;s}{(0)}\Big)^{(ir,l)}_q(x,Y)$ for $d_0=d'_0=3-3\ep$ 
and $\al_0=y_0=1$.}
\end{center}
\end{table}

%
%

\begin{table}[htbp]
\renewcommand{\arraystretch}{1.2} 
\begin{center}
\begin{tabular}{|c|c|c|c|c|c|}
\hline\hline
colour & $x$ & $Y$ & $1/\ep^{2}$ &  $1/\ep^{1}$ &  finite
\\
\hline\hline
\multirow{2}{*}{$1$} 
 & 1.000 & 1.000
 & -6.126 $\pm$ 0.002
 & 0.812 $\pm$ 0.007
 & 7.619 $\pm$ 0.025
\\ 
 & 0.667 & 0.750
 & -11.839 $\pm$ 0.003
 & -12.094 $\pm$ 0.018
 & 6.585 $\pm$ 0.089
\\ 
\hline 
\multirow{2}{*}{$\frac{\Nf \TR}{\CA}$} 
 & 1.000 & 1.000
 & 3.005 $\pm$ 0.000
 & 7.201 $\pm$ 0.002
 & 3.080 $\pm$ 0.017
\\ 
 & 0.667 & 0.750
 & 3.517 $\pm$ 0.001
 & 11.052 $\pm$ 0.007
 & 16.471 $\pm$ 0.045
\\ 
\hline\hline
\end{tabular}
\caption{Coefficients of the Laurent expansion of 
$\Big(\IcSCS{ir;s}{(0)}\Big)^{(ir,l)}_g(x,Y)$ for $d_0=d'_0=3-3\ep$ 
and $\al_0=y_0=1$.}
\end{center}
\end{table}

%
%

\begin{table}[htbp]
\renewcommand{\arraystretch}{1.2} 
\begin{center}
\begin{tabular}{|c|c|c|c|c|}
\hline\hline
colour & $x$ & $1/\ep^{2}$ &  $1/\ep^{1}$ &  finite
\\
\hline\hline
\multirow{2}{*}{$1$} 
 & 1.000
 & 3.437 $\pm$ 0.001
 & -6.996 $\pm$ 0.004
 & -61.629 $\pm$ 0.015
\\ 
 & 0.667
 & 5.886 $\pm$ 0.001
 & -6.784 $\pm$ 0.005
 & -100.401 $\pm$ 0.019
\\ 
\hline\hline
\end{tabular}
\caption{Coefficients of the Laurent expansion of 
$\Big(\IcC{irs}{}\IcSCS{ir;s}{(0)}\Big)_q(x)$ for $d_0=d'_0=3-3\ep$ 
and $\al_0=y_0=1$.}
\end{center}
\end{table}

%
%

\begin{table}[htbp]
\renewcommand{\arraystretch}{1.2} 
\begin{center}
\begin{tabular}{|c|c|c|c|c|}
\hline\hline
colour & $x$ & $1/\ep^{2}$ &  $1/\ep^{1}$ &  finite
\\
\hline\hline
\multirow{2}{*}{$1$} 
 & 1.000
 & 4.048 $\pm$ 0.001
 & -6.973 $\pm$ 0.004
 & -66.612 $\pm$ 0.016
\\ 
 & 0.667
 & 6.633 $\pm$ 0.001
 & -5.975 $\pm$ 0.005
 & -103.489 $\pm$ 0.019
\\ 
\hline 
\multirow{2}{*}{$\frac{\Nf \TR}{\CA}$} 
 & 1.000
 & -1.556 $\pm$ 0.000
 & -1.601 $\pm$ 0.001
 & 8.364 $\pm$ 0.011
\\ 
 & 0.667
 & -1.826 $\pm$ 0.000
 & -3.445 $\pm$ 0.001
 & 2.731 $\pm$ 0.011
\\ 
\hline\hline
\end{tabular}
\caption{Coefficients of the Laurent expansion of 
$\Big(\IcC{irs}{}\IcSCS{ir;s}{(0)}\Big)_g(x)$ for $d_0=d'_0=3-3\ep$ 
and $\al_0=y_0=1$.}
\end{center}
\end{table}

%
%

\begin{table}[htbp]
\renewcommand{\arraystretch}{1.2} 
\begin{center}
\begin{tabular}{|c|c|c|c|c|c|}
\hline\hline
colour & $x$ & $Y$ & $1/\ep^{2}$ &  $1/\ep^{1}$ &  finite
\\
\hline\hline
\multirow{2}{*}{$1$} 
 & 1.000 & 1.000
 & 8.311 $\pm$ 0.003
 & 11.149 $\pm$ 0.010
 & 40.832 $\pm$ 0.028
\\ 
 & 0.667 & 0.750
 & 13.099 $\pm$ 0.002
 & 24.109 $\pm$ 0.011
 & 61.583 $\pm$ 0.031
\\ 
\hline\hline
\end{tabular}
\caption{Coefficients of the Laurent expansion of 
$\Big(\IcC{ir;js}{}\IcSCS{ir;s}{(0)}\Big)_{qf}(x,Y)$ for $d_0=d'_0=3-3\ep$ 
and $\al_0=y_0=1$.}
\end{center}
\end{table}

%
%

\begin{table}[htbp]
\renewcommand{\arraystretch}{1.2} 
\begin{center}
\begin{tabular}{|c|c|c|c|c|c|}
\hline\hline
colour & $x$ & $Y$ & $1/\ep^{2}$ &  $1/\ep^{1}$ &  finite
\\
\hline\hline
\multirow{2}{*}{$1$} 
 & 1.000 & 1.000
 & 10.205 $\pm$ 0.003
 & 16.358 $\pm$ 0.010
 & 51.240 $\pm$ 0.029
\\ 
 & 0.667 & 0.750
 & 15.179 $\pm$ 0.002
 & 30.471 $\pm$ 0.011
 & 75.387 $\pm$ 0.032
\\ 
\hline 
\multirow{2}{*}{$\frac{\Nf \TR}{\CA}$} 
 & 1.000 & 1.000
 & -4.454 $\pm$ 0.001
 & -14.871 $\pm$ 0.004
 & -35.687 $\pm$ 0.020
\\ 
 & 0.667 & 0.750
 & -4.825 $\pm$ 0.001
 & -17.549 $\pm$ 0.004
 & -45.158 $\pm$ 0.022
\\ 
\hline\hline
\end{tabular}
\caption{Coefficients of the Laurent expansion of 
$\Big(\IcC{ir;js}{}\IcSCS{ir;s}{(0)}\Big)_{gf}(x,Y)$ for $d_0=d'_0=3-3\ep$ 
and $\al_0=y_0=1$.}
\end{center}
\end{table}

%
%

\begin{table}[htbp]
\renewcommand{\arraystretch}{1.2} 
\begin{center}
\begin{tabular}{|c|c|c|c|c|c|}
\hline\hline
colour & $x_1$ & $x_2$ & $1/\ep^{2}$ &  $1/\ep^{1}$ &  finite
\\
\hline\hline
\multirow{2}{*}{$1$} 
 & 1.000 & 1.000
 & 0.162 $\pm$ 0.000
 & -4.108 $\pm$ 0.003
 & 5.189 $\pm$ 0.013
\\ 
 & 0.667 & 0.667
 & 4.408 $\pm$ 0.001
 & 5.732 $\pm$ 0.004
 & 18.494 $\pm$ 0.018
\\ 
\hline\hline
\end{tabular}
\caption{Coefficients of the Laurent expansion of 
$\Big(\IcC{ir;js}{(0)}\Big)_{qq}(x_1,x_2)$ for $d_0=3-3\ep$ 
and $\al_0=1$.}
\end{center}
\end{table}

%
%

\begin{table}[htbp]
\renewcommand{\arraystretch}{1.2} 
\begin{center}
\begin{tabular}{|c|c|c|c|c|c|}
\hline\hline
colour & $x_1$ & $x_2$ & $1/\ep^{2}$ &  $1/\ep^{1}$ &  finite
\\
\hline\hline
\multirow{2}{*}{$1$} 
 & 1.000 & 1.000
 & 0.859 $\pm$ 0.000
 & -3.131 $\pm$ 0.003
 & 5.070 $\pm$ 0.013
\\ 
 & 0.667 & 0.667
 & 5.333 $\pm$ 0.001
 & 8.076 $\pm$ 0.004
 & 22.304 $\pm$ 0.019
\\ 
\hline 
\multirow{2}{*}{$\frac{\Nf \TR}{\CA}$} 
 & 1.000 & 1.000
 & -1.727 $\pm$ 0.000
 & -3.683 $\pm$ 0.002
 & -3.445 $\pm$ 0.008
\\ 
 & 0.667 & 0.667
 & -2.183 $\pm$ 0.000
 & -6.871 $\pm$ 0.002
 & -14.493 $\pm$ 0.011
\\ 
\hline\hline
\end{tabular}
\caption{Coefficients of the Laurent expansion of 
$\Big(\IcC{ir;js}{(0)}\Big)_{qg}(x_1,x_2)$ for $d_0=3-3\ep$ 
and $\al_0=1$, with
$\Big(\IcC{ir;js}{(0)}\Big)_{gq}(x_1,x_2) = 
\Big(\IcC{ir;js}{(0)}\Big)_{qg}(x_2,x_1)$.}
\end{center}
\end{table}

%
%

\begin{table}[htbp]
\renewcommand{\arraystretch}{1.2} 
\begin{center}
\begin{tabular}{|c|c|c|c|c|c|}
\hline\hline
colour & $x_1$ & $x_2$ & $1/\ep^{2}$ &  $1/\ep^{1}$ &  finite
\\
\hline\hline
\multirow{2}{*}{$1$} 
 & 1.000 & 1.000
 & 1.612 $\pm$ 0.000
 & -1.855 $\pm$ 0.003
 & 5.670 $\pm$ 0.013
\\ 
 & 0.667 & 0.667
 & 6.314 $\pm$ 0.001
 & 10.780 $\pm$ 0.004
 & 27.243 $\pm$ 0.019
\\ 
\hline 
\multirow{2}{*}{$\frac{\Nf \TR}{\CA}$} 
 & 1.000 & 1.000
 & -3.677 $\pm$ 0.000
 & -8.780 $\pm$ 0.003
 & -11.184 $\pm$ 0.011
\\ 
 & 0.667 & 0.667
 & -4.589 $\pm$ 0.000
 & -15.404 $\pm$ 0.002
 & -35.158 $\pm$ 0.016
\\ 
\hline 
\multirow{2}{*}{$\frac{\Nf^2 \TR^2}{\CA^2}$} 
 & 1.000 & 1.000
 & 0.222 $\pm$ 0.000
 & 1.637 $\pm$ 0.000
 & 5.932 $\pm$ 0.003
\\ 
 & 0.667 & 0.667
 & 0.222 $\pm$ 0.000
 & 1.884 $\pm$ 0.000
 & 8.057 $\pm$ 0.004
\\ 
\hline\hline
\end{tabular}
\caption{Coefficients of the Laurent expansion of 
$\Big(\IcC{ir;js}{(0)}\Big)_{gg}(x_1,x_2)$ for $d_0=3-3\ep$ 
and $\al_0=1$.}
\label{tab:Cirjs0gg}
\end{center}
\end{table}


\section{Conclusions} 
\label{sec:conclusions} 
 
We have computed the integrals over the two-particle factorised 
phase space of the collinear-type contributions to the doubly 
unresolved counterterm of the NNLO subtraction formalism, defined 
in \Refs{Somogyi:2006da,Somogyi:2006db}. We presented those integrals 
in terms of parametric representations that are suitable for evaluation 
with sector decomposition. After evaluating them, we checked the numerical 
results with the publicly available code {\tt SecDec} \cite{Carter:2010hi}, 
always finding agreement within the numerical uncertainty of the integrations.
The soft-type contributions are presented in a companion 
paper~\cite{Somogyi:inprep}.

By these two papers, we complete the integration of the subtraction 
terms over the unresolved phase spaces, and the computation of the finite 
cross section in \eqn{eq:sigmaNNLOm+1} becomes feasible for electron-positron
annihilation into two and three jets. Although the formalism is complete, for 
a higher number of jets some more work is required, because some integrals 
were evaluated specifically for three-jet kinematics.

 
\acknowledgments{ 
This research was supported by
the LHCPhenoNet network PITN-GA-2010-264564,
the T\'AMOP 4.2.1./B-09/1/KONV-2010-0007 project,
the Hungarian Scientific Research Fund grant K-101482 and
the Swiss National Science Foundation Joint Research Project 
SCOPES IZ73Z0\_1/28079.
We are grateful to Francesco Tramontano for useful discussions.
G.S.~is grateful to the INFN Laboratori Nazionali di Frascati for the 
hospitality during the late stages of the work.} 


\begin{appendix} 


\section{Modified doubly real subtraction terms}
\label{app:modsubterms} 

We outline a simple modification to the NNLO subtraction scheme
presented in \Refs{Somogyi:2006da,Somogyi:2006db}. Parts of these
modifications were introduced previously: those relevant to the
singly unresolved approximate cross section $\dsiga{RR}{1}_{m+2}$
in \eqn{eq:sigmaNNLOm+2}, and to the approximate cross sections 
in \eqn{eq:sigmaNNLOm+1}, were introduced in \Ref{Somogyi:2008fc}, 
while those relevant to the iterated doubly unresolved approximate 
cross section $\dsiga{RR}{12}_{m+2}$ appearing in \eqn{eq:sigmaNNLOm+2} 
were presented in \Ref{Bolzoni:2010bt}.

Recall that the doubly unresolved approximate cross section can
be written symbolically as in \eqn{eq:dsigRRA2Jm} 
\beq
\dsiga{RR}{2}_{m+2} = \PS{m}[\rd p_2]\cA{2}^{(0)}\M{m+2}{(0)}\,,
\label{eq:dsigRRA2appx}
\eeq
where the doubly unresolved approximation is a sum of terms (see
\eqn{eq:dsigRRA2Jmfull}). The precise definition of these terms
involves the momentum mappings discussed in \sect{sec:psfact}. All 
such mappings lead to an exact factorisation of the $m+2$-particle 
phase space, symbolically written as
\beq
\PS{m+2}(\mom{};Q) = \PS{m}(\momt{}_{m};Q)[\rd p_{2;m}]\,.
\eeq
The only feature of the factorised phase spaces $[\rd p_{2;m}]$ that is
relevant presently is that they carry a dependence on the number of
partons, $m$, of the form
\bal
[\rd p_{2;m}^{(irs)}] &\propto
   (1-\al_{irs})^{2(m-1)(1-\ep)-1}\,,
\\
[\rd p_{2;m}^{(ir;js)}] &\propto
   (1-\al_{ir}-\al_{js})^{2(m-1)(1-\ep)}\,,
\\
[\rd p_{2;m}^{(rs)}] &\propto (1-y_{rQ}-y_{sQ} + y_{rs})^{(m-1)(1-\ep)-1}\,,
\intertext{and finally}
[\rd p_{2;m}^{(\ha{s},ir)}] =
  [\rd p_{1;m+1}^{(ir)}][\rd p_{1;m}^{(\ha{s})}] &\propto
  (1-\al_{ir})^{2m(1-\ep)-1}\,(1-y_{\ha{s}Q})^{(m-1)(1-\ep)-1}\,.
\eal
The subtraction terms, as originally defined in \Ref{Somogyi:2006da} do
not depend on the number of hard partons, thus the $m$-dependence of
the factorised phase space measures is carried over to the integrated
counterterms, where this dependence enters in a rather cumbersome way
(see e.g.,~Eqs.~(A.9) and (A.10) of \Ref{Somogyi:2006cz}).

Thus, as in \Refs{Somogyi:2008fc,Bolzoni:2010bt}, we reshuffle the 
$m$-dependence of the integrated counterterms into the subtraction 
terms themselves, where it appears in a very straightforward and harmless 
way\footnote{The modifications introduced above do not spoil any of the
cancellations that take place among the original subtraction terms.
Hence the modified counterterms are still correct regulators of all
kinematic singularities.}, through factors of $(1-\al)$ and/or $(1-y)$ 
raised to $m$-dependent powers.  We gather the results in \tab{tab:modDU}, 
where together with the subtraction terms, we give the momentum mappings 
used to define the term(s) and the expression which multiplies the original 
counterterm to produce the modified one. The function $f$ in \tab{tab:modDU} 
is defined as
\beq
f(z_0,z,p) = \Theta(z_0 - z)(1-z)^{-p}\,.
\label{eq:f_def}
\eeq
\begin{table}[t]
\renewcommand{\arraystretch}{1.5}
\footnotesize
\begin{center}
\begin{tabular}{|c|c|c|}
\hline\hline
\multicolumn{3}{|c|}{Doubly unresolved counterterms} \\
\hline\hline
Subtraction term & Momentum mapping & Function \\
\hline
\multirow{1}{3.9cm}{\centering
$\cC{irs}{(0,0)}$}
&
\multirow{1}{5.0cm}{\centering
$\mom{} \cmap{irs} \momt{(irs)}_{m}$} 
&
$f(\al_0,\al_{irs},d(m,\ep))$
\\
\hline
\multirow{1}{3.9cm}{\centering
$\cC{ir;js}{(0,0)}$}
&
\multirow{1}{5.0cm}{\centering
$\mom{} \cmap{ir;js} \momt{(ir;js)}_{m}$}
&
$f(\al_0,\al_{ir}+\al_{js},d(m,\ep))$
\\
\hline
$\cSCS{ir;s}{(0,0)}$, $\cC{irs}{}\cSCS{ir;s}{(0,0)}$,
&
\multirow{2}{5.0cm}{\centering
$\mom{} \cmap{ir} \momh{(ir)}_{m+1} \smap{\ha{s}} 
\momt{(\ha{s},ir)}_{m}$}
&
\multirow{2}{4.9cm}{\centering
$f(\al_0,\al_{ir},d(m,\ep))$\\
$\times f(y_0,y_{\ha{s}Q},d'(m,\ep))$
}
\\
$\cC{ir;js}{}\cSCS{ir;s}{(0,0)}$ & & \\
\hline
$\cS{rs}{(0,0)}$, $\cSCS{ir;s}{}\cS{rs}{(0,0)}$,
&
\multirow{3}{5.0cm}{\centering
$\mom{} \smap{rs} \momt{(rs)}_{m}$}
&
\multirow{3}{4.9cm}{\centering
$f(y_0, y_{rQ}+y_{sQ}-y_{rs},d'(m,\ep))$} \\
$\cC{irs}{}\cS{rs}{(0,0)}$, $\cC{ir;js}{}\cS{rs}{(0,0)}$, & & \\
$\cC{irs}{}\cSCS{ir;s}{}\cS{rs}{(0,0)}$ & & \\
\hline\hline
\end{tabular}
\normalsize
\caption{\label{tab:modDU}
The modified doubly unresolved subtraction terms are obtained
from the original counterterms (first column) by multiplication with an
appropriate function (last column). Also shown are the momentum
mappings used to define the subtraction terms (middle column).}
\end{center}
\end{table}
It is important to note that the functions $d(m,\ep)$, $d'(m,\ep)$
and constants $\al_0$, $y_0$ in \tab{tab:modDU} are the same as
those in all other modified subtraction terms, discussed in
\Refs{Somogyi:2008fc,Bolzoni:2010bt}. 

The form of the exponents $d(m,\ep)$ and $d'(m,\ep)$ is actually fixed
by the prescription adopted in  \Ref{Somogyi:2008fc} (see in particular Eqs.~(3.2), 
(3.12) and (3.13) therein) and the requirement that the modified
subtraction terms should still correctly regularise all kinematic
singularities. In fact, we must have
\beq
d(m,\ep) =  2m(1-\ep) - 2d_0\,,
\quad\mbox{and}\quad
d'(m,\ep) =  m(1-\ep) - d'_0\,,
\label{eq:ddp}
\eeq
where $d_0$ and $d'_0$ are the same constants which appear in
eqs.~(3.2), (3.12) and (3.13) of \Ref{Somogyi:2008fc} i.e.,
\beq
d_0 = D_0 + d_1\ep\,,
\quad\mbox{and}\quad
d'_0 = D'_0 + d'_1\ep\,,
\label{eq:d0s}
\eeq
where $D_0 ,D'_0 \ge 2$ are integers, while $d_1, d'_1$ are real.


\section{Spin-averaged splitting kernels} 
\label{app:P0} 
 
In this Appendix, we list the spin-averaged splitting kernels. 
Although some of these already appeared elsewhere, for the sake
of completeness, we give all the splitting functions of our 
computations.

%
%
 
\subsection{Two-parton kernels} 
\label{app:P0_2}
 
The azimuthally averaged two-parton splitting kernels are well known,
\bal 
P^{(0)}_{gg}(z) &= 
	2\CA \left[\frac{1}{z}+\frac{1}{1-z}-2+z-z^2\right]\,, 
\label{eq:P0gg} 
\\
P^{(0)}_{q\qb}(z;\ep) &= 
	\TR \left[1-\frac{2}{1-\ep}\left(z-z^2\right)\right]\,, 
\label{eq:P0qq} 
\\
P^{(0)}_{qg}(z;\ep) &= 
	\CF \left[\frac{2}{z}-2 + (1-\ep)z\right]\,. 
\label{eq:P0qg} 
\eal 
In our convention the ordering of the labels on the splitting kernels 
is usually meaningless, but in \eqn{eq:P0qg} $z$ means the momentum 
fraction of the second label, i.e.,~$P^{(0)}_{gq}(z;\ep) = 
P^{(0)}_{qg}(1-z;\ep)$. The other two cases are symmetric with respect 
to $z \leftrightarrow 1-z$. 
 
%
%

\subsection{Three-parton kernels}  
\label{app:P0_3}

The spin average of the splitting kernels was computed in \Ref{Catani:1999ss},
however the forms presented there for gluon splittings are not suitable
for us, as it was explained in \Ref{Somogyi:2006da}: in the gluon
splitting kernels $\hP_{g_i q_r \qb_s}$ and $\hP_{g_i g_r g_s}$, the
terms that depend on the transverse momenta must always be written
in the form $\kT{j}^\mu\kT{k}^\nu/\kT{j}\cdot\kT{k}$ ($k$ can be equal
to $j$). Otherwise the collinear behaviour of the counterterm cannot be
matched with that of the singly collinear counterterm in the
singly unresolved region of phase space. The correct azimuth dependence
can be achieved by substitutions according to the following replacements,
\beq
\bsp
\kT{j}^\mu \kT{j}^\nu &\to
	\Big(-z_j (1-z_j) s_{jkl} + z_j s_{kl}\Big)
	\frac{\kT{j}^\mu \kT{j}^\nu}{\kT{j}^2} \,,
\\
2\,\kT{j}^\mu \kT{k}^\nu &\to
	\Big(s_{jk} + 2\,z_j z_k s_{jkl} - z_i s_{j(ik)} - z_j s_{i(jk)}\Big)
	\frac{\kT{j}^\mu \kT{k}^\nu}{\kT{j}\cdot \kT{k}}\,,
\esp
\eeq
where $\{k,l\} = \{i,r,s\}\setminus \{j\}$ and $j$ can be $i$, $r$ or $s$. 
With these forms azimuthal averaging amounts to the simple substitutions,
\beq
\left\langle
\frac{\kT{j}^\mu \kT{k}^\nu}{\kT{j}\cdot \kT{k}}
\right\rangle = 
	-\frac{1}{2(1-\ep)}\,,
\eeq
where $\langle\ldots\rangle$ denote spin averaging, and $j=k$ is also
allowed.

For quark splitting into unequal and equal quark flavours we have,
\beq
P_{q_i \qb'_r q'_s}^{(0)}(\{z_{j,kl},s_{jk}\};\ep) = 
	P_{q_i \qb'_r q'_s}(\{z_{j,kl},s_{jk}\};\ep)
\eeq
and
\beq
P_{q_i \qb_r q_s}^{(0)}(\{z_{j,kl},s_{jk}\};\ep) = 
	2 P_{q_i \qb'_r q'_s}(\{z_{j,kl},s_{jk}\};\ep)
	+ P_{q_i \qb'_r q'_s}^{(\ID)}(\{z_{j,kl},s_{jk}\};\ep)\,,
\eeq
respectively (hence \eqn{eq:decomp}), where
%
%

\beq 
\bsp 
\frac{1}{s_{irs}^2} P_{q_i \qb'_r q'_s}(\{z_{j,kl},s_{jk}\};\ep) &= 
	\CF \TR \bigg\{ {1 \over s_{irs} s_{rs}} 
	\bigg[ {z_{i,rs} \over z_{r,is} + z_{s,ir}} 
	- {s_{ir} z_{s,ir} + s_{is} z_{r,is} \over s_{rs} (z_{r,is} + z_{s,ir})} 
	+ {s_{ir} s_{is} \over s_{irs} s_{rs}} 
\\ & 
	+ {s_{irs} \over s_{rs}} {z_{r,is} z_{s,ir} \over (z_{r,is} + z_{s,ir})^2} 
	-{z_{r,is} z_{s,ir} \over z_{r,is} + z_{s,ir}}  
	+{1-\ep \over 2} \bigg(z_{r,is} + z_{s,ir} - {s_{rs} \over s_{irs}}\bigg)\bigg] 
\\ & 
	+ (r \leftrightarrow s)  
	\bigg\} 
\esp  
\label{A-eq:Pqiqbrqs} 
\eeq 
and
%
%
\beq 
\bsp 
\frac{1}{s_{irs}^2} P_{q_i \qb'_r q'_s}^{(\ID)}(\{z_{j,kl},s_{jk}\};\ep) &= 
	\CF\left(\CF-{\CA \over 2}\right) \bigg\{ {1-\ep \over s_{irs}^2} 
	\left({2 s_{is} \over s_{rs}}  - \ep \right) 
	+ {1 \over s_{rs} s_{irs}} \bigg[ {1 + z_{r,is}^2 \over z_{i,rs} + z_{r,is}}  
\\ & 
	- {2 z_{s,ir} \over z_{r,is} + z_{s,ir}} 
	- \ep \left( {(z_{r,is} + z_{s,ir})^2 \over z_{i,rs} + z_{r,is}} 
	+ 1 + z_{r,is} - {2 z_{s,ir} \over z_{r,is} + z_{s,ir}}\right) 
\\ & 
	- \ep^2 (z_{r,is} + z_{s,ir})\bigg] - {1 \over s_{ir} s_{rs}} {z_{r,is} \over 2} 
	\bigg[ {1 + z_{r,is}^2 \over (z_{i,rs} + z_{r,is})(z_{r,is} + z_{s,ir})}  
\\ & 
	- \ep \left(1 + 2 {z_{i,rs} + z_{r,is} \over z_{r,is} + z_{s,ir}}\right) 
	- \ep^2\bigg] 
	+ (i \leftrightarrow s)  
	\bigg\} 
\,.
\esp 
\eeq 

For splitting into a quark and a gluon pair we find,
\beq
P_{q_i g_r g_s}^{(0)}(\{z_{j,kl},s_{jk}\};\ep) = 
	\CF P_{q_, g_r g_s}^{(\AB)}(\{z_{j,kl},s_{jk}\};\ep)
	+ \CA P_{q_i g_r g_s}^{(\NAB)}(\{z_{j,kl},s_{jk}\};\ep)\,,
\label{eq:PqggAB-NAB}
\eeq
where 
%
%
 
\beq 
\bsp 
\frac{1}{s_{irs}^2} \CF P_{q_i g_r g_s}^{(\AB)}(\{z_{j,kl},s_{jk}\};\ep) &= 
	\CF^2 \bigg\{ \left({1-z_{s,ir} \over s_{irs} s_{ir}} 
	+ {1-z_{r,is} \over s_{irs} s_{is}} 
	+ {1 - z_{r,is} - z_{s,ir} \over s_{ir} s_{is}}\right)
\\ &\times
\bigg[ {1\over z_{r,is}}  \bigg( {2 \over z_{r,is} + z_{s,ir}} - 2 
	+ (1-\ep) z_{s,ir}\bigg) + 1 -{\ep(1+\ep) \over 2}\bigg]
\\ & 
+ {1-\ep \over s_{irs}^2} 
	\bigg[ \ep - {s_{irs} \over s_{ir}} (1+\ep) (3 - z_{r,is} - 2 z_{s,ir}) 
	- {s_{is} \over s_{ir}} (1-\ep)\bigg]  
\\ & 
	+ (r \leftrightarrow s)  
	\bigg\} 
\esp 
\eeq 
%
%
and 
\beq 
\bsp 
\frac{1}{s_{irs}^2} \CA &P_{q_i g_r g_s}^{(\NAB)}(\{z_{j,kl},s_{jk}\};\ep) = 
\\ & 
	\CA \CF \bigg\{ {1 \over s_{irs} s_{rs}} 
	\bigg[ (1-\ep) \bigg( {s_{ir} z_{s,ir} + s_{is} z_{r,is}  
	\over s_{rs} (z_{r,is} + z_{s,ir})} - {s_{ir} s_{is} \over s_{irs} s_{rs}} 
	- {s_{irs} \over s_{rs}} 
	{z_{r,is} z_{s,ir} \over (z_{r,is} + z_{s,ir})^2} \bigg) 
\\ & 
	- z_{i,rs} \left( {4 \over z_{r,is} + z_{s,ir}} - {1 \over z_{r,is}}\right)\bigg]  
	- {1 \over s_{irs} s_{ir}} {(1 - z_{r,is})^2 
	+ (1 - z_{s,ir})^2 \over 2 z_{r,is} (z_{r,is} + z_{s,ir})} 
\\ & 
	- {1 \over s_{ir} s_{is}} 
	{z_{i,rs} \over 2 z_{r,is}} {1 + z_{i,rs}^2 \over z_{r,is} + z_{s,ir}} 
	+ {1 \over 2 s_{ir} s_{rs}} \left[ {1 + z_{i,rs}^2 \over z_{s,ir}} 
	+ {1 + (1 - z_{s,ir})^2  
	\over z_{r,is} + z_{s,ir}}\right] + {(1-\ep)^2 \over 2 s_{irs}^2} 
\\ & 
	+ {1 \over s_{irs} s_{rs}} \left[(1 - \ep)\left({2 \over z_{r,is}} 
	- {1 \over z_{r,is} + z_{s,ir}}\right) 
	{(z_{r,is} - z_{s,ir})^2 \over 4} - 1\right] + {1 \over 2 s_{irs} s_{ir}} 
\\ & \times 
	\left[{1 + (1 - z_{s,ir})^2 \over z_{r,is}} 
	- {4 - 2 z_{s,ir} + z_{s,ir}^2 - z_{r,is}  \over z_{r,is} + z_{s,ir}}\right] 
	+ {\ep \over 2} \bigg[ {1 \over s_{irs} s_{ir}} \bigg( (1 - z_{s,ir}) 
\\ & \times	 
	\left({z_{r,is} \over z_{s,ir}} 
	+ {z_{s,ir} \over z_{r,is}} - \ep\right) - {z_{r,is}^2  
	(1 - z_{s,ir}) \over z_{s,ir} (z_{r,is} + z_{s,ir})}\bigg) 
	+ {z_{i,rs} \over s_{ir} s_{is}} 
	\left({z_{r,is} \over z_{s,ir}} + {1+\ep \over 2}\right) 
\\ & 
	- {1 \over s_{ir} s_{rs}} \left({(z_{r,is} + z_{s,ir})^2 \over z_{s,ir}} 
	+ {z_{s,ir}^2 \over z_{r,is} + z_{s,ir}}\right)\bigg]  
	+ (r \leftrightarrow s)  
	\bigg\} 
\,.
\esp 
\label{eq:PqggNAB}
\eeq 
We call attention to the normalisation (in colour space) of the abelian 
and non-abelian parts of the splitting function in \eqn{eq:PqggAB-NAB}, which 
is the same as in \Ref{Somogyi:2005xz}. 
Notice that a factor of $\CF$, respectively $\CA$, is made explicit in 
the definition of the abelian, respectively non-abelian, part as compared 
to the complete splitting function. 
However, we prefer to define $[ \IcC{irs}{(0)} ]_{f_i f_r f_s}^{(\AB)}$ 
and $[ \IcC{irs}{(0)} ]_{f_i f_r f_s}^{(\NAB)}$ to be dimensionless in 
colour space, hence the factors of $\CF$ and $\CA$ are {\em not} made explicit 
in the definition of these functions, see \eqn{eq:ICirsAB-NAB}.
Then, \eqn{eq:ICirs0} must be interpreted with some care when computing
$[ \IcC{irs}{(0)} ]_{f_i f_r f_s}^{(\AB)}$ or 
$[ \IcC{irs}{(0)} ]_{f_i f_r f_s}^{(\NAB)}$. In particular, we must remember 
to include the factors of $\CF$ and $\CA$ explicitly with 
$P_{f_i f_r f_s}^{(\AB)}$ and $P_{f_i f_r f_s}^{(\NAB)}$, i.e., we must set 
$P_{f_i f_r f_s}^{(0)} \to \CF P_{f_i f_r f_s}^{(\AB)}$ or 
$P_{f_i f_r f_s}^{(0)} \to \CA P_{f_i f_r f_s}^{(\NAB)}$ to obtain the correct 
normalisation.
 
For gluon splitting into a gluon and a quark pair we have
\beq
P_{g_i q_r \qb_s}^{(0)}(\{z_{j,kl},s_{jk}\};\ep) = 
	\CF P_{g_i q_r \qb_s}^{(\AB)}(\{z_{j,kl},s_{jk}\};\ep)
	+\CA P_{g_i q_r \qb_s}^{(\NAB)}(\{z_{j,kl},s_{jk}\};\ep)
\,,
\label{eq:PgqqAB-NAB}
\eeq 
where

%
%
\beq 
\bsp 
\frac{1}{s_{irs}^2} \CF P_{g_i q_r \qb_s}^{(\AB)}(\{z_{j,kl},s_{jk}\};\ep) &= 
	\CF \TR \bigg\{ \left[{1 \over s_{ir} s_{is}} - {1 \over s_{irs}^2} 
	- {2 \over s_{irs} s_{ir}} 
	\left(1 - {1 - \ep \over 2} {s_{ir} + s_{is} \over s_{irs}}\right)\right]  
\\ & 
	- {1 \over 1 - \ep} {1 \over s_{irs} s_{ir} s_{is}} 
	\bigg[ 2 z_{r,is} z_{s,ir} s_{irs} + z_{i,rs} s_{rs} - z_{r,is} s_{is} 
	- z_{s,ir} s_{ir} 
\\ & 
	- (1 - \ep)(z_{i,rs} s_{rs} - z_{i,rs} (1 - z_{i,rs}) s_{irs})\bigg] 
	+ (r \leftrightarrow s)  
	\bigg\} 
\esp 
\eeq 
%
%
and
\beq 
\bsp 
\frac{1}{s_{irs}^2} \CA &P_{g_i q_r \qb_s}^{(\NAB)}(\{z_{j,kl},s_{jk}\};\ep) =
\\ & 
	\CA \TR \bigg\{ {1 \over s_{irs} s_{rs}} \bigg({1 \over z_{r,is} + z_{s,ir}} 
	- 1 - {s_{ir} z_{s,ir} + s_{is} z_{r,is} \over s_{rs} (z_{r,is} + z_{s,ir})} 
	+ {s_{ir} s_{is} \over s_{rs} s_{irs}}  
\\ & 
	+ {s_{irs} \over s_{rs}} {z_{r,is} z_{s,ir} \over (z_{r,is} + z_{s,ir})^2}  
	+ {1 \over 2 z_{i,rs}}\bigg) 
	+ {1 \over s_{ir} s_{rs}} \left({z_{r,is} \over 2 z_{i,rs}} 
	- {z_{r,is} \over 2(z_{r,is} + z_{s,ir})}\right) 
\\ & 
	+ {1 \over s_{irs} s_{is}} {1 - z_{r,is} \over 2} \left( {1 \over z_{i,rs}} 
	+ {1 \over z_{r,is} + z_{s,ir}}\right) - {1 \over 2 s_{ir} s_{is}} 
	- {1 - \ep \over 2 s_{irs}^2} 
\\ & 
	- {1 \over s_{irs} s_{rs}} {1 \over 1 - \ep} {2 z_{r,is} z_{s,ir} 
	\over z_{i,rs} (z_{r,is} + z_{s,ir})} 
	- {1 \over s_{irs} s_{ir} s_{rs}} \bigg[ z_{r,is} s_{is} - z_{r,is} 
	(1 - z_{r,is}) s_{irs} 
\\ & 
	- {1 \over 1 - \ep} {2 z_{r,is}^2 \over z_{i,rs} (z_{r,is} + z_{s,ir})} 
	(z_{s,ir} s_{ir}  
	- z_{s,ir} (1 - z_{s,ir}) s_{irs}) + \bigg( {2 z_{r,is} (z_{s,ir} - z_{i,rs}) 
	\over z_{i,rs}  
	(z_{r,is} + z_{s,ir})}  
\\ & 
	+ 1 - \ep \bigg) {1 \over 2 (1 - \ep)}(2 z_{r,is} z_{s,ir} s_{irs} 
	+ z_{i,rs} s_{rs} - z_{r,is}  
	s_{is} - z_{s,ir} s_{ir})\bigg] + {1 \over s_{irs} s_{ir} s_{is}} 
\\ & \times 
	\bigg[(2 z_{r,is} z_{s,ir} s_{irs} + z_{i,rs} s_{rs} - z_{r,is} s_{is} 
	- z_{s,ir} s_{ir}) 
	{1 \over 2 (1 - \ep)} - {1 \over 2} (z_{i,rs} s_{rs}  
\\ & 
	- z_{i,rs} (1 - z_{i,rs})s_{irs})\bigg] 
		+ (r \leftrightarrow s)  
		\bigg\} 
\,.
\esp 
\eeq 
Note again that the normalisation (in colour space) of the abelian 
and non-abelian parts of the splitting function in \eqn{eq:PgqqAB-NAB} 
matches that of \Ref{Somogyi:2005xz}, i.e., a factor of $\CF$ and 
$\CA$ is made explicit in the definition of the abelian and non-abelian 
piece respectively. Hence, the comments below \eqn{eq:PqggNAB} apply in 
this case as well.

Finally for splitting into three gluons we have
%
%
\beq 
\bsp 
\frac{1}{s_{irs}^2} &P^{(0)}_{g_i g_r g_s}(\{z_{j,kl},s_{jk}\};\ep) = 
\\ & 
	\CA^2 {1 \over s_{irs}^2} \bigg\{ {(1 - \ep) \over 4 s_{rs}^2} 
	\left(2 {s_{is} z_{r,is} - s_{ir} z_{s,ir}  
	\over z_{r,is} + z_{s,ir}} + {z_{r,is} - z_{s,ir} \over z_{r,is} 
	+ z_{s,ir}} s_{rs}\right)^2  
	+{3 \over 4} (1 - \ep) 
\\ & 
	+ {2 s_{irs} z_{r,is} z_{s,ir} \over s_{rs} z_{i,rs} (1 - z_{i,rs})} 
	- {s_{irs} \over s_{rs}}  
	{1 \over z_{i,rs}} \left[ {2(1 - z_{i,rs}) + 4 z_{i,rs}^2 \over 1 - z_{i,rs}} 
	- {1 - 2 z_{i,rs}  
	(1 - z_{i,rs}) \over z_{r,is} (1 - z_{r,is})}\right] 
\\ & 
	+ {s_{irs} \over s_{ir} s_{rs}} \bigg[ z_{r,is} 
	{s_{irs} z_{i,rs} (1 - z_{i,rs}) - s_{rs} z_{i,rs}  
	\over z_{s,ir} (1 - z_{s,ir})} (1 - 2 z_{s,ir})  
\\ & 
	+ z_{r,is} {s_{irs} z_{s,ir} (1 - z_{s,ir}) - s_{ir} z_{s,ir}  
	\over z_{i,rs} (1 - z_{i,rs})} (1 - 2 z_{i,rs}) 
\\ & 
	- {s_{irs} \over 2} \left({4 z_{i,rs} z_{s,ir} + 2 z_{r,is} (1 - z_{r,is}) 
	- 1 \over (1 - z_{i,rs}) 
	(1 - z_{s,ir})} - {1 - 2 z_{r,is} (1 - z_{r,is}) \over z_{i,rs} z_{s,ir}}\right) 
\\ & 
	- \left({2 z_{s,ir} (1 - z_{s,ir}) \over z_{i,rs} (1 - z_{i,rs})} - 3\right) 
	{2 s_{irs} z_{i,rs} z_{s,ir} - s_{ir} z_{s,ir} - s_{rs} z_{i,rs} 
	+ s_{is} z_{r,is} \over 2} 
	\bigg] 
\\ & 
	+ (\mbox{5 permutations}) 
	\bigg\} 
\,.
\esp 
\label{A-eq:Pgigrgs} 
\eeq 
 
 
\section{Integrating the triple collinear counterterm} 
\label{app:Cirs} 

%
%
 
\subsection{Master integrals} 
 
The triple collinear momentum mapping leads to the exact factorisation
of  phase space as given by \eqn{eq:PSfact_Cirs}, where the
two-particle factorised phase space can be written as 
\beq 
\bsp 
[\rd p_{2;m}^{(irs)}(p_r,p_s,\ti{p}_{irs};Q)] &= 
	\rd \al\, (1-\al)^{2(m-1)(1-\ep)-1} 
	\,\frac{s_{\wti{irs}Q}}{2\pi} 
	\,\PS{3}(p_i,p_r,p_s; p_{(irs)}) 
\\&\times 
	\Theta(\al)\Theta(1-\al)\,,
\label{eq:dp_Cirs} 
\esp 
\eeq 
with $p_{(irs)}^\mu = (1-\al) \ti{p}_{irs}^\mu + \al Q^\mu$.
When writing \eqn{eq:ICirs0}, we have used the fact that the spin 
correlations present at the level of the factorisation formula vanish 
upon azimuthal integration via the usual arguments, since 
$\kT{j,k}^\mu$ ($j$, $k = i$, $r$ and $s$) as defined in
\Ref{Somogyi:2006da} is orthogonal to $\ti{p}_{irs}^\mu$.   
Therefore, the integrals of the spin-dependent and spin-averaged
splitting functions are equal.
  
The spin-averaged triple collinear functions $P_{f_i f_r f_s}^{(0)}$ depend
on six variables: three two-particle invariants: $s_{ir}$, $s_{is}$,
$s_{rs}$ and three momentum fractions: $\tzz{i}{rs}$, $\tzz{r}{is}$ and
$\tzz{s}{ir}$, and are given in \App{app:P0}.  When counting the number
of independent kinematical structures in \eqnss{A-eq:Pqiqbrqs}{A-eq:Pgigrgs}, 
we make two observations. Firstly, of the six variables, only four are 
independent, since $s_{ir}+s_{is}+s_{rs}=s_{irs}$ and 
$\tzz{i}{rs}+\tzz{r}{is}+\tzz{s}{ir}=1$.  Secondly, from 
\eqn{eq:dp_Cirs} it is clear that the factorised phase-space 
measure $[\rd p_{2;m}^{(irs)}(p_r,p_s,\ti{p}_{irs};Q)]$ is fully 
symmetric under permutations of the indices $\{i,r,s\}$. This permutation 
symmetry relates integrals of different terms, 
\beq 
\int_2 [\rd p_{2;m}^{(irs)}(p_r,p_s,\ti{p}_{irs};Q)] P(p_i,p_r,p_s) = 
	\int_2 [\rd p_{2;m}^{(irs)}(p_r,p_s,\ti{p}_{irs};Q)]  
	P(p_{\sigma(i)},p_{\sigma(r)},p_{\sigma(s)})\,, 
\eeq 
where $\sigma\in S_3$, with $P$ an arbitrary function of momenta (a term 
in the splitting functions). This further reduces the number of 
independent structures. 
 
Making use of the constraints among variables and the $S_3$ permutation 
symmetry, we find 46 structures to integrate\footnote{This basic set of 
integrals is not unique, nor do we claim that they are linearly independent.},
that can be grouped into five classes, 
\beq 
\bsp 
P_1^{(j,k,l,m)} &= t_{ir}^j t_{is}^k       t_{rs}^l     z_{r,is}^m 
\,,\\ 
P_2^{(j,k,l,m)} &= t_{ir}^j t_{is}^k     z_{r,is}^l     z_{s,ir}^m 
\,,\\ 
P_3^{(j,k,l,m)} &= t_{ir}^j t_{is}^k     z_{s,ir}^l (1-z_{r,is})^m 
\,,\\ 
P_4^{(j,k,l,m)} &= t_{ir}^j t_{rs}^k (1-z_{r,is})^l     z_{s,ir}^m 
\,,\\ 
P_5^{(j,k,l,m)} &= t_{is}^j t_{rs}^k (1-z_{r,is})^l (1+z_{s,ir})^m 
\,,
\esp 
\eeq 
where we introduced the scaled two-particle invariants $t_{kl} =
s_{kl}/s_{irs}$ ($k$, $l = i$, $r$, or $s$).  The exponents take values
as given in \tabss{tab:I2C1coeffs}{tab:I2C5coeffs}.  With these five
classes, we can give new forms of the spin-averaged splitting functions
that lead to the same integrated triple collinear subtraction terms,
\beq 
\frac{1}{s_{irs}^2} \frac{1}{(\bom{T}_{irs}^2)^2} P^{(0)}_{f_if_rf_s} \to  
	a_{f_i f_r f_s} \sum_{n=1}^5 \sum_{j,k,l,m} 
	c_{f_i f_r f_s;n}^{(0),j,k,l,m}  P_n^{(j,k,l,m)}
\,, 
\eeq 
where the constants $a_{f_i f_r f_s}$ are given in \eqn{eq:acoeffs},
while $c_{f_i f_r f_s;i}^{(0),j,k,l,m}$ in \tabss{tab:I2C1coeffs}
{tab:I2C5coeffs}. Thus we are left with five types of integrals,
\beq 
\bsp 
\cI_{2\cC{}{},n}^{(j,k,l,m)}(x_{\wti{irs}},\ep;\al_0,d_0) =  
	\left(\frac{(4\pi)^2}{S_\ep} Q^{2\ep}\right)^2 
	&\int_2 [\rd p_{2;m}^{(irs)}(p_r,p_s,\ti{p}_{irs};Q)] 
	\frac{1}{s_{irs}^2} P_n^{(j,k,l,m)} 
\\&\times 
	f(\al_0,\al,d(m,\ep))\,,\qquad (n = 1,\ldots, 5)\,. 
\esp 
\label{a-eq:MI2Ci} 
\eeq 
%
 
%
%
 
\subsection{Explicit representations} 
 
In order to write the integrals in \eqn{a-eq:MI2Ci} explicitly, we must 
choose a specific parametrisation of the factorised phase-space measure. 
To begin, we choose the scaled two-particle invariants $t_{kl}\equiv 
s_{kl}/s_{irs}$, ($k,l=i,r$ or $s$), and $v_r$, $v_s$, defined by 
\beq 
v_r = \frac{z_{r,is}-z_r^{(-)}}{z_r^{(+)}-z_r^{(-)}}\,, 
\qquad 
v_s = \frac{z_{s,ir}-z_s^{(-)}}{z_s^{(+)}-z_s^{(-)}}\,, 
\eeq 
as integration variables. The momentum fractions take values between 
\beq 
z_r^{(+)} = (1-t_{is})\frac{\al+(1-\al)x_{\wti{irs}}}{2\al+(1-\al)x_{\wti{irs}}}\,, 
\qquad 
z_r^{(-)} = (1-t_{is})\frac{\al}{2\al+(1-\al)x_{\wti{irs}}} 
\eeq 
and similarly for $z_s^{(\pm)}$ with $r\leftrightarrow s$. Clearly,
$v_r$ and $v_s$  are simply momentum fractions rescaled to take values
between zero and one. In terms  of these variables, we have 
\beq 
z_{k,il} = (1-t_{il})\frac{\al+(1-\al)x_{\wti{irs}} v_k}
	{2\al+(1-\al)x_{\wti{irs}}}\,, 
\qquad 
k,l = r,s\,. 
\eeq 
Using the variables $t_{kl}$ and $v_k$, the factorised phase-space measure 
reads 
\beq 
\bsp 
[\rd p_{2;m}^{(irs)}(p_r,p_s,\ti{p}_{irs};Q)] &= 
	\left(\frac{S_\ep}{(4\pi)^2} Q^{-2\ep}\right)^2  
	\frac{\Gamma^2(1-\ep)}{\pi \Gamma(1-2\ep)} (Q^2)^2\, x_{\wti{irs}}\,  
	\rd \al\, (1-\al)^{2(m-1)(1-\ep)-1}  
\\ &\times	 
	y_{irs}^{1-2\ep} 
	\rd t_{ir}\, \rd t_{is}\, \rd t_{rs}\, 
	\rd v_{r}\, \rd v_{s}\, \rd y_{irs} 
	\delta(\al(\al+(1-\al)x_{\wti{irs}})-y_{irs}) 
\\ &\times 
	\delta(1-t_{ir}-t_{is}-t_{rs}) 
	(1-t_{ir})(1-t_{is}) 
	[(t_{rs}^{(+)}-t_{rs})(t_{rs}-t_{rs}^{(-)})]^{-\frac{1}{2}-\ep} 
\\ &\times 
	\Theta(1-t_{ir})\Theta(t_{ir})\Theta(1-t_{is})\Theta(t_{is}) 
	\Theta(t_{rs}^{(+)}-t_{rs})\Theta(t_{rs}-t_{rs}^{(-)}) 
\\ &\times 
	\Theta(1-v_r)\Theta(v_r)\Theta(1-v_s)\Theta(v_s)\,, 
\esp 
\label{a-eq:dP2-Cirs-1} 
\eeq 
where 
\beq 
t_{rs}^{(\pm)} = (1-t_{ir})(1-t_{is}) \tau^{(\pm)} 
\qquad\mbox{with}\qquad 
\tau^{(\pm)} = \left[\sqrt{v_r(1-v_s)} \pm \sqrt{v_s(1-v_r)}\right]^2\,.  
\eeq 
\eqn{a-eq:dP2-Cirs-1} is not yet in a very useful form for computing the 
integrals because it contains constraints in the form of 
nontrivial $\Theta$ functions. In order to isolate the singular behaviour 
we map the region of integration onto the unit hypercube such that 
physical singularities are on the borders. Solving the constraints in a 
particular way, we find a parametrisation over the unit hypercube where 
all physical singularities are on the border except $z_s=0$, which 
introduces a line singularity (see below).  We find 
\beq 
\bsp 
[\rd p_{2;m}^{(irs)}(p_r,p_s,\ti{p}_{irs};Q)] &= 
	2^{-4\ep}\,\left(\frac{S_\ep}{(4\pi)^2}\,Q^{-2\ep}\right)^2 \, 
	\frac{\Gamma^2(1-\ep)}{\pi \Gamma(1-2\ep)}\,(Q^2)^2\, x_{\wti{irs}}\,  
	\rd \al\, (1-\al)^{2(m-1)(1-\ep)-1}  
\\&\times	 
	\rd t_{is}\, \rd \tau_{rs}\, 
	\rd v_{r}\, \rd w_{s}\, \rd y_{irs}\, 
	\delta(\al(\al+(1-\al)x_{\wti{irs}})-y_{irs})  
\\&\times 
	y_{irs}^{1-2\ep}\, 
	[t_{is} (1-t_{is})]^{1-2\ep}\, 
	[\tau_{rs}(1-\tau_{rs}) v_r(1-v_r)]^{-\ep} 
\\&\times 
	[w_s(1-w_s)]^{-\frac{1}{2}-\ep}\, 
	(1-\tau_{rs}+\tau_{rs} t_{is})^{-2+2\ep} 
\\&\times 
	\Theta(1-t_{is})\,\Theta(t_{is})\,\Theta(1-\tau_{rs})\,\Theta(\tau_{rs}) 
\\&\times 
	\Theta(1-v_r)\,\Theta(v_r)\,\Theta(1-w_s)\,\Theta(w_s)\,. 
\esp 
\label{eq:dP2-Cirs} 
\eeq 
In terms of the variables $t_{is}$, $\tau_{rs}$, $v_r$ and $w_s$ we have 
\beq 
t_{ir} = {(1-\tau_{rs})(1-t_{is}) \over 1-\tau_{rs}+\tau_{rs} t_{is}}\,, 
\qquad 
t_{rs} = {\tau_{rs} t_{is} (1-t_{is}) \over 1-\tau_{rs}+\tau_{rs} t_{is}}\,, 
\label{eq:tjl-def} 
\eeq 
and 
\beq 
\bsp 
z_{r,is} &= (1-t_{is}) {\al+(1-\al)x_{\wti{irs}} v_r \over 
	2\al+(1-\al)x_{\wti{irs}}}\,, 
\\ 
z_{s,ir} &= {t_{is} \over 1-\tau_{rs}+\tau_{rs} t_{is}} 
\\&\times 
	{\al+(1-\al)x_{\wti{irs}}[\tau_{rs}(1-v_r) + v_r(1-\tau_{rs}) 
	-2\sqrt{\tau_{rs}(1-\tau_{rs})v_r(1-v_r)} 
	(1-2w_s)] \over 2\al+(1-\al)x_{\wti{irs}}}\,. 
\label{eq:zl-def} 
\esp 
\eeq 
We see that $z_s=0$ corresponds to $\al=0$, $w_s=0,1$ and 
$\tau_{rs}=v_r$, hence the line singularity. Since the only integral 
involving $1/z_s$ is $\cI_{2\cC{}{},4}$, this is the only place where 
the line singularity has to be resolved. 
 
Using the parametrisation of \eqn{eq:dP2-Cirs} and 
\eqns{eq:tjl-def}{eq:zl-def}, we find the following explicit 
parametric integral representations of the basic integrals. 
\beq 
\bsp 
& 
\cI_{2\cC{}{},1}^{(\KKone,\KKtwo,\KKthree,\LL)}(x_{\wti{irs}},\ep;\al_0,d_0) =  
	2^{-4\ep}\,{\Gamma^{2}(1-\ep) \over \pi \Gamma(1-2\ep)}\, x_{\wti{irs}} 
	\int_{0}^{\al_{0}} \rd{\al}\, \int_{0}^{1}  
	\rd{t_{is}}\, \rd{\tau_{rs}}\, \rd{v_r}\, \rd{w_s}\,  
\\&\qquad\times		 
	\al^{-1-2\ep} \, 
	(1-\al)^{2d_0-3+2\ep} \, 
	(\al+(1-\al)x_{\wti{irs}})^{-1-2\ep}\, 
	(2\al+(1-\al)x_{\wti{irs}})^{-\LL} 
\\&\qquad\times		 
	t_{is}^{\KKtwo+\KKthree+1-2\ep} \, 
	(1-t_{is})^{\KKone+\KKthree+\LL+1-2\ep} \, 
	\tau_{rs}^{\KKthree-\ep}\, 
	(1-\tau_{rs})^{\KKone-\ep}\, 
	v_r^{-\ep}\, 
	(1-v_r)^{-\ep} 
\\&\qquad\times	 
	w_s^{-{1 \over 2}-\ep}\, 
	(1-w_s)^{-{1 \over 2}-\ep} \, 
	(1-\tau_{rs}+\tau_{rs}\, 
	t_{is})^{-\KKone-\KKthree-2+2\ep}\, 
	(\al+(1-\al)x_{\wti{irs}} v_r)^\LL\,, 
\esp 
\label{eq:MI-3C-1} 
\eeq 
\beq 
\bsp 
& 
\cI_{2\cC{}{},2}^{(\KKone,\KKtwo,-1,\LL)}(x_{\wti{irs}},\ep;\al_0,d_0) =  
	2^{-4\ep}\, {\Gamma^{2}(1-\ep) \over \pi \Gamma(1-2\ep)}\, x_{\wti{irs}} 
	\int_{0}^{\al_{0}} \rd{\al}\, \int_{0}^{1}  
	\rd{t_{is}}\, \rd{\tau_{rs}}\, \rd{v_r}\, \rd{w_s}\,  
\\&\quad\times	 
	\al^{-1-2\ep} \, 
	(1-\al)^{2d_0-3+2\ep} \, 
	(\al+(1-\al)x_{\wti{irs}})^{-1-2\ep}\, 
	(2\al+(1-\al)x_{\wti{irs}})^{1-\LL} 
\\&\quad\times	 
	t_{is}^{\KKtwo+\LL+1-2\ep} \, 
	(1-t_{is})^{\KKone-2\ep} \, 
	\tau_{rs}^{-\ep}\, 
	(1-\tau_{rs})^{\KKone-\ep}\, 
	v_r^{-\ep}\, 
	(1-v_r)^{-\ep} 
\\&\quad\times	 
	w_s^{-{1 \over 2}-\ep}\, 
	(1-w_s)^{-{1 \over 2}-\ep} \, 
	(1-\tau_{rs}+\tau_{rs}\, 
	t_{is})^{-\KKone-\LL-2+2\ep}\, 
	(\al+(1-\al)x_{\wti{irs}} v_r)^{-1} 
\\&\quad\times	 \!
	\left\{\al + (1-\al) x_{\wti{irs}}  
		[(1-\tau_{rs}) v_r + \tau_{rs} (1-v_r)  
		-2\sqrt{\tau_{rs}(1-\tau_{rs})v_r(1-v_r)} (1-2w_s)]\right\}^\LL 
	, 
\esp 
\label{eq:MI-3C-2} 
\eeq 
\beq 
\bsp 
& 
\cI_{2\cC{}{},3}^{(\KKone,\KKtwo,\LLone,\LLtwo)}(x_{\wti{irs}},\ep;\al_0,d_0) =  
	2^{-4\ep}\,{\Gamma^{2}(1-\ep) \over \pi \Gamma(1-2\ep)}\, x_{\wti{irs}} 
	\int_{0}^{\al_{0}} \rd{\al}\, \int_{0}^{1}  
	\rd{t_{is}}\, \rd{\tau_{rs}}\, \rd{v_r}\, \rd{w_s}\,  
\\&\quad\times		 
	\al^{-1-2\ep} \, 
	(1-\al)^{2d_0-3+2\ep} \, 
	(\al+(1-\al)x_{\wti{irs}})^{-1-2\ep}\, 
	(2\al+(1-\al)x_{\wti{irs}})^{-\LLtwo-\LLone} 
\\&\quad\times	 
	t_{is}^{\KKtwo+\LLone+1-2\ep} \, 
	(1-t_{is})^{\KKone+1-2\ep} \, 
	\tau_{rs}^{-\ep}\, 
	(1-\tau_{rs})^{\KKone-\ep}\, 
	v_r^{-\ep}\, 
	(1-v_r)^{-\ep} 
\\&\quad\times	 
	w_s^{-{1 \over 2}-\ep}\, 
	(1-w_s)^{-{1 \over 2}-\ep} \, 
	(1-\tau_{rs}+\tau_{rs} t_{is})^{-\KKone-\LLone-2+2\ep} 
\\&\quad\times		 
	[(1+t_{is})\al + (1-\al)x_{\wti{irs}}(1-(1-t_{is}) v_r)]^{\LLtwo} 
\\&\quad\times \!
	\left\{\al + (1-\al) x_{\wti{irs}}  
		[(1-\tau_{rs}) v_r + \tau_{rs} (1-v_r)  
		-2\sqrt{\tau_{rs}(1-\tau_{rs})v_r(1-v_r)} (1-2w_s)]\right\}^{\LLone} 
	\!, 
\esp 
\label{eq:MI-3C-3} 
\eeq 
\beq 
\bsp 
& 
\cI_{2\cC{}{},4}^{(\KKone,\KKtwo,-1,-1)}(x_{\wti{irs}},\ep;\al_0,d_0) =  
	2^{-4\ep}\, {\Gamma^{2}(1-\ep) \over \pi \Gamma(1-2\ep)}\, x_{\wti{irs}} 
	\int_{0}^{\al_{0}} \rd{\al}\, \int_{0}^{1}  
	\rd{t_{is}}\, \rd{\tau_{rs}}\, \rd{v_r}\, \rd{w_s}\,  
\\&\quad\times		 
	\al^{-1-2\ep} \, 
	(1-\al)^{2d_0-3+2\ep} \, 
	(\al+(1-\al)x_{\wti{irs}})^{-1-2\ep}\, 
	(2\al+(1-\al)x_{\wti{irs}})^{2} 
\\&\quad\times	 
	t_{is}^{\KKtwo-2\ep} \, 
	(1-t_{is})^{\KKone+\KKtwo+1-2\ep} \, 
	\tau_{rs}^{\KKtwo-\ep}\, 
	(1-\tau_{rs})^{\KKone-\ep}\, 
	v_r^{-\ep}\, 
	(1-v_r)^{-\ep} 
\\&\quad\times	 
	w_s^{-{1 \over 2}-\ep}\, 
	(1-w_s)^{-{1 \over 2}-\ep} \, 
	(1-\tau_{rs}+\tau_{rs} t_{is})^{-\KKone-\KKtwo-1+2\ep} 
\\&\quad\times	 
	[(1+t_{is})\al + (1-\al)x_{\wti{irs}}(1-(1-t_{is}) v_r)]^{-1} 
\\&\quad\times	\!
	\left\{\al + (1-\al) x_{\wti{irs}}  
		[(1-\tau_{rs}) v_r + \tau_{rs} (1-v_r)  
		-2\sqrt{\tau_{rs}(1-\tau_{rs})v_r(1-v_r)} (1-2w_s)]\right\}^{-1} 
\esp 
\label{eq:MI-3C-4} 
\eeq 
and 
\beq 
\bsp 
& 
\cI_{2\cC{}{},5}^{(-1,-1,-1,-1)}(x_{\wti{irs}},\ep;\al_0,d_0) =  
	2^{1-4\ep}\, {\Gamma^{2}(1-\ep) \over \pi \Gamma(1-2\ep)}\, x_{\wti{irs}} 
	\int_{0}^{\al_{0}} \rd{\al}\, \int_{0}^{1}  
	\rd{t_{is}}\, \rd{\tau_{rs}}\, \rd{v_r}\, \rd{w_s}\,  
\\&\quad\times		 
	\al^{-1-2\ep} \, 
	(1-\al)^{2d_0-3+2\ep} \, 
	(\al+(1-\al)x_{\wti{irs}})^{-1-2\ep}\, 
	(2\al+(1-\al)x_{\wti{irs}}) 
\\&\quad\times	 
	t_{is}^{-1-2\ep} \, 
	(1-t_{is})^{-2\ep} \, 
	\tau_{rs}^{-1-\ep}\, 
	(1-\tau_{rs})^{-\ep}\, 
	v_r^{-\ep}\, 
	(1-v_r)^{-\ep}\, 
	w_s^{-{1 \over 2}-\ep}\, 
	(1-w_s)^{-{1 \over 2}-\ep}  
\\&\quad\times	 
	(1-\tau_{rs}+\tau_{rs} t_{is})^{-1+2\ep}\, 
	[(1+t_{is})\al + (1-\al)x_{\wti{irs}}(1-(1-t_{is}) v_r)]^{-1} 
\\&\quad\times		 
	\Big\{1+{t_{is} \over 1-\tau_{rs}+\tau_{rs} t_{is}} 
\\&\quad\times	\!
	{\al + (1-\al) x_{\wti{irs}} [(1-\tau_{rs}) v_r + \tau_{rs} (1-v_r)  
	-2\sqrt{\tau_{rs}(1-\tau_{rs})v_r(1-v_r)} (1-2w_s)] \over  
	2\al+(1-\al)x_{\wti{irs}}}\Big\}^{-1} 
	\!. 
\esp 
\label{eq:MI-3C-5} 
\eeq 

The expressions for $\cI_{2\cC{}{},n}$, $n = 1$, $2$ and $3$ in
\eqnss{eq:MI-3C-1}{eq:MI-3C-3} are suitable for  evaluation with
general purpose sector decomposition codes as they stand.  In
$\cI_{2\cC{}{},5}$, \eqn{eq:MI-3C-5}, the terms in the braces
require some care. Although this factor is just $1/(1+z_s)$, which is finite 
and hence in principle it can be simply carried through sector 
decomposition as it is, nevertheless depending on the precise internal 
implementation, it may lead to undefined arithmetic expressions 
in the (symbolic) computation. We find that no such trouble arises either 
with our private implementation or with {\tt SecDec}, if we gather the two 
terms in the braces over a common denominator.
 
$\cI_{2\cC{}{},4}$ requires special attention, as could be anticipated 
by the presence of the factor $1/z_s$. There are two problems: firstly,
the presence of the square roots in the vanishing denominator in the
braces on the last line of \eqn{eq:MI-3C-4} interferes with the
treatment of overlapping singularities. (This could be solved by
simply making the expression square-free.) Secondly, as we have
indicated, there is a line singularity inside the integration region at
$\tau_{rs}=v_r$. We address both issues by deriving an alternative
representation for $\cI_{2\cC{}{},4}$, which is free of both square
roots and line singularities.  The price to pay is that the new
representation is quite cumbersome,
\beq 
\bsp 
& 
\cI_{2\cC{}{},4}^{(\KKone,\KKtwo,-1,-1)}(x_{\wti{irs}},\ep;\al_0,d_0) =  
	-2\ep {\Gamma^{2}(1-\ep) \over \Gamma(1-2\ep)} x_{\wti{irs}} 
	\int_{0}^{\al_{0}} \rd{\al}\, \int_{0}^{1}  
	\rd{t_{is}}\, \rd{\tau_{rs}}\, \rd{s}\, \rd{u}\,  
	\al^{-1-2\ep}  
\\&\qquad\times		 
	(1-\al)^{2d_0-3+2\ep}  
	(\al+(1-\al)x_{\wti{irs}})^{-1-2\ep} 
	(2\al+(1-\al)x_{\wti{irs}})^{4-2\ep} 
	t_{is}^{\KKtwo-2\ep}  
\\&\qquad\times	 
	(1-t_{is})^{\KKone+\KKtwo+1-2\ep}  
	\tau_{rs}^{\KKtwo-\ep} 
	(1-\tau_{rs})^{\KKone-\ep} 
	s^{-2\ep} 
	u^{-2\ep}  
	(1-\tau_{rs}+\tau_{rs} t_{is})^{-\KKone-\KKtwo-1+2\ep} 
\\&\qquad\times	 
	\Big[2 (1+t_{is}) \al (\al+(1-\al)x_{\wti{irs}}) (1+u) 
\\&\qquad\qquad 
		+ (1-\al)^2 (1-\tau_{rs}+\tau_{rs} u)x_{\wti{irs}}^2 
		+ t_{is} (1-\al)^2 (\tau_{rs}+u-\tau_{rs} u)x_{\wti{irs}}^2\Big]^{1-2\ep} 
\\&\qquad\times 
	\Big\{\Big[(1+t_{is})^2 \al (\al+(1-\al)x_{\wti{irs}})  
		+ t_{is} (1-\al)^2 x_{\wti{irs}}^2\Big] (2\al+(1-\al)x_{\wti{irs}})^4 
		(1-s)^2 u^2 
\\&\qquad\qquad 
		+ \Big[\al (\al+(1-\al)x_{\wti{irs}}) s (1-u)^2 
			+  (2\al+(1-\al)x_{\wti{irs}})^2 u\Big]
\\&\qquad\qquad \times	 
		s \Big[2 (1+t_{is}) \al (\al+(1-\al)x_{\wti{irs}}) (1+u) 
\\&\qquad\qquad\qquad  
		+ (1-\al)^2 (1-\tau_{rs}+\tau_{rs} u)x_{\wti{irs}}^2 
		+ t_{is} (1-\al)^2 (\tau_{rs}+u-\tau_{rs} u)x_{\wti{irs}}^2\Big]^2
		\Big\}^{-1+\ep} \,. 
\esp 
\label{eq:MI-3C-4ALT2} 
\eeq 
The derivation of this alternate  form relies on rewriting the integration
over $v_r$ and $w_s$,
\beq 
\bsp 
I_{v_r,w_s} &= \int_0^1 \rd v_r\,\rd w_s\, v_r^{-\ep} (1-v_r)^{-\ep} 
	w_s^{-{1 \over 2}-\ep} (1-w_s)^{-{1 \over 2}-\ep} 
\\&\qquad\times 
	\left[\al(1+t_{is}) + (1-\al)(1-(1-t_{is})v_r) x_{\wti{irs}}\right]^{-1} 
\\&\qquad\times 
	\Big\{\al + (1-\al) x_{\wti{irs}}  
		[(1-\tau_{rs}) v_r + \tau_{rs} (1-v_r)  
\\&\qquad\qquad 
		-2\sqrt{\tau_{rs}(1-\tau_{rs})v_r(1-v_r)} (1-2w_s)]\Big\}^{-1} \,,
\esp 
\label{eq:Ivrws} 
\eeq 
as an angular integral,
\beq 
I_{v_r,w_s} = 2^{1+4\ep} (1+t_{is})^{-1} (2\al + (1-\al) x_{\wti{irs}})^{-2} 
	{1 \over \Omega_{d-3}} \int \rd\Omega_{d-1}(q) {1 \over (p_1\cdot q)(p_2\cdot q)}\,, 
\eeq 
where in a suitable frame,
\bal 
p_1^\mu &= \left(1,\ldots, 
	-{(1-t_{is})(1-\al) x_{\wti{irs}} \over (1+t_{is})(2\al + (1-\al) x_{\wti{irs}})}\right)\,, 
\\ 
p_2^\mu &= \left(1,\ldots, 
	{(1-\al) x_{\wti{irs}} \over 2\al + (1-\al) x_{\wti{irs}}} \sin\chi, 
	{(1-\al) x_{\wti{irs}} \over 2\al + (1-\al) x_{\wti{irs}}} \cos\chi\right) 
\eal 
(the $\ldots$ denotes vanishing components) and 
\beq 
q^\mu = (1, \mbox{..`angles'..}, \sin\vth\sin\vph, \sin\vth\cos\vph, \cos\vth)\,. 
\eeq 
(Here $\mbox{..`angles'..}$ represents those angular variables that are 
trivial to integrate, since the integrand does not depend on them.) 
Clearly both $p_1$ and $p_2$
are  massive and time-like, $p_1^2$, $p_2^2>0$, and $p_1\cdot p_2>0$. 
The angular integral can be written in terms of a \MB
representation according to Eq.\,(60) of \Ref{Somogyi:2011ir}.
Finally, we can perform the \MB integrations at the expense
of reintroducing two real integrations over the unit interval. We
obtain a new two-dimensional real integral representation,
\beq 
\bsp 
I_{v_r,w_s} &= 
-2^{1+4\ep} \pi \ep (2\al+(1-\al)x_{\wti{irs}})^{2-2\ep} 
\\&\times 
	\int_0^1 \rd s\,\rd u\, s^{-2\ep} u^{-2\ep} 
	\Big[2 (1+t_{is}) \al (\al+(1-\al)x_{\wti{irs}}) (1+u) 
\\&\qquad 
		+ (1-\al)^2 (1-\tau_{rs}+\tau_{rs} u)x_{\wti{irs}}^2 
		+ t_{is} (1-\al)^2 (\tau_{rs}+u-\tau_{rs} u)x_{\wti{irs}}^2\Big]^{1-2\ep} 
\\&\times 
	\Big\{\Big[(1+t_{is})^2 \al (\al+(1-\al)x_{\wti{irs}})  
		+ t_{is} (1-\al)^2 x_{\wti{irs}}^2\Big] (2\al+(1-\al)x_{\wti{irs}})^4 
		(1-s)^2 u^2 
\\&\qquad 
		+ \Big[\al (\al+(1-\al)x_{\wti{irs}}) s (1-u)^2 
			+  (2\al+(1-\al)x_{\wti{irs}})^2 u\Big]
\\&\qquad \times
		s \Big[2 (1+t_{is}) \al (\al+(1-\al)x_{\wti{irs}}) (1+u) 
\\&\qquad 
		+ (1-\al)^2 (1-\tau_{rs}+\tau_{rs} u)x_{\wti{irs}}^2 
		+ t_{is} (1-\al)^2 (\tau_{rs}+u-\tau_{rs} u)x_{\wti{irs}}^2\Big]^2
		\Big\}^{-1+\ep}\,,
\label{eq:MI-3C-4-form2} 
\esp 
\eeq 
free of square roots and singularities inside the integration region. 
 
 
\section{Integrating the double collinear counterterm} 
\label{app:Cirjs} 

%
%

\subsection{Master integrals} 
 
The double collinear momentum mapping leads to the exact factorisation
of  phase space as given by \eqn{eq:PSfact_Cirjs}, where the
two-particle factorised phase space can be written as,
\beq 
\bsp 
[\rd p_{2;m}^{(ir;js)}(p_r,p_s,\ti{p}_{ir},\ti{p}_{js};Q)] &=  
	\rd\al\, \rd\be\, (1-\al-\be)^{2(m-1)(1-\ep)}  
	\Theta(\al)\,\Theta(\be)\,\Theta(1-\al-\be)
\\& \times 
	\frac{s_{\wti{ir}Q}}{2\pi} \PS{2}(p_i,p_r; p_{(ir)}) 
	\,\frac{s_{\wti{js}Q}}{2\pi}\PS{2}(p_j,p_s; p_{(js)})\,,
\esp 
\label{eq:dp_Cirjs} 
\eeq 
where $p_{(ir)}^\mu = (1-\al-\be) \ti{p}_{ir}^\mu + \al Q^\mu$
and $p_{(js)}^\mu = (1-\al-\be) \ti{p}_{js}^\mu + \be Q^\mu$.
 
When writing \eqn{eq:ICirjs}, we have used that spin correlations
generally  present at the level of factorisation formulae cancel after
azimuthal integration,  since $\kT{i,r}^\mu$ and $\kT{j,s}^\mu$ as
defined in \Ref{Somogyi:2006da} are orthogonal to $\ti{p}_{ir}^\mu$ and 
$\ti{p}_{js}^\mu$ respecitvely.  Examining the actual form of the spin-averaged 
\AP functions, and using the symmetry of the phase-space measure
under the momentum exchanges $p_i\leftrightarrow p_r$, or
$p_j\leftrightarrow p_s$, and under exchange of the  pairs
$(p_i,p_r)\leftrightarrow (p_j,p_s)$, we see that we must compute the 
following integrals,
\beq 
\bsp 
\cI_{2\cC{}{},6}^{(k,l)}(x_{\wti{ir}},x_{\wti{js}};\ep,\al_{0},d_0) =  
	\left(\frac{(4\pi)^2}{S_\ep} Q^{2\ep}\right)^2 
	&\int_2 [\rd p_{2;m}^{(ir;js)}(p_r,p_s,\ti{p}_{ir},\ti{p}_{js};Q)] 
	\frac{1}{s_{ir}} \frac{1}{s_{js}} 
\\&\times 
	\tzz{r}{i}^{k} \tzz{s}{j}^{l} f(\al_0,\al+\be,d(m,\ep))\,, 
\esp 
\label{a-eq:MI-2C-def} 
\eeq 
where $k$, $l =-1$, $0$, $1$ and $2$. By symmetry, we need only
consider the cases when  e.g.,~$k\le l$. 
 
%
%

\subsection{Explicit representation} 
 
To write the integral in \eqn{a-eq:MI-2C-def} explicitly, we must choose  
a specific representation of the factorised phase-space measure. We choose  
the scaled two-particle invariants $y_{ir}$ and $y_{js}$  and rescaled
momentum fractions $\tvv{r}{i}$ and $\tvv{s}{j}$, 
\beq 
\tvv{r}{i} = \frac{\tzz{r}{i} - z_r^{(-)}}{z_r^{(+)} - z_r^{(-)}}\,,
\qquad 
\tvv{s}{j} = \frac{\tzz{s}{j} - z_s^{(-)}}{z_s^{(+)} - z_s^{(-)}}\,, 
\label{eq:vrivsj-def}
\eeq 
as integration variables. Here 
\beq 
z_r^{(+)} = \frac{\al+(1-\al-\be)x_{\wti{ir}}}{2\al+(1-\al-\be)x_{\wti{ir}}}\,,
\qquad 
z_r^{(-)} = \frac{\al}{2\al+(1-\al-\be)x_{\wti{ir}}}\,, 
\eeq 
with similar expressions for $z_s^{(+)}$ and $z_s^{(-)}$ with $\al \to
\be$ and  $ir \to js$. In terms of $\al$, $\be$, $v_{r,i}$ and
$v_{s,j}$ we have 
\beq 
y_{ir} = \al(\al+(1-\al-\be)x_{\wti{ir}})\,, 
\qquad 
y_{js} = \be(\al+(1-\al-\be)x_{\wti{js}})\,, 
\label{eq:ykt-def} 
\eeq 
\beq 
z_{r,i} = {\al + (1-\al-\be)x_{\wti{ir}} v_{r,i} \over 
	2\al + (1-\al-\be)x_{\wti{ir}}}\,, 
\qquad 
z_{s,j} = {\be + (1-\al-\be)x_{\wti{js}} v_{s,j} \over 
	2\be + (1-\al-\be)x_{\wti{js}}}\,, 
\label{eq:ztk-def} 
\eeq 
and for the factorised phase-space measure we obtain 
\beq 
\bsp 
[\rd p^{(ir;js)}_{2;m}&(p_r,p_s,\ti{p}_{ir},\ti{p}_{js};Q)] =  
	\left(\frac{S_\ep}{(4\pi)^2} Q^{-2\ep}\right)^2 (Q^2)^2\,
	x_{\wti{ir}Q}\, x_{\wti{js}Q} 
\\ &\times 
	\rd\al\,\rd\be\,(1-\al-\be)^{2(m-1)(1-\ep)} 
	\Theta(\al)\Theta(\be)\Theta(1-\al-\be) 
\\ &\times 
	\rd y_{ir}\,\rd v_{r,i} [y_{ir} v_{r,i} (1-v_{r,i})]^{-\ep} 
	\delta\Big(\al(\al+(1-\al-\be)x_{\wti{ir}}) - y_{ir}\Big)
\\ &\times 
	\rd y_{js}\,\rd v_{s,j} [y_{js} v_{s,j} (1-v_{s,j})]^{-\ep} 
	\delta\Big(\be(\be+(1-\al-\be)x_{\wti{js}}) - y_{js}\Big)
\\ &\times	 
	\Theta(v_{r,i})\Theta(1-v_{r,i})\Theta(v_{s,j})\Theta(1-v_{s,j})\,. 
\esp 
\label{eq:PSfact_Cirjs_expl} 
\eeq 

Using the parametrisation of \eqn{eq:PSfact_Cirjs_expl} and  
the expressions in \eqns{eq:ykt-def}{eq:ztk-def}, we find the 
following parametric integral representation for the master integral,
\beq 
\bsp 
\cI_{2\cC{}{},6}^{(\KKtwo,\KKthree)}&(x_{\wti{ir}},x_{\wti{js}};\ep,\al_{0},d_0) = 
x_{\wti{ir}} x_{\wti{js}}
\int_{0}^{1} \rd{\al}\, \rd{\be}\, 
\Theta(\al_0-\al-\be) 
\int_{0}^{1} \rd{v}\, \rd{u}\,
\\ &\times 
\al^{-1-\ep}\, \be^{-1-\ep}\, (1-\al-\be)^{2d_0-2(1-\ep)} \,
v^{-\ep}\,(1-v)^{-\ep}\,u^{-\ep}\,(1-u)^{-\ep}
\\ &\times 
[\al + (1-\al-\be) x_{\wti{ir}}]^{-1-\ep}\,
[\be + (1-\al-\be) x_{\wti{js}}]^{-1-\ep}  
\\ &\times 
	\left({\al + (1-\al-\be)x_{\wti{ir}} v  
		\over 2\al + (1-\al-\be)x_{\wti{ir}}}\right)^{\KKtwo} 
	\left({\be + (1-\al-\be)x_{\wti{js}} u  
		\over 2\be + (1-\al-\be)x_{\wti{js}}}\right)^{\KKthree}\,. 
\esp 
\label{eq:MI-2C} 
\eeq 
\eqn{eq:MI-2C} is directly suitable for treatment with  
general purpose sector decomposition codes. 
 
 
\section{Integrating the soft collinear counterterms} 
\label{app:CSirs} 
 
%
%
 
\subsection{Master integrals} 
 
The consecutive collinear and soft mappings lead to the exact
factorisation of  phase space as given by \eqn{eq:PSfact_CSirs}, 
where the one-particle factorised phase spaces can be written in 
the following form. For the collinear mapping we have 
\beq
[\rd p_{1;m+1}^{(ir)}(p_r,\ha{p}_{ir};Q)] =
	\rd\al (1-\al)^{2m(1-\ep)-1} 
	\,\frac{s_{\wti{ir}Q}}{2\pi} \,\PS{2}(p_i,p_r;p_{(ir)})
	\,\Theta(\al)\Theta(1-\al)\,, 
\label{eq:PSfact_CSirs_coll} 
\eeq 
where $p_{(ir)}^\mu = (1-\al)\ha{p}_{ir}^\mu + \al Q^\mu$.
For the soft mapping we find 
\beq 
[\rd p_{1;m}^{(\ha{s})}(\ha{p}_s,K;Q)] =
	\rd y (1-y)^{(m-1)(1-\ep)} 
	\,\frac{Q^2}{2\pi} \,\PS{2}(\ha{p}_s,K;Q)
	\,\Theta(y) \Theta(1-y)\,, 
\label{eq:PSfact_CSirs_soft} 
\eeq
where $y\equiv y_{\ha{s}Q}$ and the momentum $K$ is massive with 
$K^2 = (1-y)Q^2$. As the notation above indicates, $\al$ and $y$ 
are integration variables. 

To integrate \eqnss{eq:ICSirs}{eq:ICirjsCSirs} over the factorised phase
spaces of  \eqns{eq:PSfact_CSirs_coll}{eq:PSfact_CSirs_soft}, we can 
pass to the azimuthally averaged \AP splitting functions, since the 
azimuthal correlations generally present in
\eqnss{eq:ICSirs}{eq:ICirjsCSirs} vanish  after integration by the usual
argument (the transverse  momentum $\kTtt{i,r}^\mu$ in the splitting
kernels is defined to be orthogonal  to the parent momentum
$\ti{p}_{ir}^\mu$). The \AP functions can be expressed as linear
combinations of powers of momentum fractions, so we need to compute
integrals of the form,
\beq	 
\bsp 
\left(\frac{(4\pi)^2}{S_\ep} Q^{2\ep}\right)^2 
&\int_2 [\rd p_{1;m+1}^{(ir)}(p_{r},\ha{p}_{ir};Q)] 
\,[\rd p_{1;m}^{(\ha{s})}(\ha{p}_s;Q)]\, 
\left\{ 
\frac12 \calS_{jk}(s) 
\,,\; 
\frac{2}{s_{(ir)s}}\frac{1-\tzz{s}{ir}}{\tzz{s}{ir}} 
\,,\; 
\frac{2}{s_{js}}\frac{\tzz{j}{s}}{\tzz{s}{j}} 
\right\}  
\\&\times 
\frac{1}{s_{ir}}\tzz{r}{i}^{l}\,  
f(\al_0,\al,d(m,\ep)) f(y_0,y,d(m,\ep))\,. 
\esp 
\eeq 

We perform the integration over $[\rd p_{1;m+1}^{(ir)}]$ first, so we
need to write the integrands in terms of  $\ha{p}_s^\mu$ and tilded momenta.
The singly collinear mapping of the momenta rescales the 
invariants which do not involve the collinear pair, 
\beq
y_{jk} = (1-\al)^2 y_{\ha{j}\ha{k}}
\qquad\mbox{and}\qquad
y_{jQ} = (1-\al) y_{\ha{j}Q}\,,
\eeq
while those which involve the collinear momenta become 
\beq
y_{(ir)k} = (1-\al) 
[(1-\al) y_{\wha{ir}\ha{k}} + \al\, y_{\ha{k}Q}]
\qquad\mbox{and}\qquad
y_{(ir)Q} = (1-\al) y_{\wha{ir}Q} + 2 \al\,.
\eeq 
The successive singly soft mapping of the momenta rescales those 
two-particle invariants that do not involve the soft parton, 
\beq
y_{\ha{j}\ha{k}} = (1-y) y_{\ti{j}\ti{k}}\,,
\eeq
while those involving the soft parton $s$ become $y_{\ha{j}\ha{s}}
= y_{\ti{j}\ha{s}}$. Finally, we have
\beq
y_{\ha{j}Q} = (1-y) y_{\ti{j}Q} + y_{\ti{j}\ha{s}}\,,
\eeq 
for $j\ne s$, while $y_{\ha{s}Q}=y$ is simply an integration variable.
Therefore, the integrands are expressed as follows. For the eikonal  
factor, we find 
\beq 
\calS_{jk}(s) = 
	\frac{1-y}{(1-\al)^2} \,\calS_{\wti{j}\wti{k}}(\ha{s})\,, 
\label{eq:eikshjs} 
\eeq 
if $j$ and $k$ are distinct from $(ir)$, while 
\beq 
\frac{1}{2}\calS_{(ir)k}(s) =  
	\frac{1-y}{(1-\al)^2} 
	\frac{(1-\al) s_{\wti{ir}\ti{k}} + \al s_{\ti{k}Q}} 
	{[(1-\al) s_{\wti{ir}\ha{s}} + \al s_{\ha{s}Q}] s_{\ti{k}\ha{s}}} 
	+\frac{\al}{(1-\al)^2} 
	\frac{s_{\ti{k}\ha{s}}} 
	{[(1-\al) s_{\wti{ir}\ha{s}} + \al s_{\ha{s}Q}] s_{\ti{k}\ha{s}}}\,,
\label{eq:Sirks} 
\eeq 
if e.g.,~$j$ coincides with $(ir)$. Finally, we have 
\beq 
\frac{2}{s_{(ir)s}}\frac{1-\tzz{s}{ir}}{\tzz{s}{ir}} =  
\frac{2}{s_{(ir)s}}\frac{s_{(ir)Q}}{s_{sQ}} = 
\frac{2}{(1-\al)^2} 
\frac{2\al+(1-\al)[(1-y) s_{\wti{ir}Q}+s_{\wti{ir}\ha{s}}]} 
	{[(1-\al)s_{\wti{ir}\ha{s}} + \al s_{\ha{s}Q}] s_{\ha{s}Q}} \,, 
\eeq 
and 
\beq 
\frac{2}{s_{js}} \frac{\tzz{j}{s}}{\tzz{s}{j}} =
\frac{2}{s_{js}} \frac{s_{jQ}}{s_{sQ}} = 
\frac{2}{(1-\al)^2 s_{\ti{j}\ha{s}}}
\frac{(1-y)s_{\ti{j}Q} + s_{\ti{j}\ha{s}}}{s_{\ha{s}Q}}\,. 
\eeq 
Hence, we define five soft collinear master integrals,
\beq 
\bsp 
& 
\cI_{2\cSCS{}{},1}^{(l)}(x_{\wti{ir}}, 
	\Yt{ir}{j},\Yt{ir}{k},\Yt{j}{k};\ep,\al_{0},y_0,d_0,d'_0) = 
\\&\qquad = 
	\left(\frac{(4\pi)^2}{S_\ep} Q^{2\ep}\right)^2 
	\int_2 [\rd p_{1;m+1}^{(ir)}(p_{r},\ha{p}_{ir};Q)] 
	\,[\rd p_{1;m}^{(\ha{s})}(\ha{p}_s;Q)]\, 
\\&\qquad\times 
	\frac{1}{2}\frac{1-y}{(1-\al)^2} \,\calS_{\wti{j}\wti{k}}(\ha{s}) 
	\frac{1}{s_{ir}}\tzz{r}{i}^{l}\,  
\\&\qquad\times 
	f(\al_0,\al,d(m,\ep)) f(y_0,y,d(m,\ep))\,, 
\label{eq:I2CS1}
\esp 
\eeq 
\beq 
\bsp 
& 
\cI_{2\cSCS{}{},2}^{(l)}(x_{\wti{ir}}, 
	\Yt{ir}{k};\ep,\al_{0},y_0,d_0,d'_0) = 
\\&\qquad = 
	\left(\frac{(4\pi)^2}{S_\ep} Q^{2\ep}\right)^2 
	\int_2 [\rd p_{1;m+1}^{(ir)}(p_{r},\ha{p}_{ir};Q)] 
	\,[\rd p_{1;m}^{(\ha{s})}(\ha{p}_s;Q)]\, 
\\&\qquad\times 
	\frac{1-y}{(1-\al)^2} 
	\frac{(1-\al) s_{\wti{ir}\ti{k}} + \al s_{\ti{k}Q}} 
	{[(1-\al) s_{\wti{ir}\ha{s}} + \al s_{\ha{s}Q}] s_{\ti{k}\ha{s}}}
	\frac{1}{s_{ir}}\tzz{r}{i}^{l}\,  
\\&\qquad\times 
	f(\al_0,\al,d(m,\ep)) f(y_0,y,d(m,\ep))\,, 
\esp 
\eeq 
\beq 
\bsp 
& 
\cI_{2\cSCS{}{},3}^{(l)}(x_{\wti{ir}}; 
	\ep,\al_{0},y_0,d_0,d'_0) = 
\\&\qquad = 
	\left(\frac{(4\pi)^2}{S_\ep} Q^{2\ep}\right)^2 
	\int_2 [\rd p_{1;m+1}^{(ir)}(p_{r},\ha{p}_{ir};Q)] 
	\,[\rd p_{1;m}^{(\ha{s})}(\ha{p}_s;Q)]\, 
\\&\qquad\times 
	\frac{\al}{(1-\al)^2} 
	\frac{s_{\ti{k}\ha{s}}} 
	{[(1-\al) s_{\wti{ir}\ha{s}} + \al s_{\ha{s}Q}] s_{\ti{k}\ha{s}}}
	\frac{1}{s_{ir}}\tzz{r}{i}^{l}\,  
\\&\qquad\times 
	f(\al_0,\al,d(m,\ep)) f(y_0,y,d(m,\ep))\,, 
\esp 
\eeq 
\beq 
\bsp 
& 
\cI_{2\cSCS{}{},4}^{(l)}(x_{\wti{ir}}; 
	\ep,\al_{0},y_0,d_0,d'_0) = 
\\&\qquad = 
	\left(\frac{(4\pi)^2}{S_\ep} Q^{2\ep}\right)^2 
	\int_2 [\rd p_{1;m+1}^{(ir)}(p_{r},\ha{p}_{ir};Q)] 
	\,[\rd p_{1;m}^{(\ha{s})}(\ha{p}_s;Q)]\, 
\\&\qquad\times 
	\frac{2}{(1-\al)^2} 
	\frac{2\al + (1-\al)[(1-y) s_{\wti{ir}Q} + s_{\wti{ir}\ha{s}}]} 
	{[(1-\al)s_{\wti{ir}\ha{s}} + \al s_{\ha{s}Q}] s_{\ha{s}Q}} 
	\frac{1}{s_{ir}}\tzz{r}{i}^{l}\,  
\\&\qquad\times 
	f(\al_0,\al,d(m,\ep)) f(y_0,y,d(m,\ep))
\esp 
\eeq 
and 
\beq 
\bsp 
& 
\cI_{2\cSCS{}{},5}^{(l)}(x_{\wti{ir}}, 
	\Yt{ir}{j};\ep,\al_{0},y_0,d_0,d'_0) = 
\\&\qquad = 
	\left(\frac{(4\pi)^2}{S_\ep} Q^{2\ep}\right)^2 
	\int_2 [\rd p_{1;m+1}^{(ir)}(p_{r},\ha{p}_{ir};Q)] 
	\,[\rd p_{1;m}^{(\ha{s})}(\ha{p}_s;Q)]\, 
\\&\qquad\times 
	\frac{2}{(1-\al)^2 s_{\ti{j}\ha{s}}} 
	\frac{(1-y)s_{\ti{j}Q} + s_{\ti{j}\ha{s}}} 
	{s_{\ha{s}Q}} 
	\frac{1}{s_{ir}}\tzz{r}{i}^{l}\,  
\\&\qquad\times 
	f(\al_0,\al,d(m,\ep)) f(y_0,y,d(m,\ep))\,. 
\label{eq:I2CS5}
\esp 
\eeq 
We need to compute these integrals for $l=-1,0,1,2$. 

%
%

\subsection{Explicit representations} 
 
\eqnss{eq:I2CS1}{eq:I2CS5} are very similar to the iterated
collinear -- soft collinear integrals we computed in
\Ref{Bolzoni:2010bt}, and we employ the techniques of
\Ref{Bolzoni:2010bt} for computing the integrals
$\cI_{2\cSCS{}{},n}^{(l)}$. 
We begin by recalling the specific representation of the factorised 
phase-space measures in \eqns{eq:PSfact_CSirs_coll}{eq:PSfact_CSirs_soft}. 
For the singly collinear measure in \eqn{eq:PSfact_CSirs_coll}, we 
choose the scaled two-particle invariant $y_{ir}$ and the rescaled 
momentum fraction $v_{r,i}$,
\beq
v_{r,i} = \frac{\tzz{r}{i} - z_r^{(-)}}{z_r^{(+)} - z_r^{(-)}}\,,
\eeq
as integration variables. Here
\beq
z_r^{(+)} = \frac{\al + (1-\al) x_{\wha{ir}}}{2\al + (1-\al) x_{\wha{ir}}}\,,
\qquad
z_r^{(-)} = \frac{\al}{2\al + (1-\al) x_{\wha{ir}}}\,.
\eeq
In terms of $\al$ and $v_{r,i}$, we have
\beq
y_{ir} = \al(\al + (1-\al) x_{\wha{ir}})\,,
\qquad\mbox{and}\qquad
z_{r,i} = \frac{\al + (1-\al) x_{\wha{ir}} v_{r,i}}{2\al + (1-\al) x_{\wha{ir}}}\,,
\eeq
while the phase-space measure $[\rd p_{1;m+1}^{(ir)}(p_r,\ha{p}_{ir};Q)]$ 
reads
\beq 
\bsp 
[\rd p_{1;m+1}^{(ir)}(p_r,\ha{p}_{ir};Q)] &= 
	\frac{S_\ep}{(4\pi)^2} (Q^2)^{1-\ep}
	x_{\wha{ir}}\, \rd \al\, \rd y_{ir}\, \rd v_{r,i}\, 
	(1-\al)^{2m(1-\ep)-1} 
	[y_{ir} v_{r,i} (1-v_{r,i})]^{-\ep}  
\\ &\times 
	\delta\Big(\al(\al+(1-\al)x_{\wha{ir}}) - y_{ir}\Big)
	\Theta(\al) \Theta(1-\al) 
	\Theta(v_{r,i}) \Theta(1-v_{r,i})\,.
\esp 
\eeq 
Turning to $[\rd p_{1;m}^{(\ha{s})}(\ha{p}_s,K;Q)]$ in \eqn{eq:PSfact_CSirs_soft},
we use the (scaled) energy and the angular variables of $\ha{p}_s^\mu$ in some 
specific Lorentz frame to write the two-particle phase space $\PS{2}(\ha{p}_s,K;Q)$ 
explicitly. In particular, we work in the rest frame of $Q^\mu$, where 
$Q^\mu = \sqrt{s}(1,\ldots)$ and $\ha{p}_s^\mu$ is parametrised as,
\beq 
\ha{p}_s^\mu = 
	\ha{E}_s(1,\mbox{..`angles'..},\sin\vth\sin\vph\sin\eta, 
	\sin\vth\sin\vph\cos\eta,\sin\vth\cos\vph,\cos\vth)\,.
\eeq 
The specific orientation of this frame is chosen below according to what is 
most convenient for computing each integral. However, independent of orientation,
in terms of the scaled energy-like variable,
\beq 
\eps_{\ha{s}} = \frac{2\ha{p}_s\cdot Q}{Q^2} = \frac{2\ha{E}_s}{\sqrt{s}}\,, 
\eeq 
and the angular variables $\vth$, $\vph$ and $\eta$, the two-particle phase space  
$\PS{2}(\ha{p}_s,K;Q)$ reads 
\beq 
\bsp 
\PS{2}(\ha{p}_s,K;Q) &= 
	\frac{(Q^2)^{-\ep}}{(4\pi)^2} S_\ep (-2^{2\ep}\ep) 
	\rd \eps_{\ha{s}}\, \eps_{\ha{s}}^{1-2\ep} 
	\delta(y - \eps_{\ha{s}}) 
\\ &\times 
	\rd(\cos\vth)\,\rd(\cos\vph)\,\rd(\cos\eta)\,
	(\sin\vth)^{-2\ep}\,(\sin\vph)^{-1-2\ep}\,(\sin\eta)^{-2-2\ep}\,. 
\esp 
\label{eq:dPS2pshat}
\eeq 
We are now ready to display the master integrals. 


  
\paragraph{Integrated soft collinear counterterm for $i$, $j$, $k$ distinct.} 
This case leads to the integral $\cI_{2\cSCS{}{},1}$.
We choose the orientation of the frame such that the momenta appearing in the 
integrand take the following forms,
\beq
\bsp
\ti{p}_j^\mu &= \ti{E}_j(1,\ldots,1)\,,\qquad 
\ti{p}_k^\mu = \ti{E}_k(1,\ldots,\sin\chi_{k},\cos\chi_{k})\,, 
\label{eq:CSirs-frame-1}
\\ 
\ti{p}_{ir}^\mu &= \ti{E}_{ir}(1,\ldots,\sin\phi_{ir}\sin\chi_{ir}, 
	\cos\phi_{ir}\sin\chi_{ir},\cos\chi_{ir})\,.
\esp
\eeq
Then, expressing the various invariants in terms of the integration  
variables in the chosen frame of \eqn{eq:CSirs-frame-1}, we find 
\bal 
y_{\ti{j}\ha{s}}  
&=  
	{1\over 2}y_{\ti{j}Q} y  
	(1 - \cos\vth)\,, 
\label{eq:ytjhs}
\\ 
y_{\ti{k}\ha{s}}  
&=  
	{1\over 2}y_{\ti{k}Q} y  
	(1 - \sin\chi_{k} \sin\vth \cos\vph - \cos\chi_{k} \cos\vth)\,, 
\label{eq:ytkhs}
\\ 
y_{\wti{ir}\ha{s}}  
&=  
	{1\over 2}x_{\wti{ir}} y  
	(1 - \sin\phi_{ir} \sin\chi_{ir} \sin\vth \sin\vph \cos\eta 
	- \cos\phi_{ir} \sin\chi_{ir} \sin\vth \cos\vph - \cos\chi_{ir} \cos\vth)\,, 
\label{eq:ytirhs}
\eal 
where 
\beq
\bsp
\cos\chi_{k} = \cos\chi(\Yt{j}{k})
\,,\qquad
\cos\chi_{ir} = \cos\chi(\Yt{j}{ir})
\\
\cos\phi_{ir} = 
	{\Yt{j}{k} + \Yt{j}{ir} - \Yt{k}{ir} - 2 \Yt{j}{k} \Yt{j}{ir} \over  
	\sin\chi(\Yt{j}{k})\,\sin\chi(\Yt{j}{ir})}\,, 
\label{eq:a2YCSirs} 
\esp
\eeq
with
\beq
\cos\chi(Y) = 1 -2Y
\,,\qquad
\sin\chi(Y) = 2\sqrt{Y(1-Y)}
\label{eq:coschi2Y}
\,.
\eeq 
 
The factor $y_{\ti{k}\ha{s}}$ in the denominator of the eikonal factor
vanishes at $\cos\vph=1$ and $\cos\vth = \cos\chi_{k}$. Hence, the
integrand has a line singularity that we remove by performing partial
fractioning as in \Refs{Bolzoni:2010bt,Anastasiou:2010pw},  
\beq 
\frac{s_{\ti{j}\ti{k}}}{s_{\ti{j}\ha{s}} s_{\ti{k}\ha{s}}} =  
	\frac{1}{Q^2}\, 4\Yt{j}{k}  
	\left(\COEFFJ {y_{\ti{j}Q} \over 2 y_{\ti{j}\ha{s}}}  
	+ \COEFFK {y_{\ti{k}Q} \over 2 y_{\ti{k}\ha{s}}} \right) 
	\left({2 y_{\ti{j}\ha{s}} \over y_{\ti{j}Q}}  
	+ {2 y_{\ti{k}\ha{s}} \over y_{\ti{k}Q}} \right)^{-1}\,. 
\eeq 
Then we find 
\beq 
\bsp 
& 
\cI_{2\cSCS{}{},1}^{(\LLone)}(x_{\wti{ir}},\Yt{ir}{j},\Yt{ir}{k},\Yt{j}{k};
	\ep,\al_{0},y_0,d_0,d'_0) = 
	-\left({-2^{2\ep}\ep \over 2\pi}\right) 4 \Yt{j}{k}
\\ &\qquad\times 
\int_{0}^{y_0} \rd{y}\, y^{1-2\ep}\,(1-y)^{d'_0-1+\ep}  
\\ &\qquad\times	 
	\int_{-1}^{1}\rd{(\cos\vth)}\, 
	\rd{(\cos\vph)}\, 
	\rd{(\cos\eta)}\, 
	(\sin\vth)^{-2\ep}\,(\sin\vph)^{-1-2\ep}\, 
	(\sin\eta)^{-2-2\ep}
\\ &\qquad\times	 
	\int_{0}^{\al_0} \rd{\al}\, \al^{-1-\ep} (1-\al)^{2d_0-3}  
	\{\al+(1-\al)[(1-y)x_{\wti{ir}} + y_{\wti{ir}\ha{s}}]\}^{-1-\ep} 
\\ &\qquad\times	 
	\int_{0}^{1} \rd{v}\, v^{-\ep} (1-v)^{-\ep}  
	\left({\al+(1-\al)[(1-y)x_{\wti{ir}} + y_{\wti{ir}\ha{s}}]v \over 
	2\al+(1-\al)[(1-y)x_{\wti{ir}} + y_{\wti{ir}\ha{s}}]}\right)^{\LLone}
\\ &\qquad\times	 
[(1-y)x_{\wti{ir}} + y_{\wti{ir}\ha{s}}] 
	\left(\COEFFJ {y_{\ti{j}Q} \over 2 y_{\ti{j}\ha{s}}}  
	+ \COEFFK {y_{\ti{k}Q} \over 2 y_{\ti{k}\ha{s}}} \right) 
	\left({2 y_{\ti{j}\ha{s}} \over y_{\ti{j}Q}}  
	+ {2 y_{\ti{k}\ha{s}} \over y_{\ti{k}Q}} \right)^{-1} 
\,,
\esp 
\label{eq:MI-CS-1} 
\eeq 
where $y_{\ti{j}\ha{s}}$, $y_{\ti{k}\ha{s}}$ and $y_{\wti{ir}\ha{s}}$  
are understood to be functions of the integration variables as given in  
\eqnss{eq:ytjhs}{eq:ytirhs}. We also used $y_{\wti{ir}Q} = x_{\wti{ir}}$, 
and performed the integration over $\eps_{\ha{s}}$ with the $\delta$-function 
in \eqn{eq:dPS2pshat}.  
The choice of frame in
\eqn{eq:CSirs-frame-1} is convenient for integrating  the first term in
the partial fraction, while the second term is more straightforward to 
integrate in a frame where $k$ and $j$ are exchanged. Upon performing
this exchange, we  find that the functional form of the master integral
is unchanged, and we must simply  interchange $\Yt{j}{ir}$ with
$\Yt{k}{ir}$ to obtain one term from the other.  

The integral $\cI_{2\cSCS{}{},1}^{(l)}$ as given in its most general
form  in \eqn{eq:MI-CS-1} first appears in electron-positron
annihilation into four or more jets at NNLO. For a three-jet
computation, the $\eta$ integral is trivial, since momentum 
conservation forces the three final state momenta to be coplanar, and
hence we find  $\phi_{ir}=0$ or $\pi$ in \eqn{eq:CSirs-frame-1}.
We display the resulting simplifications. Choosing
$\phi_{ir}=\pi$, \eqn{eq:CSirs-frame-1} becomes 
\beq
\bsp 
& \ti{p}_j^\mu = \ti{E}_j(1,\ldots,1)\,,\qquad 
	\ti{p}_k^\mu = \ti{E}_k(1,\ldots,\sin\chi_{k},\cos\chi_{k})\,, 
\\ & 
	\ti{p}_{ir}^\mu = \ti{E}_{ir}(1,\ldots,-\sin\chi_{ir},\cos\chi_{ir})\,,
\label{eq:CSirs-frame-3j}
\esp
\eeq 
which leads to 
\bal 
y_{\ti{j}\ha{s}}  
&=  
	{1\over 2}y_{\ti{j}Q} y  
	(1 - \cos\vth)\,, 
\label{eq:ytijhas}
\\ 
y_{\ti{k}\ha{s}}  
&=  
	{1\over 2}y_{\ti{k}Q} y  
	(1 - \sin\chi_{k} \sin\vth \cos\vph - \cos\chi_{k} \cos\vth)\,, 
\label{eq:ytikhas}
\\ 
y_{\wti{ir}\ha{s}}  
&=  
	{1\over 2}x_{\wti{ir}} y  
	(1 + \sin\chi_{ir} \sin\vth \cos\vph - \cos\chi_{ir} \cos\vth)\,,
\label{eq:tixi}
\eal 
with
$\cos\chi_{k} = \cos\chi (\Yt{j}{k})$ and
$\cos\chi_{ir} = \cos\chi (\Yt{j}{ir})$ as in \eqn{eq:a2YCSirs}.
Then the integration over $\eta$ can be performed using  
\beq 
\int_{-1}^{1} \rd(\cos\eta)\, (\sin\eta)^{-2-2\ep} = -\frac{2^{-2\ep}}{\ep} 
	\frac{\Gamma^2(1-\ep)}{\Gamma(1-2\ep)}\,.
\eeq 
Due to momentum conservation, only two out of the five kinematic variables 
in \eqnss{eq:ytijhas}{eq:tixi} are independent. E.g.,~in terms of $\Yt{j}{k}$ 
and $\Yt{j}{ir}$, we have
\bal 
y_{\ti{j}Q} 
&\equiv x_{\ti{j}} = 
	{(2\Yt{j}{k}-1)\sqrt{\Yt{j}{ir}(1-\Yt{j}{ir})} 
	+(2\Yt{j}{ir}-1)\sqrt{\Yt{j}{k}(1-\Yt{j}{k})} \over 
	\Yt{j}{k}\sqrt{\Yt{j}{ir}(1-\Yt{j}{ir})} 
	+\Yt{j}{ir}\sqrt{\Yt{j}{k}(1-\Yt{j}{k})}}\,, 
\\
y_{\ti{k}Q} 
&\equiv x_{\ti{k}} = 
	{\sqrt{\Yt{j}{ir}(1-\Yt{j}{ir})} \over 
	\Yt{j}{k}\sqrt{\Yt{j}{ir}(1-\Yt{j}{ir})} 
	+\Yt{j}{ir}\sqrt{\Yt{j}{k}(1-\Yt{j}{k})}}\,, 
\\
x_{\wti{ir}} 
&=  
	{\sqrt{\Yt{j}{k}(1-\Yt{j}{k})} \over 
	\Yt{j}{k}\sqrt{\Yt{j}{ir}(1-\Yt{j}{ir})} 
	+\Yt{j}{ir}\sqrt{\Yt{j}{k}(1-\Yt{j}{k})}}\,.
\eal 
Note that $\Yt{k}{ir}$ is also easily expressed with $\Yt{j}{k}$ 
and $\Yt{j}{ir}$,
\beq
\Yt{k}{ir} 
= 
        \Yt{j}{k} + \Yt{j}{ir} - 2 \Yt{j}{k} \Yt{j}{ir} 
        + 2 \sqrt{\Yt{j}{k} (1-\Yt{j}{k})  
          \Yt{j}{ir} (1-\Yt{j}{ir})}\,. 
\eeq 
The physical region in $\Yt{j}{k}$ and $\Yt{j}{ir}$ is simply given  
by $0<\Yt{j}{k},\Yt{j}{ir}<1$ and $\Yt{j}{k}+\Yt{j}{ir}>1$. 
 
In the case of two-jet production, this integral does not appear at all
because we do not have three independent hard final state momenta. 



\paragraph{Integrated soft collinear counterterm for $j=(ir)$.} 
This case leads to the integrals $\cI_{2\cSCS{}{},2}^{(l)}$ and
$\cI_{2\cSCS{}{},3}^{(l)}$. For $\cI_{2\cSCS{}{},2}^{(l)}$
the integrand again has a line singularity, which we can
remove via partial fractioning. Using
\beq 
\bsp 
& 
\frac{(1-\al) s_{\wti{ir}\ti{k}} + \al s_{\ti{k}Q}} 
{[(1-\al) s_{\wti{ir}\ha{s}} + \al s_{\ha{s}Q}] s_{\ti{k}\ha{s}}} =  
	\frac{1}{Q^2}\,4\frac{\al + (1-\al) x_{\wti{ir}} \Yt{ir}{k}}{y^2} 
\\&\qquad\times 
	\left(\COEFFIR {1 \over 2 \al + (1-\al) x_{\wti{ir}} 
	{2 y_{\wti{ir}\ha{s}} \over y x_{\wti{ir}}}} 
	+ \COEFFK {y y_{\ti{k}Q} \over 2 y_{\ti{k}\ha{s}}} \right) 
	\left(2 \al + (1-\al) x_{\wti{ir}} {2 y_{\wti{ir}\ha{s}} \over y x_{\wti{ir}}}  
	+ {2 y_{\ti{k}\ha{s}} \over y y_{\ti{k}Q}} \right)^{-1} 
\,,
\esp 
\eeq 
we find the following parametric integral representation,
\beq 
\bsp 
& 
\cI_{2\cSCS{}{},2}^{(\LLone)}
(x_{\wti{ir}},\Yt{ir}{k};\ep,\al_{0},y_0,d_0,d'_0) =
	-{\Gamma^{2}(1-\ep) \over 2\pi \Gamma(1-2\ep)} 4 \int_{0}^{y_0} \rd{y}\,  
	y^{-1-2\ep} (1-y)^{d'_0-1+\ep}  
\\ &\qquad\times 
	\int_{-1}^{1}\rd{(\cos\vth)}\, 
	\rd{(\cos\vph)}\, 
	(\sin\vth)^{-2\ep}\,(\sin\vph)^{-1-2\ep}\, 
	[(1-y)x_{\wti{ir}} + y_{\wti{ir}\ha{s}}] 
\\ &\qquad\times	 
	\int_{0}^{\al_0} \rd{\al}\, \al^{-1-\ep} (1-\al)^{2d_0-3}  
	\{\al+(1-\al)[(1-y)x_{\wti{ir}} + y_{\wti{ir}\ha{s}}]\}^{-1-\ep} 
\\ &\qquad\times	 
	\int_{0}^{1} \rd{v}\, v^{-\ep} (1-v)^{-\ep}  
	[\al + (1-\al) x_{\wti{ir}} \Yt{ir}{k}] 
	\left(\COEFFIR {1 \over 2 \al + (1-\al) x_{\wti{ir}} {2 y_{\wti{ir}\ha{s}} 
	\over y x_{\wti{ir}}}} 
	+ \COEFFK {y y_{\ti{k}Q} \over 2 y_{\ti{k}\ha{s}}} \right) 
\\ &\qquad\times		 
	\left(2 \al + (1-\al) x_{\wti{ir}} {2 y_{\wti{ir}\ha{s}} \over y x_{\wti{ir}}}  
	+ {2 y_{\ti{k}\ha{s}} \over y y_{\ti{k}Q}} \right)^{-1} 
	\left({\al+(1-\al)[(1-y)x_{\wti{ir}} + y_{\wti{ir}\ha{s}}]v \over 
	2\al+(1-\al)[(1-y)x_{\wti{ir}} + y_{\wti{ir}\ha{s}}]}\right)^{\LLone}
\,.
\esp 
\label{eq:MI-CS-2} 
\eeq 
The two terms in the
partial fraction are evaluated most conveniently in two different frames.
In the first one
\beq 
\ti{p}_{ir}^\mu = \ti{E}_{ir}(1,\ldots,1)\,,\qquad 
\ti{p}_k^\mu = \ti{E}_k(1,\ldots,\sin\chi_{k},\cos\chi_{k})\,, 
\label{eq:CSirsMI-2-frame-IR} 
\eeq 
and hence 
\bal 
y_{\wti{ir}\ha{s}}  
&=  
	{1\over 2}x_{\wti{ir}} y  
	(1 - \cos\vth)\,, 
\\
y_{\ti{k}\ha{s}}  
&=  
	{1\over 2}y_{\ti{k}Q} y  
	(1 - \sin\chi_{k} \sin\vth \cos\vph - \cos\chi_{k} \cos\vth)\,, 
\eal 
with $\cos\chi_{k} = \cos\chi(\Yt{ir}{k})$.
In this frame, the $y_{\wti{ir}\ha{s}}=0$ singularity is at the border of 
integration. Instead, in the second frame, the $y_{\wti{k}\ha{s}}=0$  
singularity is at the border,
\beq 
\ti{p}_{k}^\mu = \ti{E}_{k}(1,\ldots,1)\,,\qquad 
\ti{p}_{ir}^\mu = \ti{E}_{ir}(1,\ldots,\sin\chi_{ir},\cos\chi_{ir})\,, 
\label{eq:CSirsMI-2-frame-K} 
\eeq 
(note that clearly $\cos\chi_{ir}=\cos\chi_{k}$) and hence 
\bal 
y_{\ti{k}\ha{s}}  
&=  
	{1\over 2}y_{\ti{k}Q} y  
	(1 - \cos\vth)\,, 
\\
y_{\wti{ir}\ha{s}}  
&=  
	{1\over 2}x_{\wti{ir}} y  
	(1 - \sin\chi_{ir} \sin\vth \cos\vph - \cos\chi_{ir} \cos\vth)\,, 
\eal 
with $\cos\chi_{ir} = \cos\chi(\Yt{ir}{k})$.

As for $\cI_{2\cSCS{}{},3}^{(l)}$, it is most conveniently
evaluated in the  frame defined by \eqn{eq:CSirsMI-2-frame-IR}, 
where we can integrate over $\vph$ using 
\beq 
\frac{\Gamma^2(1-\ep)}{2\pi\Gamma(1-2\ep)}
\int_{-1}^{1}\rd (\cos\vph)\, (\sin\vph)^{-1-2\ep} = 
	2^{-1-2\ep}\,. 
\eeq 
Then we obtain the following explicit integral representation,
\beq 
\bsp 
& 
\cI_{2\cSCS{}{},3}^{(\LLone)}(x_{\wti{ir}};\ep,\al_{0},y_0,d_0,d'_0;l) = 
	-2^{2\ep} \int_{0}^{y_0} \rd{y}\, y^{-2\ep} (1-y)^{d'_0-2+\ep}  
	\int_{-1}^{1}\rd{(\cos\vth)}\, 
	(\sin\vth)^{-2\ep}\, 
\\ &\qquad\times 
	[(1-y)x_{\wti{ir}} + y_{\wti{ir}\ha{s}}] 
	\int_{0}^{\al_0} \rd{\al}\, \al^{-\ep} (1-\al)^{2d_0-3}  
	\{\al+(1-\al)[(1-y)x_{\wti{ir}} + y_{\wti{ir}\ha{s}}]\}^{-1-\ep} 
\\ &\qquad\times	 
	\int_{0}^{1} \rd{v}\, v^{-\ep} (1-v)^{-\ep}  
	{1 \over 2 \al + (1-\al) x_{\wti{ir}} {2 y_{\wti{ir}\ha{s}} \over y x_{\wti{ir}}}} 
	\left({\al+(1-\al)[(1-y)x_{\wti{ir}} + y_{\wti{ir}\ha{s}}]v \over 
	2\al+(1-\al)[(1-y)x_{\wti{ir}} + y_{\wti{ir}\ha{s}}]}\right)^{\LLone}\,. 
\esp 
\label{eq:MI-CS-3} 
\eeq 
%


 
\paragraph{Integrated triple collinear -- soft collinear counterterm.} 
For the integral $\cI_{2\cSCS{}{},4}^{(l)}$ we use the frame where
\beq 
\ti{p}_{ir}^\mu = \ti{E}_{ir}(1,\ldots,1)\,, 
\label{eq:CSirsMI-4-frame} 
\eeq 
and hence 
\bal 
y_{\wti{ir}\ha{s}}  
&=  
	{1\over 2}x_{\wti{ir}} y  
	(1 - \cos\vth)\,. 
\eal 
Then we obtain the following explicit representation,
\beq 
\bsp 
& 
\cI_{2\cSCS{}{},4}^{(\LLone)}(x_{\wti{ir}};\ep,\al_{0},y_0,d_0,d'_0) = 
	2^{1+2\ep} \int_{0}^{y_0} \rd{y}\,  
	y^{-1-2\ep} (1-y)^{d'_0-2+\ep}  
\\ &\qquad\times	 
	\int_{-1}^{1}\rd{(\cos\vth)}\, 
	(\sin\vth)^{-2\ep}\, 
	[(1-y)x_{\wti{ir}} + y_{\wti{ir}\ha{s}}] 
\\ &\qquad\times 
	\int_{0}^{\al_0} \rd{\al}\, \al^{-1-\ep} (1-\al)^{2d_0-3}  
	\{\al+(1-\al)[(1-y)x_{\wti{ir}} + y_{\wti{ir}\ha{s}}]\}^{-1-\ep} 
\\ &\qquad\times	 
	\int_{0}^{1} \rd{v}\, v^{-\ep} (1-v)^{-\ep}  
	{2\al + (1-\al)[(1-y)x_{\wti{ir}} + y_{\wti{ir}\ha{s}}] \over 
	2 \al + (1-\al) x_{\wti{ir}} {2 y_{\wti{ir}\ha{s}} \over y x_{\wti{ir}}}} 
\\ &\qquad\times	 
	\left({\al+(1-\al)[(1-y)x_{\wti{ir}} + y_{\wti{ir}\ha{s}}]v \over 
	2\al+(1-\al)[(1-y)x_{\wti{ir}} + y_{\wti{ir}\ha{s}}]}\right)^{\LLone}\,. 
\esp 
\label{eq:MI-CS-4} 
\eeq 
%


 
\paragraph{Integrated double collinear -- soft collinear counterterm.} 
Finally the integral $\cI_{2\cSCS{}{},5}^{(l)}$ is most conveniently  
evaluated in the frame where 
\beq 
\ti{p}_{j}^\mu = \ti{E}_{j}(1,\ldots,1)\,,\qquad 
\ti{p}_{ir}^\mu = \ti{E}_{ir}(1,\ldots,\sin\chi_{ir},\cos\chi_{ir})\,, 
\label{eq:CSirsMI-5} 
\eeq 
and hence 
\bal 
y_{\wti{j}\ha{s}}  
&=  
	{1\over 2}y_{\ti{j}Q} y  
	(1 - \cos\vth)\,, 
\\
y_{\wti{ir}\ha{s}}  
&=  
	{1\over 2}x_{\wti{ir}} y  
	(1 - \sin\chi_{ir} \sin\vth \cos\vph - \cos\chi_{ir} \cos\vth)\,, 
\eal 
with $\cos\chi_{ir} = \cos\chi(\Yt{ir}{j})$.
The explicit expression for the master integral reads 
\beq 
\bsp 
& 
\cI_{2\cSCS{}{},5}^{(\LLone)}
(x_{\wti{ir}},\Yt{ir}{j};\ep,\al_{0},y_0,d_0,d'_0) = 
	{\Gamma^{2}(1-\ep) \over 2\pi \Gamma(1-2\ep)} 4 \int_{0}^{y_0} \rd{y}\,  
	y^{-1-2\ep} (1-y)^{d'_0-2+\ep}  
\\ &\qquad\times 
	\int_{-1}^{1}\rd{(\cos\vth)}\, 
	\rd{(\cos\vph)}\, 
	(\sin\vth)^{-2\ep}\,(\sin\vph)^{-1-2\ep}\, 
	[(1-y)x_{\wti{ir}} + y_{\wti{ir}\ha{s}}] 
\\ &\qquad\times	 
	\int_{0}^{\al_0} \rd{\al}\, \al^{-1-\ep} (1-\al)^{2d_0-3}  
	\{\al+(1-\al)[(1-y)x_{\wti{ir}} + y_{\wti{ir}\ha{s}}]\}^{-1-\ep} 
\\ &\qquad\times	 
	\int_{0}^{1} \rd{v}\, v^{-\ep} (1-v)^{-\ep}  
	{y y_{\ti{j}Q} \over 2 y_{\ti{j}\ha{s}}} 
	\left(1 - y + {y_{\ti{j}\ha{s}} \over y_{\ti{j}Q}}\right) 
	\left({\al+(1-\al)[(1-y)x_{\wti{ir}} + y_{\wti{ir}\ha{s}}]v \over 
	2\al+(1-\al)[(1-y)x_{\wti{ir}} + y_{\wti{ir}\ha{s}}]}\right)^{\LLone}\,.
\esp 
\label{eq:MI-CS-5} 
\eeq 
 
The parametric forms for the integrals $\cI_{2\cSCS{}{},n}^{(l)}$
presented in this section are directly suitable for evaluation using 
sector decomposition, after introducing the new variables
\beq
z_1 = \frac{1-\cos\vth}{2}
\qquad\mbox{and}\qquad
z_2 = \frac{1-\cos\vph}{2}
\,.
\eeq

\end{appendix}



\providecommand{\href}[2]{#2}\begingroup\raggedright\endgroup


\end{document}